\documentclass[twocolumn, 
        showkeys, 
        jpb,
        superscriptaddress,
        floatfix,
        nofootinbib]{revtex4-2}% Math and Physical Sciences Numbered Reference Style

\usepackage[utf8]{inputenc}
\usepackage{amsmath}
\usepackage{graphicx}
\usepackage{amsfonts}
\usepackage{amssymb}
\usepackage{stmaryrd}
\usepackage[normalem]{ulem}
\usepackage{xcolor}
\usepackage{float}
\usepackage[colorlinks=true, allcolors=blue]{hyperref}
\usepackage[capitalise]{cleveref}
\usepackage[english]{babel}
\usepackage{csquotes}
\usepackage{blindtext}
\usepackage{comment}
\usepackage[letterpaper,top=2cm,bottom=2cm,left=2cm,right=2cm,marginparwidth=1.75cm]{geometry}
\usepackage{xr}
\usepackage{siunitx}
\usepackage{titletoc}

\usepackage{stix}

\usepackage{xr-hyper}
\usepackage{hyperref}

\newcommand{\nucleotide}{nt}

\newcommand{\minute}{\ensuremath{\mathrm{min}}} 
\newcommand{\hour}{\ensuremath{\mathrm{hour}}} 
\newcommand{\milli}{\ensuremath{\mathrm{m}}}  
\newcommand{\meter}{\ensuremath{\mathrm{m}}}  
\newcommand{\micro}{\ensuremath{\mu}}  
\newcommand{\molar}{\ensuremath{\mathrm{M}}}  
\newcommand{\nano}{\ensuremath{\mathrm{n}}}  
\newcommand{\calorie}{\ensuremath{\mathrm{cal}}}  
\newcommand{\kilo}{\ensuremath{\mathrm{kilo}}}    
\newcommand{\liter}{\ensuremath{\mathrm{L}}}     
\newcommand{\second}{\ensuremath{\mathrm{s}}}    
\newcommand{\per}{\ensuremath{\,/\,}}     % Spaced slash for "per"
\newcommand{\mole}{\ensuremath{\mathrm{mol}}}  % Upright "mol" unit
\newcommand{\rcf}{\ensuremath{\times\, g}}    
\newcommand{\ampere}{\ensuremath{\mathrm{A}}}    
\newcommand{\volt}{\ensuremath{\mathrm{V}}}    
\newcommand{\percent}{\ensuremath{\,\%}}     

\newcommand{\eced}{\ensuremath{\epsilon_c / \epsilon_d}}

\newcommand{\degreeCelsius}{\ensuremath{^\circ\mathrm{C}}}

\usepackage{xcolor}
\definecolor{CNRSMediumBlue}{RGB}{83,148,184}

\begin{document} 

%%%%%%%%%%%%%%%% TITLE AND AUTHORS %%%%%%%%%%%%%%%%
\title{
	Topological Defect Engineering enables Size and Shape Control in Self-Assembly
}

\author{Lara Koehler}
\email{These authors contributed equally to this work}
\affiliation{Université Paris-Saclay, CNRS, LPTMS, 91405, Orsay, France}
\affiliation{Max Planck Institute for the Physics of Complex Systems, Dresden, Germany}
\author{Markus Eder}
\email{These authors contributed equally to this work}
\affiliation{Department of Bioscience, School of Natural Sciences, Technical University of Munich, Garching, Germany}
\author{Vincent Ouazan-Reboul}
\affiliation{Université Paris-Saclay, CNRS, LPTMS, 91405, Orsay, France}
\author{Jude Ann Vishnu}
\affiliation{Université Paris-Saclay, CNRS, LPTMS, 91405, Orsay, France}
\author{Christoph Karfusehr}
\affiliation{Department of Bioscience, School of Natural Sciences, Technical University of Munich, Garching, Germany}
\affiliation{Max Planck School Matter to Life, Jahnstraße 29, Heidelberg, D-69120, Germany}
\author{Alexander Hebel}
\affiliation{Department of Bioscience, School of Natural Sciences, Technical University of Munich, Garching, Germany}
\author{Andrey Zelenskiy}
\affiliation{Université Paris-Saclay, CNRS, LPTMS, 91405, Orsay, France}
\author{Pierre Ronceray}
\affiliation{Aix Marseille Universit\'e, CNRS, CINAM, Turing Center for Living Systems, 13288 Marseille, France}
\author{Friedrich C. Simmel}
\affiliation{Department of Bioscience, School of Natural Sciences, Technical University of Munich, Garching, Germany}
\affiliation{Max Planck School Matter to Life, Jahnstraße 29, Heidelberg, D-69120, Germany}
\author{Martin Lenz}
\email{martin.lenz@universite-paris-saclay.fr}
\affiliation{Université Paris-Saclay, CNRS, LPTMS, 91405, Orsay, France}
\affiliation{PMMH, CNRS, ESPCI Paris, PSL University, Sorbonne Université, Université Paris-Cit\'e, F-75005, Paris, France}
%%%%%%%%%%%%%%%%% END OF PREAMBLE %%%%%%%%%%%%%%%%

%%%%%%%%%%%%%%%% START OF MAIN TEXT %%%%%%%%%%%%%%%

% Insert the title and author list
\maketitle

% Abstract, in bold
% There are strict length limits, and not all formats have abstracts.
% Consult the journal instructions to authors for details.
% Do not cite any references in the abstract.
%\begin{abstract} \bfseries \boldmath
%The self-assembly of complex structures from engineered subunits is a major goal of nanotechnology, but controlling their size becomes increasingly difficult in larger assemblies.  
%\end{abstract}

\textbf{
Equilibrium self-assembly is a powerful way to build nano- and microscale structures out of interacting subunits~\cite{glotzer2007anisotropy,pinheiro2011challenges}. The size and shape of such structures must be controlled in many biological and technological functions~\cite{caspar1962physical,fu2013single}, posing significant practical challenges as current strategies require multiple subunit types or the precise control of their shape and mechanics~\cite{hagan2021equilibrium}.
Here we introduce an alternative approach that circumvents these obstacles. Our method uses subunits whose interactions promote crystals, but also favor crystalline defects.
We show theoretically that the magnitude of these interactions, which is well controlled in experiments~\cite{SantaLucia:1998uz,Zadeh:2011aa}, governs the self-assembly through topological restrictions on the scope of the defects.
Using DNA origami, we demonstrate both size and shape control in two-dimensional disk- and fiber-like assemblies. Our basic concept of defect engineering operates well beyond these examples, and provides a broadly applicable framework to control self-assembly.
}
\smallskip

Size-controlled assembly can be achieved using DNA origami~\cite{rothemund2006folding, Douglas:2009dd, dey2021origamireview} and DNA tiles at the sub-micrometer scale~\cite{Wei:2012,Ke:2012}, as well as larger colloidal systems interacting through attached DNA strands~\cite{Park:2008kx,Nykypanchuk:2008cp,Mirkin:2023aa} or through shape recognition mediated by depletion interactions~\cite{Biancaniello:2005,McMullen:2022,sacanna2010lock,Tigges:2016}.
%Forming large assemblies requires pushing these techniques to their limits.
In a first approach known as programmable assembly, a set of several distinct subunits bind with one another according to a predefined pattern of interactions~\cite{Lund:2005aa,Liu:2005aa,tikhomirov2017fractal,evans2017physical}. The outer surface of a complete assembly is designed not to interact with any remaining subunits, leading to size control. The large numbers of distinct subunits involved can however lead to high costs and low yields~\cite{huntley2016information,Jacobs:2025}.
A second strategy, self-closing assembly, uses identical subunits that bind at an angle. A collection of many such subunits forms a ring, a cylinder or a sphere whose closure terminates the assembly process~\cite{wagenbauer2017gigadalton,Sigl:2021,videbaek2024economical,hayakawa2022geometrically,karfusehr2024cellcontainer}. As a result, any flexibility or small uncertainty in the subunit shapes can lead to the formation of diverse, potentially off-target structures~\cite{Fotin:2004,hayakawa2022geometrically}.
A last pathway relies on deformable subunits whose shapes change upon binding to one another. In properly designed systems, these deformations increase as more and more subunits are added to the assembly, which eventually inhibits further growth~\cite{lenz2017geometrical,hall2023building,hackney2023dispersed,berengut2020self}.
Deformable subunits are unfortunately difficult to manufacture, and their elastic properties must be tightly controlled to avoid the formation of infinite assemblies.

\begin{figure}[b]% Do NOT use \begin{figure*}
    \centering
    \includegraphics[width=\linewidth]{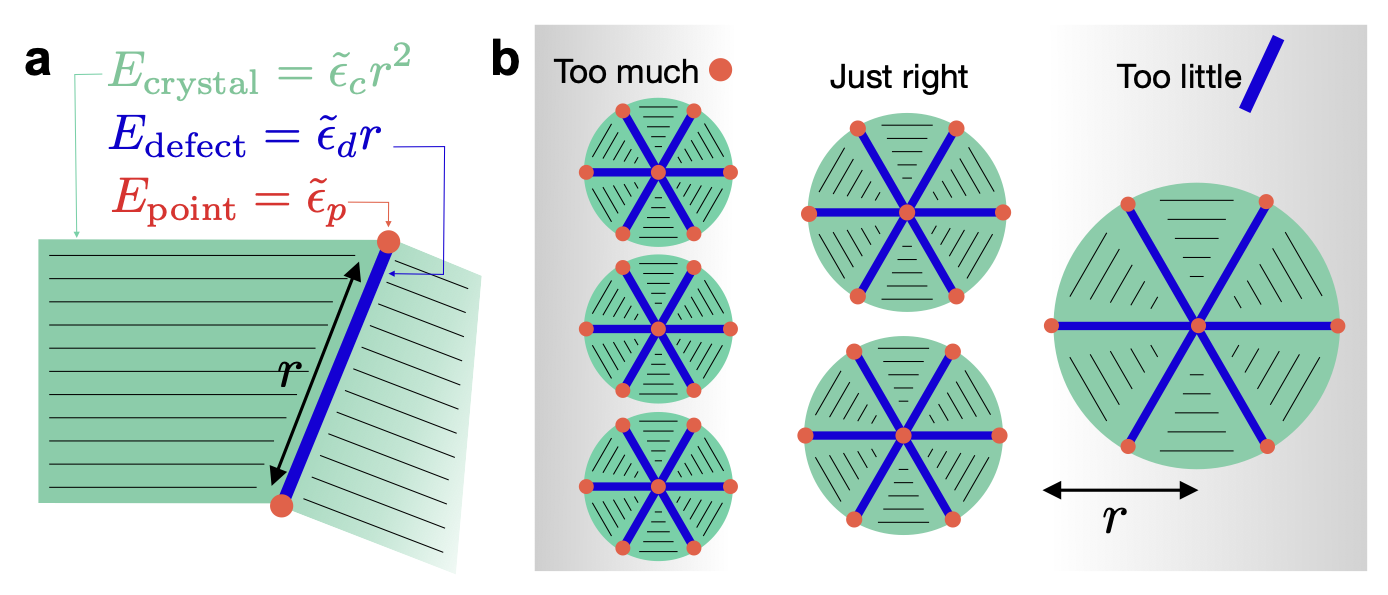}
    \caption{
    \textbf{Crystals with energetically favorable defects may not grow to arbitrarily large sizes.}\\
    \small
    (\textbf{a})~Our proposed subunits form crystals with binding free energy $\tilde{\epsilon}_c <0$, but also favor defective grain boundaries with $\tilde{\epsilon}_d < \tilde{\epsilon}_c$. Singular points at the end of these defects generically incur a cost $\tilde{\epsilon}_p>0$.
    Denoting the typical size scale of the assembly by $r$, the associated free energies respectively scale as $\tilde{\epsilon}_c r^2$, $\tilde{\epsilon}_d r^1$ and $\tilde{\epsilon}_p r^0$.
    (\textbf{b})~Because these quantities scale differently with $r$, small assemblies are dominated by energetically costly singular points and large ones by crystalline interactions. The favorable defect interactions may dominate in assemblies with intermediate sizes, making them the most stable. Expressing $r$ in units of the subunit size, this qualitative result is recovered by minimizing the free energy per subunit $\approx -|\tilde{\epsilon}_c|-|\tilde{\epsilon}_d|/r+|\tilde{\epsilon}_p|/r^2$, yielding a preferred radius $r^*\approx|\tilde{\epsilon}_p/\tilde{\epsilon}_d|$.    }
    \label{fig:intro_concept}
\end{figure}

Our proposed route to size- and shape-controlled assembly bypasses these limitations by using a single type of lattice-forming subunits whose shape or elasticity need not be finely controlled. Instead, the regulation of the assembly size and shape relies on the competition between two binding energies, which in DNA-based systems can be controlled to a high precision. The two energies favor two incompatible structures, resulting in geometrical frustration. While frustration is known to shape the properties of systems as diverse as structural glasses, supercooled liquids, metallic alloys, block copolymers, magnetic systems, surfactants and liquid crystals~\cite{Sadoc:2008,Berthier:2011a}, its application to self-assembly is relatively recent and typically involves deformable or shape-incompatible subunits~\cite{Grason:2016}. By contrast, here we harness a type of frustration akin to that found in spin glasses~\cite{Anderson:1978}.
As illustrated in Fig.~\ref{fig:intro_concept}a, the first of our two competing energies is associated with subunits organizing into a crystalline lattice. The second describes defects in the crystal, specifically grain boundaries between domains with different orientations. Our strategy requires that unlike in usual crystals, the defect energy be more favorable than that of the crystal. As illustrated in the simple design of Fig.~\ref{fig:intro_concept}b, the resulting competition between localized and bulk contributions to the energy results in the emergence of an optimal assembly size.

%%%%%%%%%%%%%%% SECTION 1 %%%%%%%%%%%%%%%%%%%%%%
%\subsection*{Theory of Defect-Induced Size Control}
\vspace{0.5cm}
\noindent \textbf{Theory of defect-induced size control}

\begin{figure*}
    \centering
    \includegraphics[width=\linewidth]{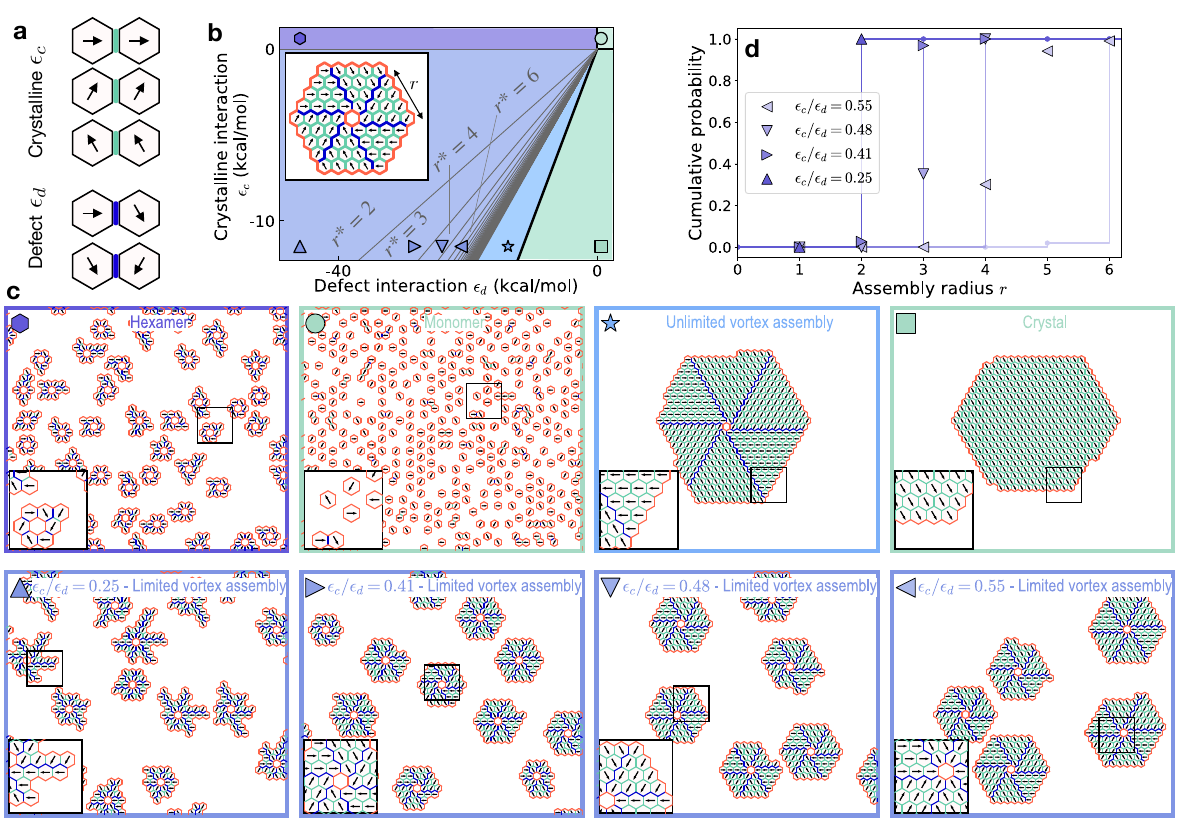}
    \caption{
    \textbf{Modeling demonstrates size control in randomly assembling subunits.}\\
    \small
    (\textbf{a})~We consider hexagonal subunits with anisotropic two-body interactions visualized by \emph{colored patches}.  All relative binding orientations not pictured in this panel are strongly penalized in our model.
	(\textbf{b})~Analytical phase diagram derived from Eqs.~\eqref{eq:cam_and_crystal_energies}. \emph{Colored symbols (triangle, square, hexagon, star, circle)} refer to the markings on the top-left of each snapshot of panel c. \emph{Inset:} vortex assembly structure considered in our analytical calculations.
	(\textbf{c})~Final structures of simulated assemblies for the parameter values marked by \emph{symbols} in panel b. Long protrusions can form at the corner of the smallest assemblies, but are increasingly penalized for smaller values of $\epsilon_c/\epsilon_d$ as discussed in the Supplementary Information. See Table~S3 for detailed parameters.
	(\textbf{d})~Distributions of final assembly sizes at the end of the simulations (\emph{symbols})~\cite{methods} and comparison with an ideal-gas-of-cluster theory applied to Eq.~\eqref{eq:camembert_energy} (\emph{solid lines}, see Supplementary Information). The moderate deviations observed for the largest assemblies are likely due to equilibrium entropic effects as well as nucleation kinetics (Fig.~S10).
    }
    \label{fig:theory}
\end{figure*}

\noindent To formalize our proposal, we consider a set of identical hexagonal subunits that can bind to their neighbors through the short-range orientation-dependent interaction rules illustrated in Fig.~\ref{fig:theory}a. The subunits bind through a crystalline or a defect interaction with binding (free) energies $\epsilon_c$ and $\epsilon_d$. Consistent with the schematics of Fig.~\ref{fig:intro_concept}b, topological constraints limit the fraction of defect interactions in large assemblies (Supplementary Sec.~V).

To determine the most favorable assembly structure, we express the energy of two idealized cases. We thus consider a piece of crystal in the shape of a regular hexagon (which optimizes the crystalline energy), and a six-fold symmetric assembly whose defects impose a vortex-like pattern of subunit orientations (inset of Fig.~\ref{fig:theory}b).  To minimize the total energy of a fixed number of subunits, we equivalently minimize the average energy per subunit $e$. Denoting the assembly radius by $r$, this quantity reads (see Supplementary Sec.~I)
\begin{subequations}\label{eq:cam_and_crystal_energies}
\begin{align}
e^{\mathrm{(crystal)}}(r) &= \frac{ 9 r^2+ 3r}{3r(r+1)+1}\epsilon_c \label{eq:crystal_energy}\\
e^{\mathrm{(vortex)}}(r) &= \frac{3\epsilon_c r^2 + (-3\epsilon_c+4\epsilon_d)r-2\epsilon_d}{r(r+1)}. \label{eq:camembert_energy}
\end{align}
\end{subequations}
Here the crystalline case includes isolated monomeric subunits as its $r=0$ case, while the $r=1$ vortex assembly is the largest possible structure that includes only defect interactions, namely a ring-like hexamer. For large ($r\gg 1$) assemblies, Eq.~\eqref{eq:camembert_energy} recapitulates the discussion in the caption of Fig.~\ref{fig:intro_concept} with $\tilde{\epsilon}_c=3\epsilon_c$,
$\tilde{\epsilon}_d=4\epsilon_d-6\epsilon_c$ and $\tilde{\epsilon}_p=6(\epsilon_c-\epsilon_d)$.

The most favorable assembly as a function of the binding energies is shown in the phase diagram of Fig.~\ref{fig:theory}b. Its nature depends only on the ratio and signs of $\epsilon_c$ and $\epsilon_d$. 

Unlike the crystal energy of Eq.~\eqref{eq:crystal_energy}, Eq.~\eqref{eq:camembert_energy} can display a minimum at a finite radius $r^*$. The resulting finite-size vortex assemblies dominate the left-hand-side of the phase diagram. For attractive crystalline interactions ($\epsilon_c<0$), the size of these assemblies diverges as the system approaches the $\epsilon_c=2\epsilon_d/3$ line. This suggests that any assembly size can be energetically favored under suitable conditions, which could enable size control over arbitrarily large assemblies.

To test the robustness of our conclusions against thermal fluctuations and assess the possible dominance of structures beyond our crystal and ideal vortex assembly, we numerically simulate the self-assembly process. We run a lattice Monte Carlo simulated annealing down to a fixed finite temperature~\cite{methods}, 
and successfully recover all predicted structures in the expected range of parameters (Fig.~\ref{fig:theory}c). The assembly size distribution is consistent with a straightforward finite-temperature extension of our theory (Fig.~\ref{fig:theory}d). Vortex assemblies also form in off-lattice simulations and in the absence of simulated annealing, and they are robust to the presence of heterogeneous or faulty interactions (Supplementary Sec.~II, Supplementary Movies~SM4-SM7).

%%%%%%%%%%%%%%% SECTION 2 %%%%%%%%%%%%%%%%%%%%%%
%\subsection*{DNA Origami Implementation}
\vspace{0.5cm}
\noindent \textbf{DNA origami implementation}

\begin{figure*}
    \centering
    \includegraphics[width=\linewidth]{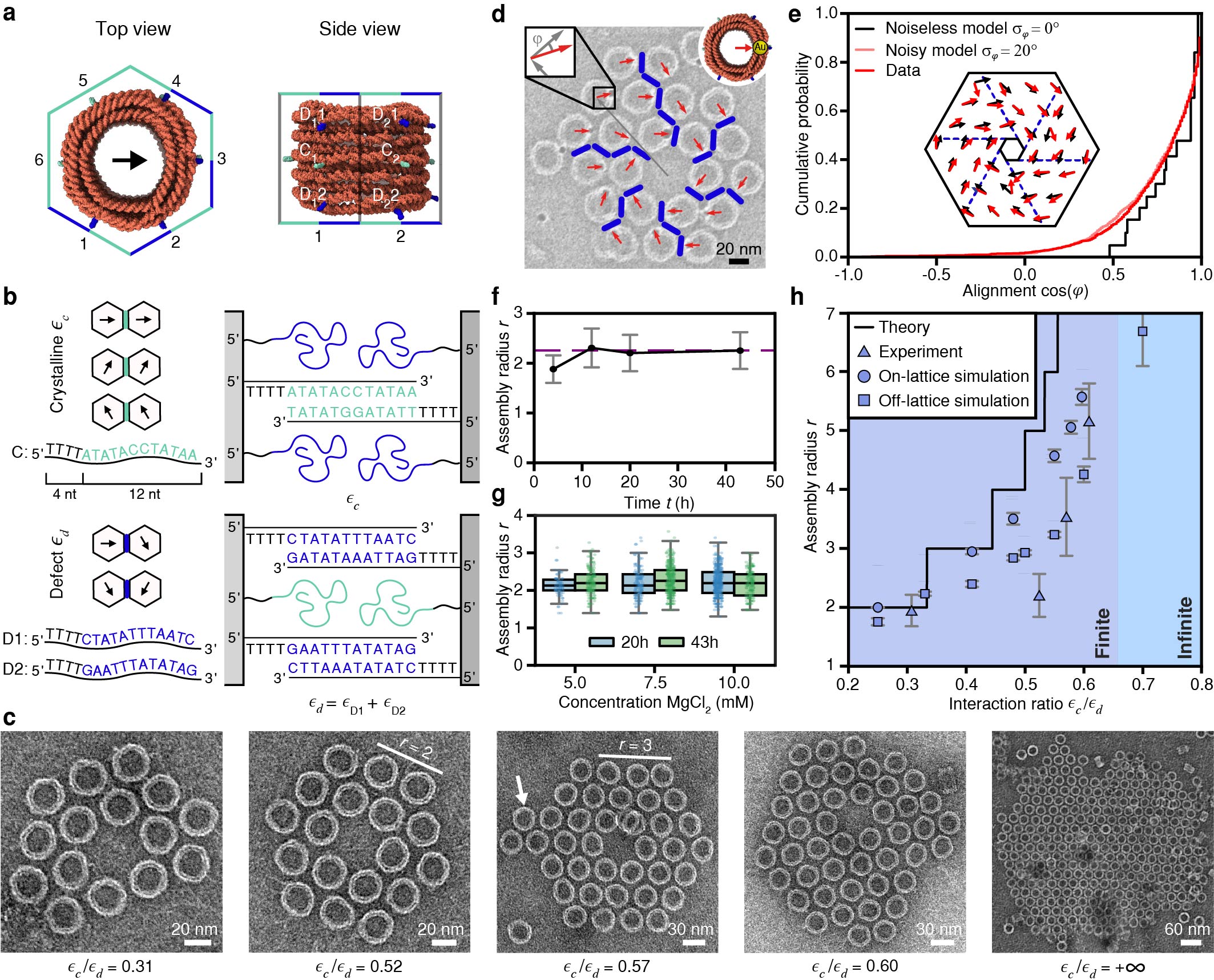}
    \caption{
    \textbf{DNA origami nanocylinders form size-controlled assemblies.}\\
    \small
	(\textbf{a})~OxDNA molecular model of our nanocylinders. \emph{Light} (\emph{dark}) \emph{blue} linkers provide crystalline (defect) interactions. Sides~5 and 6 never form defect interactions and do not carry defect linkers (\emph{all-green sides of the hexagonal box}).
    (\textbf{b})~\emph{Top row: }In a crystalline interaction, complementary crystalline linkers on opposite nanocylinder sides (\emph{e.g.}, 2 and 5) hybridize. Defect linkers remain unhybridized (\emph{wavy blue lines}). \emph{Bottom row: }In a defect interaction (\emph{e.g.}, between sides 2 and 4) two pairs of defect linkers hybridize while crystalline linkers are unhybridized (\emph{wavy green lines}).
    (\textbf{c})~TEM micrographs of the assemblies formed by our nanocylinders illustrate the dependence of their radius on $\epsilon_c/\epsilon_d$. \emph{White arrow:} a defect-interaction-induced appendage reminiscent of our simulations (Fig.~\ref{fig:theory}c; more examples in Fig.~S15b)
    (\textbf{d})~Labeling of side~3 with gold nanoparticles reveals subunit orientations consistent with our vortex assembly design (\emph{red arrows}; see Fig.~S15 for the original micrograph). Apparent defect interaction sites are outlined in \emph{dark blue} ($\epsilon_c/\epsilon_d= 0.51$).
	(\textbf{e})~The empirical distribution of orientations closely resembles an ideal $r=3$ vortex assembly structure with noise added ($\epsilon_c/\epsilon_d= 0.51$).
	(\textbf{f})~The assembly size distribution is robust to variations in incubation time. Here $[\text{MgCl}_2]=5\,{\milli\molar}$, $\epsilon_c/\epsilon_d= 0.51$.
	(\textbf{g})~The assembly size distribution is robust to variations in salt concentration ($\epsilon_c/\epsilon_d= 0.51$).
	(\textbf{h}).~The assembly radius depends on the binding free energies as predicted by theory and observed in both on- and off-lattice simulations.
	See~\cite{methods} for statistical information.
    \label{fig:exp1}
    }
\end{figure*}

\noindent We implement our proposed six-fold subunit design using hollow DNA origami nanocylinders~\cite{wickham2020barrel}. Our subunits interact through a six-fold pattern of 16 nucleotide (nt)  long single-stranded DNA linkers each comprising a $4\,${\nucleotide} oligo\-thymidine anchor domain at the 5' end for flexibility, followed by a $12\,${\nucleotide} sticky domain designed to hybridize with its complementary sequence through base-pairing (Fig.~\ref{fig:exp1}a).
These sticky domains can be modified without redesigning the nanocylinder itself, allowing us to easily and quickly alter the strength and geometry of our interactions.
Their length and sequence enable fine control over the hybridization free energy (Supplementary Sec.~IV\,A) and ensure that it is associated with a melting temperature within the experimentally accessible range $25^\circ {\rm C} \leq T\leq 40^\circ {\rm C}$. The linkers are organized in three vertical layers along the height of the cylinder. The middle layer comprises one linker in each of the six binding directions and serves to implement a crystalline interaction $\epsilon_c$ ranging from $-13$ to $-9\,\text{kcal}\cdot\text{mol}^{-1}$ (approximately $-21$ to $-15k_BT$ per bond with $T=36^\circ \rm C$; see Fig.~\ref{fig:exp1}b and~\cite{methods}). We implement each defect interaction using two pairs of linkers from the top and bottom layer. This allows us to probe the range $0.31\leqslant \epsilon_c/\epsilon_d\leqslant 0.61$ relevant for size-controlled assembly while retaining experimentally accessible melting temperatures.

We assemble our nanocylinders by incubating them at constant temperature $T=36^\circ \rm C$ for 20 hours in a buffer containing $[\text{MgCl}_{2}]=5\,\text{mM}$, resulting in structures highly reminiscent of our theoretical predictions~(Fig.~\ref{fig:exp1}c and Fig.~S7). %\ref{fig:gallery}). 
We further investigate the orientations of individual nanocylinders by labeling them at a single position with a $5\,\text{nm}$ DNA-grafted gold nanoparticle. We find that they are consistent with our predicted vortex assemblies (Fig.~\ref{fig:exp1}d) and that adding an orientational noise with standard deviation $\sigma_\varphi=20^\circ$ to the ideal orientations yields almost perfect agreement with the data (Fig.~\ref{fig:exp1}e, Kolmogorov-Smirnov test p-value $p= 0.65$). Taken together, these results demonstrate that our DNA nanocylinders indeed form our predicted vortex assemblies.

We next quantitatively examine our size control mechanism by studying the distribution of assembly sizes~\cite{methods}. In Fig.~\ref{fig:exp1}f we find that it does not markedly depend on the incubation time after 10 hours, suggesting that they are at equilibrium. We additionally vary the MgCl$_2$ concentration in our sample based on the fact that an increased salt concentration increases non-specific interactions in DNA origami (Fig.~\ref{fig:exp1}g). This does not lead to a significant change of our assembly sizes, indicating that our structures are primarily based on specific hybridization interactions. Finally, Fig.~\ref{fig:exp1}h demonstrates that consistent with our theoretical predictions, the interaction ratio $\epsilon_c/\epsilon_d$ controls the size of assemblies with mean radii ranging from 1.9 to 5.2. This corresponds to a number of subunits ranging from 17 to 96, implying a control of the assembly mass over an almost 6-fold range. Our largest assemblies' subunit count thus outperforms the state of the art in artificial single-subunit, self-limited assemblies by more than 60\% (see Table~S2 for detailed comparisons).

%%%%%%%%%%%%%%%%%%% SECTION 3 %%%%%%%%%%%%%%%%%%%%%%%%%%%%%%%%%%%%%%%%%%%%%

\newpage
%\subsection*{Shape Control through Defect Engineering}
\vspace{0.5cm}
\noindent \textbf{Shape control through defect engineering}

\noindent Beyond the design and control of vortex assemblies, defect engineering is a general and versatile self-assembly strategy. While its full potential remains to be explored, here we demonstrate additional possible designs and point out some of the associated challenges. We also show in the Supplementary Information that it can be harnessed to address many technical limitations of the simple yet relatively fragile vortex assembly, from its shape fluctuations to the sensitivity of its radius to the interaction parameters.

\begin{figure*}
    \centering
	\includegraphics[width=.99\linewidth]{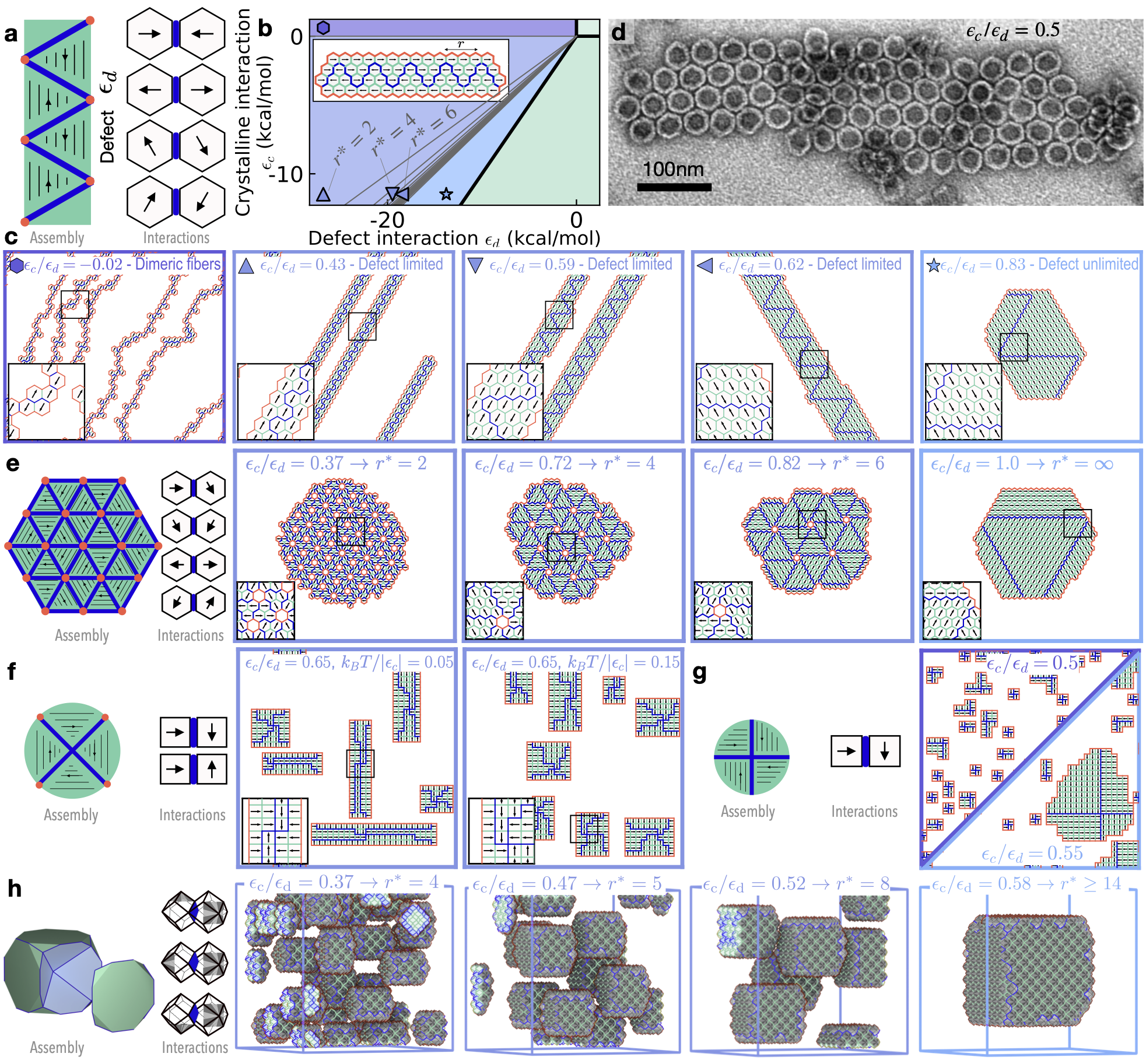}
    \caption{
    \textbf{Defect engineering provides a rich design space.}\\
    \small
    Here subunits with identical orientations form crystalline interactions with energy $\epsilon_c$ throughout.
    (\textbf{a})~Defect interactions and conceptual schematic for fiber-like assemblies.
    (\textbf{b})~Corresponding phase diagram derived from Eqs.~\eqref{eq:crystal_energy} and \eqref{eq:fiber}. It differs from Fig.~\ref{fig:theory}b only through the slopes of the diagonal lines. \emph{Inset:} Structure of an ideal fiber.
    (\textbf{c})~Simulations conducted as in Fig.~\ref{fig:theory}c.
    (\textbf{d}) TEM micrograph of a fiber formed as in Fig.~\ref{fig:exp1} but with the new defect interactions. 
    (\textbf{e})~Patterned bulk design.
	(\textbf{f})~Rectangular size-limited design where entropic effects control the aspect ratio of the assembly.
    (\textbf{g})~A non-size-controlled design.
	(\textbf{h})~Three-dimensional size control. The expected assembly is a solid (non-hollow) truncated cube (shown here as an exploded view). The interacting subunits are rhombic dodecahedra whose orientation is indicated by a \emph{dark grey arrowhead}. These arrows point away from a hedgehog topological defect at the center of the assembly (see Fig.~S3 for more details). The simulation results show disjoint, size-controlled assemblies (see Movie~SM1 for a 3D visualization).
	See Table~S3 for simulation parameters and the Supplementary Information for the predictions of the optimal assembly sizes $r^*$.}
    \label{fig:shape}
\end{figure*}

In our vortex design, defect interactions cause the subunit orientation to rotate clockwise when crossing a defect line~(Fig.~\ref{fig:theory}a). In Fig.~\ref{fig:shape}a we instead consider defect interactions that reverse the arrow orientation. This again favors triangle-shaped crystalline domains, but as shown in the inset of Fig.~\ref{fig:shape}b and Fig.~S1e these form an alternating fiber-like pattern with an energy per subunit
\begin{equation}\label{eq:fiber}
    e^\mathrm{(fiber)}(r)=\frac{3\epsilon_c r^2 + (4\epsilon_d-3\epsilon_c)r+2(\epsilon_c-\epsilon_d)}{r(r+1)}.
\end{equation}
As in the vortex case, this energy can display a minimum for a finite size $r$ of the triangular domains. Figure~\ref{fig:shape}b shows the resulting phase diagram, which we further validate through numerical simulations in Fig.~\ref{fig:shape}c. Finally, we implement this design using our six-fold DNA nanocylinder platform, and indeed observe the formation of fiber-like assemblies (Fig.~\ref{fig:shape}d and S20).%\ref{fig:gallery}).

Further reassignments of the defect interactions allow for multiple other size-controlled assemblies. The design of Fig.~\ref{fig:shape}e thus rearranges our triangular crystalline domains into a patterned bulk
whose superlattice spacing is controlled by the interaction ratio $\epsilon_c/\epsilon_d$, similar to the vortex assembly radius. Its structure is significantly more robust than that of the vortex and makes it easier to form large domains, for two main reasons. First, it is much less susceptible to stochastic nucleation events, as new topological defects can be templated onto existing assemblies. Second, the size of its triangular crystalline domains is highly constrained by several layers of neighboring domains, resulting in a dramatic collective suppression of their size fluctuations (Supplementary Sec.~IV). The regularly spaced holes resulting from such periodic designs could be functionalized to template photonic materials with large lattice spacings.
A variant of the vortex assembly of Fig.~\ref{fig:theory} based on square subunits displays size control but features flexible defect lines (Fig.~\ref{fig:shape}f). These lines form diagonal paths on a square lattice, implying that their length depends solely on their total horizontal and vertical amplitude. This results in a high degeneracy, whereby all directed defect lines that start and end at the corners of a rectangular assembly yield the same overall energy. When binding energies are low enough to be comparable to the thermal energy $k_BT$, the entropic contribution from this conformational freedom dominates. As shown in Fig.~\ref{fig:shape}f, increasing $k_BT/\epsilon_c$ while keeping $\epsilon_c/\epsilon_d$ fixed thus favors squares over elongated rectangular assemblies (Supplementary Information and Fig.~S2d-f).
An alternative square design shown in Fig.~\ref{fig:shape}g does not lead to size control and serves as a cautionary example of the subtleties of defect engineering. In this case the local geometry of the interactions implies a vanishing $\tilde{\epsilon}_p$ in the language of Fig.~\ref{fig:intro_concept}, and thus a function $e(r)$ that does not have a minimum at finite $r$. The condition $\tilde{\epsilon}_p>0$, while necessary, is not sufficient to enable size control. For instance, a three-fold vortex design gives rise to alternate structures that are more stable than the intended vortex assembly (Fig.~S2h). 

Figure~\ref{fig:shape}h demonstrates defect engineering in three dimensions. The assembly is comprised of six crystalline domains, each in the rough shape of a square-based pyramid. These domains come into contact through planar grain boundaries to form a truncated cube. Individual subunits are rhombic dodecahedra with $D_4$ dihedral symmetry about an axis joining two of their vertices, and can form three non-equivalent defect interactions as illustrated. Numerical simulations demonstrate successful size control through the interaction ratio $\epsilon_c / \epsilon_d$ (see also Supplementary Information and Fig.~S3d-e). This design has six crystalline domains, just like the simple vortex, but the size of each is coupled to four neighbors instead of two. This results in reduced shape and size fluctuations compared to the vortex, similar to the case of Fig.~\ref{fig:shape}e. These designs show that defect engineering is much broader than the simple example of the vortex assembly suggests, and enables the production of much more robust structures.

%\section*{Discussion}
\vspace{0.5cm}
\noindent \textbf{Discussion}

\noindent The induction of energetically costly grain boundaries has been used to design materials for centuries~\cite{Williams:2012}. Here we propose to harness modern self-assembly techniques to instead make them energetically favorable. We find that this approach results in an energetic situation that resembles classical nucleation theory in reverse. In classical nucleation, an unfavorable boundary energy works alongside a favorable bulk energy to make assemblies of intermediate sizes, the so-called critical nuclei, unstable~\cite{oxtoby1992homogeneous}. In our approach, the boundary energy stems not from the outer surface of the assembly, but from the grain boundaries within. Because it is favorable, it produces stable, finite assemblies of arbitrarily large sizes and controllable morphologies in two and three dimensions.

Our strategy plays to the strengths of DNA nanotechnologies developed over the past decades, and could be implemented in other platforms where directional, reversible interactions are controlled quantitatively \cite{Rossi:2011,Mayarani:2024}. In contrast with other size control schemes~\cite{hagan2021equilibrium}, ours can be implemented with a single subunit type and does not require fine control over subunit shape and elasticity. As in other designs however, controlling the morphology of large assemblies becomes increasingly difficult as their size $r$ increases, as the minimum in the specific energy $e(r)$ becomes more and more shallow. Under such conditions the assembly nucleation kinetics plays a non-negligible role in determining the final size of our simplest vortex assemblies, although other designs circumvent this limitation.
This issue is less likely to influence our DNA origami implementation, which benefits from comparatively long equilibration times. Instead, the slight shift between the corresponding experimental data and our predictions may be due to our relatively crude estimate of the subunit binding free energy. This estimate could be refined by taking into account other (\emph{e.g.}, entropic) effects beyond base pairing. 

A crucial requirement for our approach is the ability to favor only a few specific grain boundary interactions, resulting in topological restrictions that prevent their proliferation throughout the bulk. In our hexagonal subunit design, this is achieved by singling out two to four favorable interactions out of twenty-one~\cite{Koehler:2024} possible defect-inducing options. The implied number of unexplored choices hints at an enormous untapped design freedom. This freedom is further broadened when considering three-dimensional subunits (for instance, rhombic dodecahedra enable $\approx 10^{67}$ distinct defect interactions patterns -- see Supplementary Information), the possibility of assigning distinct energies to different defect interactions, or an independent control of the point energy $\tilde{\epsilon}_p$ through three-body or steric interactions between subunits.
In the Supplementary Information, we show that this freedom can be harnessed to mitigate many technical limitations of the basic designs presented above. Beyond these improvements, exploring the enormous defect engineering design space and understanding its structuring principles represents an exciting experimental and theoretical challenge, and could open new avenues in nanotechnology.

%%%%%%%%%%%%%%%% REFERENCES %%%%%%%%%%%%%%%

%\clearpage % Clear all remaining figures and tables then start a new page

% The list of references goes after the main text and before the acknowledgements
% When preparing an initial submission, we recommend you use BibTeX, like this:
%
%\bibliographystyle{plainnat}  % or another natbib-compatible style
\bibliographystyle{unsrtnat}
\bibliography{biblio}

%%%%%%%%%%%%%%%% ACKNOWLEDGEMENTS %%%%%%%%%%%%%%%

\section*{Acknowledgments}

\paragraph*{Funding:}

L.K. was supported by Ecole nationale des ponts et chaussées.
P.R. acknowledges funding from the French National Research Agency (ANR-16-CONV-0001), from Excellence Initiative of Aix-Marseille University A*MIDEX and from the European Union (ERC-SuperStoc-101117322).
F.C.S. was supported by the TUM Innovation Network ”Robotic Intelligence in the Synthesis of Life (RISE)”, which is financed through the Excellence Strategy of the German Federal Government. C.K. and F.C.S. acknowledge funding by the Max Planck School Matter to Life, supported by the German Federal Ministry of Education and Research (BMBF) in collaboration with the Max Planck Society. 
M.L. was supported by Marie Curie Integration Grant No. PCIG12-GA-2012-334053, “Investissements d’Avenir” LabEx PALM (Grant No. ANR-10-LABX-0039-PALM), ANR Grants No. ANR-15-CE13-0004-03, No. ANR-21-CE11-0004-02, No. ANR-22-ERCC-0004-01, and No. ANR-22-CE30-0024-01, as well as ERC Starting Grant No. 677532 and the Impulscience® program from Fondation Bettencourt-Schueller. M.L.’s group belongs to the CNRS consortium AQV. This work used HPC resources from GENCI-IDRIS (Grant 2026-AD010918055), as well as computational resources from the ``M\'esocentre'' computing center of Universit\'e Paris-Saclay, CentraleSup\'elec and \'Ecole Normale Sup\'erieure Paris-Saclay supported by CNRS and R\'egion \^Ile-de-France (\url{https://mesocentre.universite-paris-saclay.fr/}).

\paragraph*{Author Contributions:}
L.K., V.O.R. and A.Z. designed the on-lattice numerical simulations. L.K. and V.O.R. performed the on-lattice numerical simulations. L.K. performed the analytical calculations. J.A.V. designed the off-lattice numerical simulations in interaction with M.L. and performed them. L.K., P.R. and M.L. designed the concept of defect engineering. L.K., V.O.R., P.R. and M.L. designed the theoretical part of the work with help from A.Z.. M.E., C.K., and F.C.S. planned the experiments. M.E. and C.K. conceived the concrete experimental implementation. M.E. performed all experiments with contributions from C.K. for initial experiments. M.E. analysed the experimental data with contributions from A.H.. M.E. wrote all scripts for experimental data analysis, with contributions from C.K.. L.K., M.E., V.O.R. C.K., P.R., F.C.S and M.L. wrote the paper.

\paragraph*{Competing Interests:}
There are no competing interests to declare.

\paragraph*{Data and Materials Availability:}
Github repository to run the 2D simulations: \url{https://github.com/Soft-Biophysics-Group/VortexAssembly.git}

Github repository to run the 3D simulations: \url{https://github.com/Soft-Biophysics-Group/frusa_lattice_mc/tree/VortexAssembly3D}

% To include the methods in the same file 
\clearpage
\setcounter{section}{0}
\setcounter{figure}{0}
\setcounter{table}{0}
\renewcommand{\thefigure}{M\arabic{figure}}
\renewcommand{\thetable}{M\arabic{table}}
\renewcommand{\thesection}{M\arabic{section}}

\onecolumngrid
\section*{Methods}

\section{Numerical materials and methods}

Here we describe the methodology underpinning the simulations presented in the main text and in the Supplementary Information. Section~\ref{subsec:simu_protocol} describes our 2D on-lattice simulations protocol. Section~\ref{subsec:size_eval} presents the data analysis procedure used to extract the assembly sizes from these simulations. Section~\ref{subsec:three_d_simulations} describes our 3D on-lattice simulations protocol. Section~\ref{subsec:off_lattice} presents our off-lattice molecular dynamics simulations.

\subsection{Simulation and annealing protocol}
\label{subsec:simu_protocol}
We perform simulated annealing with 2D lattice subunits interacting as described in Fig.~2a of the main text. We use a triangular lattice of $40\times40$ sites populated with $400$ subunits. We perform simulated annealing by linearly increasing the inverse temperature ${1}/{k_BT}$ from $0$ to $1$ in $2000$ increments. For each temperature increment, we perform $A=4\times 10^7$ Monte Carlo steps (which corresponds to $10^5$ steps per subunit). One Monte Carlo move is attempted at each step. A move consists in either a rotation of a single subunit, or the transfer of a subunit to an unoccupied site, which we refer to as a translation. Each Monte Carlo move is accepted according to the Metropolis criterion. 

The energies involved in our simulations are most naturally expressed in units of $k_BT$, while the binding free energies in the experiments are typically listed in $\mathrm{kcal}/\mathrm{mol}$. Assuming a temperature of $36^{\circ}C=309.15 K$ as during the assembly stage in our experiments, the correspondence is $1 \mathrm{kcal}/\mathrm{mol} = 1.628\text{ } k_BT  \text{/subunit}$. Therefore, a binding energy of $-11.54 \text{ }\mathrm{kcal}/\mathrm{mol}$, which is the typical binding energy used in the experiments, corresponds to an interaction energy between two subunits of $-18.7 k_BT$. For all simulations in the size-limitation regime, we choose $\epsilon_c = -18.7 k_BT/\mathrm{subunits} = -11.54\mathrm{kcal}/\mathrm{mol}$.  

The parameter values used in the 2D simulations are shown in Tab.~S3 of the Extended Data.

\subsection{Measurement of assembly radii in simulations}
\label{subsec:size_eval}

We now show how we use the simulation results to measure the assembly size distribution. For each assembly within an annealed system, we automatically count its number of subunits $N$ and its number of interfaces between a subunit and an empty site $S$, as illustrated in Fig.~S1a of the Extended Data. As shown in Fig~2c of the main text, some vortex assemblies display linear protrusions containing extended grain boundaries, which we here refer to as branches. To measure the assembly radius $r$, we assume that all assemblies have the geometry illustrated in Fig.~S1b of the Extended Data. We use Eq.~\eqref{eq:simulation_measure} to determine $r$ and $b$, the number of subunits within the branches, from $N$ and $S$ for each assembly in the simulation as 
\begin{subequations}
\label{eq:simulation_measure}
\begin{align}
    &N = 3r(r+1)+b\\
    &S = 6(2r+1)+2b.
\end{align} 
\end{subequations}

To obtain sufficient statistics on the assembly shapes, we repeat our simulated annealing $20$ times. This yields a total of respectively $228,223,165$ and $103$ vortex assemblies, for the ratios $\epsilon_c/\epsilon_d$ listed in Fig.~2d of the main text. We then directly determine the number of assemblies of radius $r$, $m(r)$, within all those simulations, regardless of the length of their branches $b$. The averaged radius in the simulations shown in Fig.~3h of the main text is $\sum_k km(k)/\sum_km(k)$. The measured cumulative probabilities shown in Fig.~2d of the main text are given by 
\begin{equation}
\label{eq:cumulative_proba_simu}
    P_r(r)=\text{Proba}(\text{radius}\leq r) = \frac{\sum_{r'\leq r} m(r')}{\sum_{r'} m(r')},
\end{equation}

where $r$ is an integer in the idealized geometry of Fig.~S1b, but we sum over all values extracted from the simulations, integer or not.

These measurements allow us to directly assess the impact of the binding energy ratio $\epsilon_c/\epsilon_d$ on the assembly size. They also allow us to directly compare simulation, experimental and theoretical results in Fig.~3h.

\subsection{Three-dimensional numerical simulations}
\label{subsec:three_d_simulations}

The simulations shown in Fig.~4h are performed on an FCC lattice with $20 \times 20 \times 20$ sites populated by 1728 three-dimensional subunits. Each subunit is a rhombic dodecahedron, a shape which on an FCC lattice tiles three-dimensional space similar to hexagons on a triangular lattice. 
We perform a simulated annealing by decreasing $\log_{10}(k_\mathrm{B}T)$ from 2 to 0 in 240 constant increments. 
At each temperature increment we perform $A = 6.64 \times 10^8$ Monte Carlo steps, \emph{i.e.}, 384,000 steps per subunit. Each step consists in attempting one of three single-subunit moves with equal probabilities: moving a subunit to an empty site, rotating a subunit, or moving a subunit to an empty site while rotating it. We run the simulations with a crystalline energy of $\epsilon_{c} = -15 k_B T$, and four different values of the defect energy $\epsilon_d$ as detailed in Tab.~S3 of the Extended Data.

For visualization purposes, we embed the final structure into a cubic domain, as opposed to the parallelepiped-shaped FCC cell which constitutes the actual simulation domain. This conversion is performed by rendering the FCC cell accompanied by its nearest-neighbour periodic copies, and truncating the result to a cube of the same volume as the original FCC cell. One side effect of this visualization trick is that the resulting cubic domain does not fully respect the symmetries of the original cell.
As a consequence, not all subunits may be contained in the cube, and several periodic copies of the same subunit my be included in the visualization box. We also emphasize that the cubic domains shown in Fig.~4h do not have cubic periodic boundary conditions, but instead those associated with the FCC unit cell.

The parameter values used in the 3D simulations are shown in Tab.~S3 of the Extended Data.

\subsection{Off-lattice molecular dynamics simulations}
\label{subsec:off_lattice}

To test the relevance of our lattice simulations, we perform off-lattice molecular dynamics simulations of 2D vortex-assembly-forming subunits using the HOOMD-blue simulation package \cite{anderson2020HOOMDbluePythonPackage}. We describe their results in Sec.~II.C of the Supplementary Information.

We use rigid~\cite{glaser2020PressureRigidBody} subunits comprised of a spherical core of radius $1.2 \sigma$ surrounded by a six spherical interaction patches of radius $0.4\sigma$, where $\sigma$ denotes our unit length scale.
As shown in Fig.~S7a of the Supplementary Information, the patches are regularly arranged in the equatorial plane of the core at $60^\circ$ angles along its rim and at a distance $1.2\sigma$ from its center.
While the core has strictly repulsive (steric) interactions with all other spheres, the patches encode directional and selective bonding similar to the crystalline and defect interactions present in our lattice simulations and implemented through DNA hybridization in our experiments.
In Fig.~S7a of the Supporting Information we number the patches from 1 to 6, consistent with the numbering used in Fig.~3a for the experiment. As in the vortex assembly experiments and lattice simulations, patch pairs $(1, 4)$, $(2,5)$, and $(3, 6)$ interact with a crystalline energy $\epsilon_c$, while pairs $(1,3)$ and $(2, 4)$ interact with a defect energy $\epsilon_d$ given in units of $kT$. All other pairs interact with a hard-core steric repulsion. We set $\epsilon_c=5.5kT$ and use different values of $\epsilon_d$ as needed.

The total energy of a given system configuration is a sum of pairwise interactions between all possible pairs of cores and patches belonging to different particles. To specify these interactions we first define the Weeks-Chandler-Andersen potential, which models a steric repulsions between two spheres (core or patch) separated by a distance $r$:
% TODO Verify consistency of notations
\begin{equation}
   	U^{\text{WCA}}(D,r) = 
    \begin{cases}
      4 k_\mathrm{B}T
      \left[ 
        \left( \frac{D}{r} \right)^{12} 
        - \left( \frac{D}{r} \right)^{6}
        + \frac{1}{4} 
    \right]
     & \text{if } r < 2^{1/6}D \\
    0 & \text{if } r \geq 2^{1/6}D.
    \end{cases}
\end{equation}
Here $D$ denotes the distance of closest approach between two patches: $D = 2.4 \ \sigma$ between two cores, $D=1.4 \ \sigma$ between a core and a patch, $0.8 \ \sigma$ between patches with forbidden interactions, and $D=0.4 \ \sigma$ between two patches with crystalline or defect interactions.

\begin{figure}[t]
    \centering
    \includegraphics[width=0.8\linewidth]{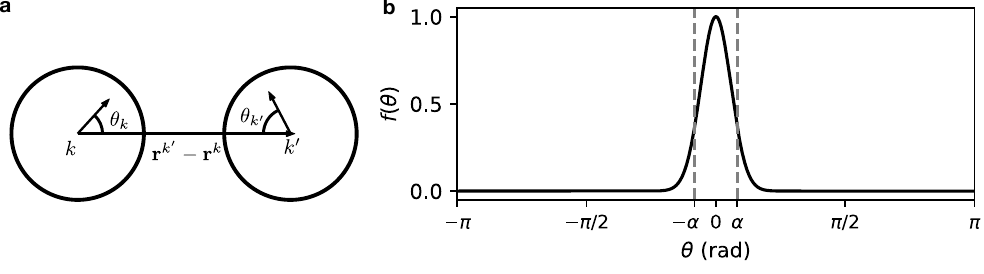}
    \caption{
    \textbf{Definition of the attractive inter-patch interactions used in our off-lattice simulations.}\\
    (\textbf{a})~Definition of the angles $\theta_k$, $\theta_{k'}$ used to parametrize the relative orientations of interacting patches in Eq.~\eqref{eq:patchy_interaction}. The arrow carried by each patch represents the direction away from the core of the subunit that it belongs to.
    (\textbf{b})~Plot of the angular envelope defined in Eq.~\eqref{eq:off_lattice_f} with the parameters $\alpha=14.89^{\circ}$ (dashed vertical lines) and $\omega = 30$ used in our simulations.
    }
    \label{fig:off_lattice_orientations}
\end{figure}

We define the total energy of the system as
\begin{equation}\label{eq:off_lattice_energy}
U
=
\sum_{i<j}
\left[
U^{ij}
+
\sum_{k=1}^6
\left(V_k^{ij}+V_k^{ji}\right)
+
\sum_{k,k'=1}^6
W_{kk'}^{ij}
\right],
\end{equation}
where the leftmost sum runs over all unique subunit pairs $(i, j)$. The three terms in the square bracket of Eq.~\eqref{eq:off_lattice_energy} respectively correspond to the core-core, core-patch and patch-patch interaction energies. We denote the position of the center of the core and the $k$th patch of subunit $i$ by $\mathbf{r}^c_i$ and $\mathbf{r}^k_i$; the terms of the sum of Eq.~\eqref{eq:off_lattice_energy} are then given by
\begin{subequations}
    \begin{align}
        U^{ij} &= U^\mathrm{WCA}\left(2.4\sigma, |\mathbf{r}^c_i-\mathbf{r}^c_j|\right),\\
        V_k^{ij} &= U^\mathrm{WCA}\left(1.4\sigma,|\mathbf{r}^c_i-\mathbf{r}^k_j|\right),\\
        W_{kk'}^{ij} &= 
        U^\mathrm{WCA}\left(0.8\sigma,|\mathbf{r}^c_i-\mathbf{r}^k_j|\right)\quad\text{for repelling patches,}\\
        W_{kk'}^{ij} &= 
        U^\mathrm{WCA}\left(0.4\sigma,|\mathbf{r}^c_i-\mathbf{r}^k_j|\right)
        +
        U_{kk'}^\mathrm{patch}\left(|\mathbf{r}^k_i-\mathbf{r}^{k'}_j|, \theta_k, \theta_{k'} \right)\quad\text{for interacting patches.}
    \end{align}
\end{subequations}
These interactions are all steric, except for the last one, which also includes a directional attractive interaction $U_{kk'}^\mathrm{patch}$ between patch $k$ of subunit $i$ and patch $k'$ of subunit $j$. This interaction is a function of three arguments. The first one is the distance between the two patches, and the last two are angles $\theta_k$ and $\theta_{k'}$ that quantify the orientation of each patch with respect to the line connecting the centers of the two patches. We define these angles in Fig.~\ref{fig:off_lattice_orientations}a.

The directional attractive interaction is defined through the ``Patchy'' anisotropic Lennard-Jones potential provided by HOOMD-Blue, yielding
\begin{equation}\label{eq:patchy_interaction}
U_{kk'}^{\mathrm{patch}}(r, \theta_k, \theta_{k'}) = \epsilon_{kk'}\, u^{\mathrm{LJ}}(r)\,  f(\theta_k)\, f(\theta_{k'}), 
\end{equation}
where $\epsilon_{kk'}$ is equal to $\epsilon_c$ ($\epsilon_d$) if the patches $k$, $k'$ interact through a crystalline (defect) interaction. The functions $u^{\mathrm{LJ}}$ and $f$ respectively denote a radial attractive Lennard-Jones potential and an angular envelope which penalizes the misalignment of the patches. They are given by
\begin{subequations}
\begin{align}
u^{\mathrm{LJ}}(r)
&=
\begin{cases}
4\times
\left[
\left(\frac{0.2 \sigma}{r}\right)^{12}
-
\left(\frac{0.2 \sigma}{r}\right)^6
\right]
 \qquad &\text{if} \  r < \sigma,\\
0 \qquad &\text{if} \ r \ge \sigma,
\end{cases}\\
f(\theta)
&=
\frac{
g(\theta) - g(\pi)
}{
g(0)-g(\pi)
}, \quad \text{with} \quad g(\theta)=\frac{1}{1+\exp\left[\omega(\cos\alpha-\cos\theta)\right]}.\label{eq:off_lattice_f}
\end{align}
\end{subequations}
The two parameters $\alpha$ and $\omega$ of the function $f$ respectively set the width of the angular window over which the attractive interaction is active, and the steepness of the fall-off at the edge of that window. We give the values of these parameters used in our simulations and plot the resulting angular interaction profile in Fig.~\ref{fig:off_lattice_orientations}b. 

We simulate $N=400$ subunits in a two-dimensional square box of size $L_x=L_y=105 \ \sigma$. The centers of the subunits are initially placed on a square lattice with a lattice spacing of $5\sigma$. We run a Brownian dynamics simulation where each subunit has translational and rotational damping constants $\gamma_t$ and $\gamma_r$ such that $\gamma_t=\gamma_r/\sigma^2$. We define the diffusional time scale $\tau=\sqrt{\gamma_t \sigma^2/k_BT}$ required for a subunit to diffuse over a distance of the order of its radius, and use a simulation time step $\Delta t=10^{-5}\tau$. We run simulations for different values of the interaction ratio $\epsilon_c / \epsilon_d \in \left[0.25, 0.33, 0.41, 0.48, 0.5, 0.55, 0.6, 0.7 \right]$ as discussed in the Supplementary Information and shown in Fig.~3h of the main text. Each run comprises $N = 4{,}000{,}500{,}000$ simulation steps, and we collect statistics over eight runs for each parameter regime. We measure the size of the aggregates obtained in off-lattice simulations by analyzing the final snapshot of each simulation using the same image analysis pipeline as in our experiments. See paragraph ``\textit{Determination of Assembly Sizes}'' of Sec.~\ref{sec:data_analysis} for a full description.

\section{Experimental materials and methods}
To experimentally verify our theoretical predictions and numerical findings, we implement a size-controlled assembly process using DNA origami nanocylinders that interact via base-pairing under predefined interaction rules. In Sec.~\ref{subsec:exp_design}, we describe the design of the DNA interaction strands that implement these rules, as well as the oxDNA simulation used to visualize the structure of the DNA origami subunit. In Sec.~\ref{subsec:exp_protocols}, we then detail the protocols for DNA origami folding, assembly, and functionalization with gold nanoparticles, which yield the experimental data for Fig.~3 of the main text. Finally, in Sec.~\ref{subsec:exp_analysis}, we outline the procedures for data acquisition and analysis.

\subsection{Design of interactions and subunit simulation}
\label{subsec:exp_design}

Here we first explain how the DNA interaction strand sequences for defect and crystalline interactions, as shown in Fig.~3b and listed in data S4, were designed and selected to ensure the precise interaction energies (and their ratios) required for size-controlled assembly. We then describe how the oxDNA simulation framework, integrated within oxView, was used to generate the visual representation of the DNA origami nanocylinder’s mean structure shown in Fig.~3a.

We emphasize that the origami structures are highly modular by design, and that switching between different realizations of a given interaction matrix requires only minor changes. The nanocylinders used in this work are folded from a 2873-nt scaffold using 136 staple oligonucleotides. In the vortex design, only 14 of these staples define interactions between neighboring cylinders: six mediate the crystalline interactions, and eight others encode the defect interactions. Changes in the interaction ratio $\epsilon_c/\epsilon_d$ can be implemented simply by replacing either the six crystalline staples or the eight defect staples. The corresponding oligonucleotide synthesis costs on the order of 10\,€. Likewise, converting the vortex design into the fiber design requires only four additional strands.

\paragraph*{Interaction strand design with NUPACK.}
Orthogonal defect and crystalline DNA interaction strand sets with specified Gibbs free energy differences ($\epsilon_d$ and $\epsilon_c$) were designed using the NUPACK Python package \cite{fornace2020nupack}. Initially, exhaustive libraries of all possible sequences of lengths 10, 11, and 12 nucleotides, along with their reverse complements, were generated. For each sequence, relevant properties such as the Gibbs free energy ($\epsilon$) of the structure ensemble (pfunc) and the secondary structure were calculated (model: dna04, temperature: \qty{36}{\degreeCelsius}, sodium: \qty{50}{\milli\molar}, magnesium: \qty{5}{\milli\molar}, concentration: \qty{10}{\nano\molar}). Uncertainties in the ensemble free energy values were evaluated through a central finite difference scheme by varying the baseline incubation temperature and magnesium concentration. Maximum experimental uncertainties were defined as $\Delta T = \qty{0.2}{\celsius}$, corresponding to the Mastercycler Nexus incubation chamber specifications, and $\Delta$[MgCl$_2$] = \qty{0.1}{\milli\molar}, representing standard pipetting errors. The resulting environmental variances were subsequently combined to yield the final propagated free energy error.

The sequence library of appropriate length was then filtered for sequences that exhibit no secondary structure with Gibbs free energy changes upon binding within a narrow tolerance interval $[\epsilon_d - \delta\epsilon_d, \epsilon_d + \delta\epsilon_d]$, centered around the intended value $\epsilon_d$ of the defect interaction. An initial tolerance of $\delta\epsilon_d = \qty{0.05}{\kilo\calorie\per\mole}$ was applied. If this resulted in too few sequences, $\delta\epsilon_d$ was incrementally increased by \qty{0.1}{\kilo\calorie\per\mole} until a sufficient number of candidates (typically a few hundred) was obtained. The same filtering criteria were subsequently applied to the original sequence library using $\epsilon_c$ to identify potential crystalline interaction strands. This process yielded sets of defect and crystalline interactions comprising $n_d$ and $n_c$ sequences, respectively.

To ensure intraset orthogonality, these sets were evaluated based on the bound fraction — defined as the ratio of DNA strands that are correctly bound to their intended reverse complementary partners to the total number of strands — if all strands in a set were mixed at a concentration of \qty{10}{\nano\molar} each, with the maximum allowed assembly size set to two. Specifically, sets were deemed orthogonal if at least 90\% of the defect interaction strands and at least 10\% of the crystalline interaction strands bound exclusively to their intended partners. Sets failing to reach these orthogonality thresholds were excluded from further analysis. After identifying orthogonal defect and crystalline sets, all possible pairings of one defect set with one crystalline set were generated to form combined sets comprising $n_d$+$n_c$ sequences. These new combined sets then underwent an additional intraset orthogonality check to prevent unintended interactions. Specifically, any set exhibiting binding domains of six or more consecutive base pairs between strands that are not intended to form complementary pairs was disregarded as potential candidate for a valid interaction strand set.

From the remaining fully processed interaction strand sets, candidates were randomly selected and manually inspected for potential issues, such as the formation of G-quadruplex structures and long single-base repetitions. Finally, these selected sets were validated using the NUPACK web interface \cite{zadeh2010nupack}. The sequence sets used in the experiments are listed in data S4.

\paragraph*{OxDNA simulation.}
The DNA origami topology and configuration files for the nanocylinder were exported from a scadnano~\cite{doty2020scadnano} design file (see Fig.~S21 of the Extended Data), imported into oxView~\cite{bohlin2022oxview} and manually pre-arranged, for structure relaxation. Forces used during relaxation were automatically generated in oxView, and a Molecular Dynamics (MD) relaxation step was run on GPUs (ox-serve with nanobase.org webserver; first Bussi-Donadio-Parrinello then Brownian thermostat). Next, a MD simulation was employed (oxDNA2 model \cite{poppleton2023oxdna}, 1e9 steps) to simulate the monomer structure. The PDB file for rendering in ChimeraX was generated with the oxDNA analysis tools command line script \cite{poppleton2021oat}. The monomer structure shown in Fig.~3a of the main text represents a mean structure.

\subsection{Experimental protocols}
\label{subsec:exp_protocols}

This subsection outlines the experimental protocols employed throughout this work. First, we describe how the DNA origami nanocylinder is folded using an annealing ramp, followed by the purification steps to remove incomplete or misfolded structures. We then detail the procedure for assembling properly folded nanocylinders at different experimental conditions (Fig.~3f). Finally, we explain how gold nanoparticles are functionalized and attached to infer the assemblies’ topological structure as exemplified in Fig.~3d.

\paragraph*{DNA Origami Folding.}
The DNA origami nanocylinders used in this study were originally developed by Wickham et al. \cite{wickham2020barrel}. Single-stranded p2873 DNA (scaffold strand, for sequence information see data S3) was provided by Prof. Hendrik Dietz’ group (\qty{100}{\nano\molar} in ddH$_2$O). Staple strands were purchased unpurified from Integrated DNA Technologies in 1× TE buffer (\qty{10}{\milli\molar} TRIS, \qty{0.1}{\milli\molar} EDTA, pH 8.0) at \qty{200}{\micro\molar} each (for sequence information see data S3 and data S4). Folding reactions were prepared with \qty{50}{\nano\molar} scaffold strand and \qty{200}{\nano\molar} staple strands in 1x FoB18 buffer (\qty{5}{\milli\molar} TRIS, \qty{1}{\milli\molar} EDTA, \qty{5}{\milli\molar} NaCl, \qty{18}{\milli\molar} MgCl$_2$). Folding reactions were annealed in a thermocycler (Eppendorf, Mastercycler nexus GX2). The used protocol was \qty{5}{\minute} incubation at \qty{65}{\degreeCelsius} followed by an annealing ramp of \qty{-0.1}{\degreeCelsius} per \qty{36}{\second} from \qty{60}{\degreeCelsius} to \qty{30}{\degreeCelsius}. Folded DNA origami samples were immediately collected for purification.

\paragraph*{DNA Origami Purification.}
Samples were purified using an ultrafiltration protocol. A centrifuge (Eppendorf, Centrifuge 5425 R) was pre-heated to \qty{32}{\degreeCelsius} containing one \qty{2}{\milli\liter} eppi 1x FoB5 washing buffer (\qty{5}{\milli\molar} TRIS, \qty{1}{\milli\molar} EDTA, \qty{5}{\milli\molar} NaCl, \qty{5}{\milli\molar} MgCl$_2$) per sample. Amicon filters (Amicon Ultra 0.5 ml Ultracel 100k) were placed into \qty{2}{\milli\liter} eppis, loaded with \qty{500}{\micro\liter} pre-heated 1x FoB5 washing buffer and centrifuged (\qty{5}{\minute}, \qty{10}{\kilo\rcf}). The supernatant was discarded. Filters were loaded with \qty{60}{\micro\liter} sample solution and \qty{440}{\micro\liter} washing buffer and centrifuged (\qty{5}{\minute}, \qty{10}{\kilo\rcf}). The supernatant was discarded. Filters were then loaded with \qty{500}{\micro\liter} washing buffer and centrifuged (\qty{5}{\minute}, \qty{10}{\kilo\rcf}). The supernatant was discarded once more. Filters were then loaded with \qty{500}{\micro\liter} of 1x FoB\textit{X} assembly buffer (\qty{5}{\milli\molar} TRIS, \qty{1}{\milli\molar} EDTA, \qty{5}{\milli\molar} NaCl, \textit{X}=[\qty{5}{\milli\molar}, \qty{7.5}{\milli\molar}, \qty{10}{\milli\molar}] MgCl$_2$) with \textit{X} being the MgCl$_2$ concentration used in the experiments as indicated in the main text. The filters were centrifuged (\qty{5}{\minute}, \qty{10}{\kilo\rcf}) one last time. Purified and buffer-adjusted samples were extracted from filters with a pipette by repeatedly aspirating and dispensing the solution to dissolve any potential pellets. 

\paragraph*{DNA origami assembly.}
The sample concentration was measured at \qty{260}{\nano\meter} (Implen, NanoPhotometer N50) and normalized to \qty{20}{\nano\molar} with assembly buffer (1x FoB5, 1x FoB7.5 or 1x FoB10). Samples were incubated in a thermocycler at \qty{36}{\degreeCelsius} for varying amounts of time as indicated in the main text and resulting assemblies are immediately applied to TEM grids.

\paragraph*{Functionalization of gold nanoparticles.}
To fabricate the DNA-grafted gold nanoparticles (DNA-AuNPs) with a freeze-thaw protocol \cite{liu2017freezethaw}, \qty{110}{\micro\liter} of \qty{90.8}{\nano\molar} AuNPs (Cytodiagnostics, d=\qty{5}{nm}) in  0.01x PBS buffer (\qty{0.1}{\milli\molar}) were mixed with \qty{30}{\micro\liter} of HPLC purified 5'-thiol modified \qty{20}{\nucleotide} poly-thymidine DNA strands (Biomers, lyophilized, resuspended to \qty{100}{\micro\molar} in ddH$_2$O) and \qty{860}{\micro\liter} of ddH$_2$O to an AuNP:DNA ratio of 1:300. The suspension was stored at \qty{-80}{\degreeCelsius} for \qty{1}{\hour} and subsequently thawed at room temperature. Functionalized nanoparticles were purified by ultrafiltration with an Amicon filter (Amicon Ultra 0.5 ml Ultracel 100k). The filter was pre-washed with \qty{500}{\micro\liter} of ddH$_2$O, centrifuged (\qty{14}{\kilo\rcf}, \qty{5}{\minute}) and the supernatant was discarded. Next, \qty{500}{\micro\liter} of DNA-AuNP sample was loaded, centrifuged (\qty{14}{\kilo\rcf}, \qty{5}{\minute}) and supernatant discarded. This step was repeated once. The sample was washed seven times by loading \qty{500}{\micro\liter} of ddH$_2$O, centrifuga{ting} (\qty{14}{\kilo\rcf}, \qty{5}{\minute}) and discarding the supernatant. Finally, the filter was placed into a new eppi in inverse orientation and centrifuged (\qty{14}{\kilo\rcf}, \qty{5}{\minute}) to retrieve purified DNA-AuNPs in ddH$_2$O. The concentration of the AuNPs was then measured at \qty{520}{\nano\meter} (Implen, NanoPhotometer N50) and the result of that measurement formed the basis of our dilution of the AuNPs to the standardized concentration used in our labeling experiments.

\subsection{Data acquisition and analysis}\label{sec:data_analysis}
\label{subsec:exp_analysis}

First, we detail how the data were acquired using a transmission electron microscope to capture images of the assemblies. Next, we explain how vortex assembly sizes were determined from these TEM micrographs using the \textit{scikit-image} Python library, thus enabling the generation of assembly size distributions and the calculation of mean assembly radii (Fig.~3h). Finally, we describe the evaluation method used to determine subunit orientation statistics for Fig.~3e.

\paragraph*{Acquisition of TEM micrographs.}
For TEM imaging, \qty{5}{\micro\liter} of sample solution were incubated on glow-discharged (Electron Microscopy Sciences, K100X;
\qty{20}{\second}, \qty{35}{\milli\ampere}, negative polarity) formvar carbon Cu400 TEM grids (Science Services) for $5$--$20$ {\minute}. %\SIrange{5}{20}{\minute}. 
The staining solution was prepared by adding \qty{1}{\micro\liter} of \qty{5}{\molar} NaOH to \qty{200}{\micro\liter} of a \qty{2}{\percent} uranyl formate solution,  
\qty{10}{\second} vortexing and subsequent centrifugation (\qty{2.5}{\minute}, \qty{21}{\kilo\rcf}).
After sample incubation, grids were washed with \qty{5}{\micro\liter} of staining solution and incubated with \qty{20}{\micro\liter} of stain for \qty{25}{\second}. 
Automated imaging series were conducted with a FEI Tecnai T12 (\qty{120}{\kilo\volt}) equipped with a Tietz TEMCAM-F416 camera, operated with SerialEM. 
Further sample imaging was conducted with a Philips CM100 (\qty{100}{\kilo\volt}) operated with AmtV600.

\paragraph*{Determination of assembly sizes.}
\label{subsec:eval_assemblies}

TEM image stacks were automatically pre-processed by monomer template matching, assembly recognition and calculation of properties with a custom Python script using the \textit{scikit-image} library \cite{scikit-image}. The cropped image of a DNA origami monomer was used as a template to match and extract all monomer instances from the first two TEM images in a stack. The average of all instances was calculated and the result used as an optimized template (now adjusted to the imaging conditions of the stack). For every image in a stack, areas occupied by monomers were matched with the optimized template and used as a mask to convert the image to a binary format. Pixels contained in the mask were set to 1, otherwise they were set to 0. Small gaps between features (pixels set to 1) were closed by using the \textit{skimage.morphology.closing()} method, creating connected areas representing assemblies. Resulting assemblies were labeled with unique identifiers by classifying pixels to be in the same assembly if they neighbor one another and have the same value).

Many of the objects in our electron microscopy images are isolated subunits or small assemblies thereof. Each of these objects is assigned a label under the assembly-detection scheme described above. To set them aside from properly assembled vortex assemblies, different assembly properties (bounding box, area, perimeter, and centroid) were extracted for each labelled assembly with the \textit{scikit.image.regionprops()} method. Assemblies were checked for the characteristic central hole, present in vortex assemblies. If hole size, hole eccentricity and monomer neighbor count were correct (a hole must have exactly six neighboring monomers), assemblies were marked as potential candidates and the number of monomers they contained was determined. Marked assemblies were individually presented with their extracted properties and their monomer count $N$ manually validated. To obtain a continuous value for the intrinsically discrete assembly size $r$ from $N$, an ideal vortex assembly was assumed and Eq.~\eqref{eq:simulation_measure}a was solved with $b=0$ to find
\begin{equation}
r = \frac{1}{2}\left(\sqrt{1+\frac{4N}{3}}-1\right),
\end{equation}
which was used in Figs.~3f-h of the main text. Images of vortex assemblies that were identified using this protocol were saved with their corresponding assembly properties and size $r$, then used in the analysis presented in the main text. For Fig.~3g sample sizes were  $N_{20\text{h}}=67$, $N_{43\text{h}}=137$ for assemblies at $5$\milli\molar, $N_{20\text{h}}=131$, $N_{43\text{h}}=278$ at $7.5${\milli\molar} and $N_{20\text{h}}=364$, $N_{43\text{h}}=141$ at $10${\milli\molar}. For Fig.~3h sample sizes were $N_{0.31}=106$, $N_{0.52}=562$, $N_{0.57}=122$ and $N_{0.61}=39$ for the experiments, and $N_{0.25}=228$, $N_{0.41}=223$, $N_{0.48}=165$ and $N_{0.55}=103$ for the simulations. Bars in Figs.~3f, g and h represent Bessel corrected standard deviations.

\paragraph*{Determination of subunit orientations.}
To evaluate the orientations of subunits labeled with gold nanoparticles (AuNP), assemblies were extracted from TEM image stacks as described above. Using a custom Python script, assembly images were individually presented for manual processing. Images were mirrored horizontally if the majority of monomers in an assembly exhibited a counter-clockwise direction of rotation. For every AuNP labeled monomer in an assembly, a vector $\mathbf{x_i}$ was drawn from the monomer center to the AuNP position. A second vector $\mathbf{y_i}$ was drawn center-to-center between the central hole in the assembly and the monomer. After processing all images, vectors $\mathbf{x_i}$ and $\mathbf{y_i}$ were normalized and their 2D cross product $\mathbf{\hat{x}_i}\times\mathbf{\hat{y}_i}=\cos\phi$ was computed. The results are displayed in Fig.~3e of the main text. $N=1375$ subunits from three independent experiments were analyzed.

% To include the supplement in the same file 
\clearpage
\setcounter{section}{0}
\setcounter{figure}{0}
\setcounter{table}{0}
\renewcommand{\thefigure}{S\arabic{figure}}
\renewcommand{\thetable}{S\arabic{table}}
\renewcommand{\thesection}{S\arabic{section}}

\onecolumngrid

\section*{Supplementary information and extended data}

%\tableofcontents
%\startcontents[fromhere]
%\printcontents[fromhere]{l}{0}{\section*{Contents}}

%\newpage
\section{Theory of defect-induced size control}

Here we study the stability and size distributions of the assemblies discussed in the main text, leading in particular to the theoretical predictions shown in Figs.~2 and 4 of the main text. In Sec~\ref{supp:phase_diagram}, we build the phase diagram of an idealized vortex assembly, including the thermodynamic prediction for the assembly radius. In Sec.~\ref{supp:other_geo} we discuss  the stability of alternative assembly geometries including those presented in Fig.~4 of the main text. Finally, in Sec.~\ref{supp:size_estimate}, we discuss the influence of a finite temperature on the assembly size distribution shown in Fig.~2d of the main text.

\subsection{Phase diagram of the vortex assembly}
\label{supp:phase_diagram}

Here we consider vortex assemblies of different radii $r$ as well as crystals, and determine which structure is most energetically favorable as a function of the interaction energies $\epsilon_c$ and $\epsilon_d$.

\paragraph*{Energy calculations.}
\label{supp:energy_calculation}

The energy of an assembly is determined by its number of crystalline interactions and defect interactions. We
consider perfectly hexagonal assemblies with an integer radius $r$. We compute the energy of a crystalline triangular wedge, corresponding to $1/6^{\mathrm{th}}$ of an assembly. We respectively denote by $N_c$ and $N_d$ the total number of crystalline and defect interactions in a triangular wedge.  The energy of a triangular wedge then reads 
\begin{equation}\label{eq:general_expression_for_the_energy}
    E(r) = N_c(r)\epsilon_c + N_d(r)\epsilon_d.
\end{equation}

\begin{figure}[p]
    \centering
    \includegraphics[width=0.93\linewidth]{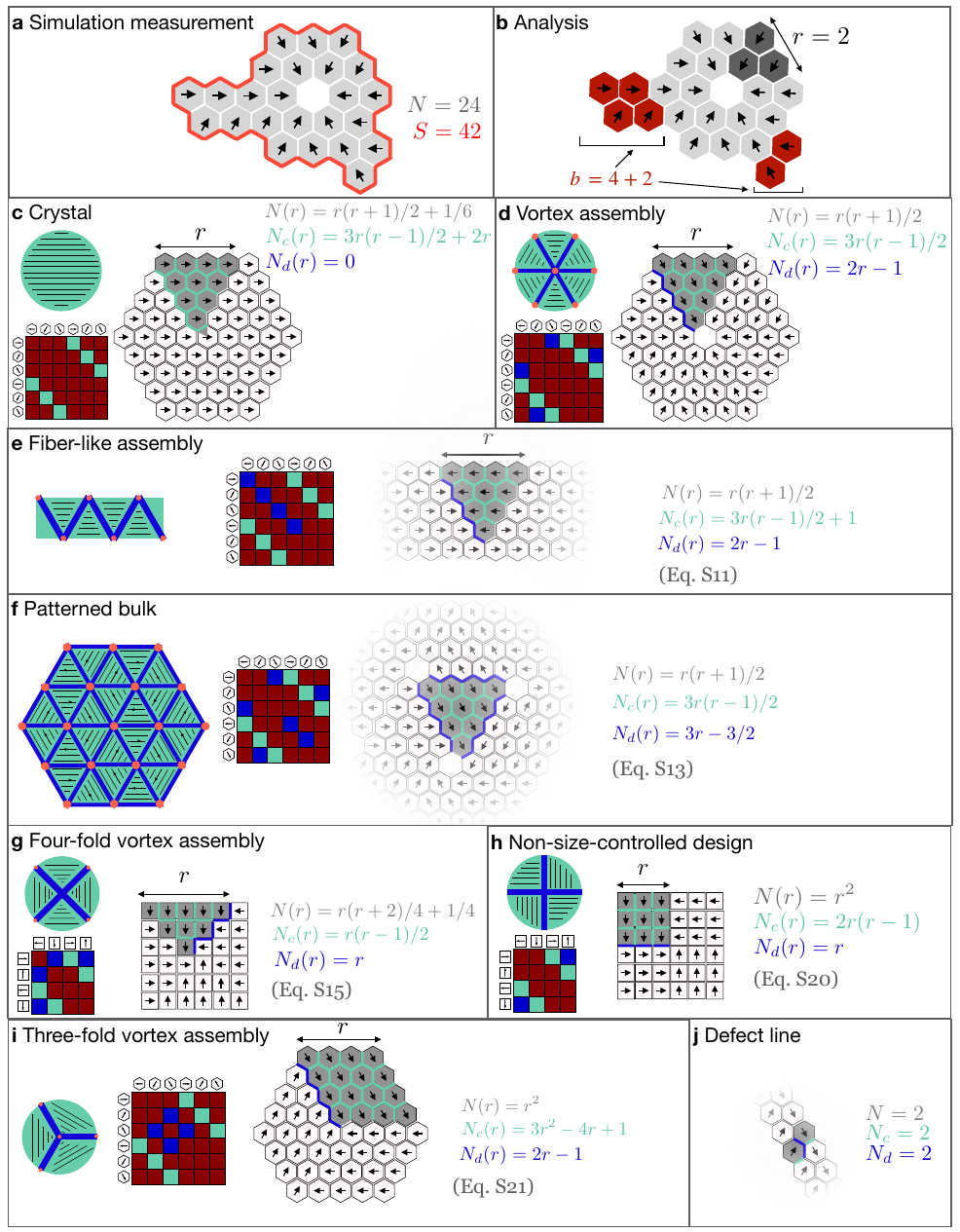}
    \caption{
    \textbf{Characterization of the assemblies' morphology and energy}\\
    (\textbf{a})~To analyze the morphology of a simulated vortex assembly, we measure its number of subunits $N$ and outer surface length $S$.
    (\textbf{b})~These numbers allow us to estimate the vortex radius $r$ (the length of the side of a triangular crystalline wedge illustrated in \emph{dark gray}) and the total number $b$ of subunits within branches (\emph{dark red}).
     (\textbf{c})~Geometry of an ideal crystalline assembly and expressions of the quantities defined in Sec.~\ref{supp:phase_diagram}. The colored matrix present the subunit's interaction map, where the crystalline interaction is colored in \emph{green}, the defect interaction in \emph{blue} (see next panels), and all other interactions in \emph{dark red}. For each pair, the orientations of the left and right subunits are respectively shown at the left and at the top of the matrix.
    (\textbf{d})~Vortex assembly, (\textbf{e})~Fiber-like assembly, (\textbf{f})~Patterned bulk (\textbf{g})~Four-fold vortex assembly (\textbf{h})~Non-size-controlled four-fold design (\textbf{i})~Three-fold vortex assembly, (\textbf{j})~width-2 fiber defect.}
    \label{fig:EnergyCalcul}
\end{figure}

We determine $N_c(r)$ and $N_d(r)$ from the schematics of Fig.~\ref{fig:EnergyCalcul}c-d and obtain
%. The crystal interactions are colored green and the defect interactions are colored in blue
%The number of crystal bonds scales like the surface area of the assembly, $r^2$ and the number of defect bonds like its perimeter, which is proportional to $r$.
\begin{subequations}
\begin{align}
&N^{(\mathrm{crystal})}_c(r) = 3r(r-1)/2 +2r\\
&N^{(\mathrm{crystal})}_d (r) = 0 \\
&N^{(\mathrm{vortex})}_c(r) = 3r(r-1)/2 \\
&N^{(\mathrm{vortex})}_d (r) = 2r-1.
\end{align}
\end{subequations}
The total number of subunits in the crystal and vortex assemblies are 
\begin{subequations}\label{sec:vortex_assembly_particle_numbers}
\begin{align}
    6N^{(\mathrm{crystal})}(r)&=  3r(r+1)+1\\
    6N^{(\mathrm{vortex})}(r)&= 3r(r+1).
\end{align}
\end{subequations}
Finally we combine Eqs.~(\ref{eq:general_expression_for_the_energy}-\ref{sec:vortex_assembly_particle_numbers}) to express energy per subunit in a vortex assembly and a crystal:
%, expressed with the rescaled variables $y_c$ an $e_1$ defined in equations~\ref{eq:yc} and~\ref{eq:e1}:
\begin{subequations}\label{sec:vortex_assembly_energies}
\begin{align}
    e^{(\mathrm{crystal})}(r) &= \frac{E^{(\mathrm{crystal})}(r)}{N^{(\mathrm{crystal})}(r)} = \frac{ (3 r^2+r)\epsilon_c }{r(r+1)+1/3}\label{eq:e(r)_crystal}\\
    e^{(\mathrm{vortex})}(r) &=\frac{3r^2 \epsilon_c + r(-3\epsilon_c + 4 \epsilon_d) - 2 \epsilon_d}{r(r+1)}\label{eq:e(r)_cam}
\end{align}
\end{subequations}

\paragraph*{Energy minimization.}
\label{supp:energy_minimization}
To determine the size of the most stable assembly we now minimize the energies Eqs.~\eqref{sec:vortex_assembly_energies} with respect to $r$. This is equivalent to minimizing the total energy of a large canonical system containing a fixed total number of subunits.
Considering relatively large assemblies, we can treat $r$ as a continuous parameter and look for the radii $r^*$ where the following derivatives vanish:
\begin{subequations}
\begin{align}
    \partial_re^{(\mathrm{crystal})}(r) &= 3\epsilon_c \frac{6r^2+6r+1}{(3r^2+3r+1)^2} \\
    \partial_r e^{(\mathrm{vortex})}(r) &= 2\frac{(3\epsilon_c-2 \epsilon_d)r^2 +\epsilon_d(2r+1)}{r^2(r+1)^2}
\end{align}
\end{subequations}
We find that the energy per crystal subunit is always monotonic. For $\epsilon_c>0$, the most stable crystal is a monomer ($r^*=0$, $e^{\mathrm{crystal}} = 0$). For $\epsilon_c<0$, a bulk crystal is most stable ($r^*=\infty$, $e^{\mathrm{crystal}} = 3\epsilon_c$). By contrast, the energy per subunit of the vortex assembly has a minimum provided that $3\epsilon_c-2\epsilon_d>0$, $\epsilon_d<0$, and $\epsilon_c>\epsilon_d$. It then reads

\begin{equation}
\label{eq:rmin_camembert}
    r^* = \frac{-\epsilon_d + \sqrt{3\epsilon_d(\epsilon_d-\epsilon_c)}}{3\epsilon_c-2\epsilon_d},
\end{equation}

When choosing the parameters to run a numerical simulation, we select an integer target assembly radius $r^*$, and determine the corresponding energy ratio by inverting Eq.~\eqref{eq:rmin_camembert}:
\begin{equation}
\label{eq:r_star_ratio}
    \frac{\epsilon_c}{\epsilon_d} = \frac{2(r^*)^2 - 2(r^*) -1}{3(r^*)^2}.
\end{equation}

\paragraph*{Phase diagram boundaries.}

For each region of parameter space, we determine whether the vortex assembly or the crystal is more stable, which allows us to construct the phase diagram shown in Fig.~2b of the main text.

\begin{itemize}

\item For $\epsilon_c>0$, $\epsilon_d>0$, vortex assemblies are unstable: their energy per subunit increases with increasing $r$, so monomers (energy $0$) are favored. No self-assembly occurs (light green region in Fig.~\ref{fig:phaseD_square_threeL}a). This situation is illustrated by the top curve in Fig.~\ref{fig:phaseD_square_threeL}c.

\item For $\epsilon_c>0$, $\epsilon_d<0$, vortex assemblies are stable with a minimum at $r^*=1$. They are more stable than crystal assemblies because $e^{(\mathrm{vortex})}(1)=\epsilon_d<0<e^{(\mathrm{crystal})}(0)$. Hexamers (vortices of radius $1$) are the most stable geometry (dark blue region, Fig.~\ref{fig:phaseD_square_threeL}a) This situation is illustrated by the top curve in Fig.~\ref{fig:phaseD_square_threeL}b.

\item For $\epsilon_c<0, \epsilon_d<0$ and $3\epsilon_c-2\epsilon_d>0$, we compare $e^{(\mathrm{vortex})}(r^*)$ with $e^{(\mathrm{crystal})}(r=\infty)=3\epsilon_c$:
\[
e^{(\mathrm{vortex})}(r^*) - e^{(\mathrm{crystal})}(r=\infty) = \frac{3(r^*)^2 \epsilon_c + r(-3\epsilon_c + 4 \epsilon_d) - 2 \epsilon_d}{r^*(r^*+1)} - 3\epsilon_c = \frac{-2r^*(3\epsilon_c - 2 \epsilon_d) - 2\epsilon_d}{r^*(r^*+1)}
\]
Using Eq.~\eqref{eq:rmin_camembert}, we get 
\[
    e^{(\mathrm{vortex})}(r^*) - e^{(\mathrm{crystal})}(r=\infty) = \frac{-2\sqrt{ -3\epsilon_d(\epsilon_c-\epsilon_d)}}{r^*(r^*+1)}<0.
\]
Hence, optimal-size vortex assemblies are always more stable than crystals in this region, and have a finite size (blue region in Fig.~\ref{fig:phaseD_square_threeL}a). This situation is illustrated by the middle curve in Fig.~\ref{fig:phaseD_square_threeL}b.

\item For $\epsilon_c<0$, $\epsilon_d<0$, $3\epsilon_c-2\epsilon_d<0$, and $\epsilon_c-\epsilon_d>0$, both assemblies extend to infinite size. Expanding at large $r$ gives
\[
\begin{aligned}
e^{(\mathrm{vortex})}(\infty) - e^{(\mathrm{crystal})}(\infty) &= 3\epsilon_c + \tfrac{1}{r}(-3\epsilon_c+4\epsilon_d) - 3\epsilon_c - \tfrac{1}{r}\epsilon_c + \mathcal{O}(1/r^2) \\
&= \tfrac{4}{r}(\epsilon_d-\epsilon_c)+\mathcal{O}(1/r^2)<0,
\end{aligned}
\]
so vortices of unlimited size dominate (light blue region in Fig.~\ref{fig:phaseD_square_threeL}a). This situation is illustrated by the bottom curve in Fig.~\ref{fig:phaseD_square_threeL}b.

\item For $\epsilon_c<0$, $\epsilon_d<0$, $\epsilon_c-\epsilon_d<0$, vortices are infinite but crystals are more stable according to the previous equation (dark green region in Fig.~\ref{fig:phaseD_square_threeL}). This situation is illustrated by the bottom curve in Fig.~\ref{fig:phaseD_square_threeL}c.
\item For $\epsilon_c<0$, $\epsilon_d>0$, $\epsilon_c-\epsilon_d<0$, vortices are unstable and infinite crystals prevail (dark green region in Fig.~\ref{fig:phaseD_square_threeL}). This situation is also illustrated by the bottom curve in Fig.~\ref{fig:phaseD_square_threeL}c.

\end{itemize}

\paragraph*{Favored assembly size.}
The most favorable vortex radius $r^*$ computed in Eq.~\eqref{eq:rmin_camembert} can take non-integer values, which are not compatible with the discrete geometry of Fig.~\ref{fig:EnergyCalcul}d. Here we explicitly determine the most favored integer radius of a vortex assembly as a function of $\epsilon_c/\epsilon_d$.

We consider a perfectly hexagonal assembly as in Fig.~\ref{fig:EnergyCalcul}d. The assembly size $r^*$ which minimizes the energy satisfies
\begin{subequations}
\label{eq:r_star_condition}
\begin{align}
    &e^{(\mathrm{vortex})}(r^*-1)>e^{(\mathrm{vortex})}(r^*)\\
    &e^{(\mathrm{vortex})}(r^*+1)>e^{(\mathrm{vortex})}(r^*),
\end{align}
\end{subequations}

From Eq.~\eqref{eq:r_star_ratio}, we observe that the non-integer $r^*$ computed above continuously increases with increasing $\epsilon_c/\epsilon_d$. We thus reason that the evolution of an integer $r^*$ proceeds by increments of 1. Determining this evolution then boils down to determining for which values of $\epsilon_c/\epsilon_d$ those increments occur.

We find that $e(r)=e(r+1)$ for $r={2}/({2-3\epsilon_c/\epsilon_d})$. When this quantity is an integer, the preferred integer assembly size increases by one. We deduce the following dependence for the integer radius $r^*$ on $\epsilon_c/\epsilon_d$ that minimizes the assembly energy:
\begin{equation}
\label{eq:thermo_size_prediction}
    r^*= 1+\mathrm{floor}\left( \frac{2}{2-3\epsilon_c/\epsilon_d } \right)
\end{equation}
We use \eqref{eq:thermo_size_prediction} for the theoretical curve of Fig.~3h of the main text. Inverting this expression now allows us to determine $\left(\epsilon_c/\epsilon_d\right)_\mathrm{lim}(r^*)$, the energy ratio for which assemblies of size $r^*$ become more favorable than those of size $r^*-1$. 
\begin{equation}
    \left(\epsilon_c/\epsilon_d\right)_\mathrm{lim}(r^*) = \frac{2}{3} \frac{r^*-2}{r^*-1}
    \label{eq:ratio_star}
\end{equation}
These ratios determine the boundaries between the stability regions of vortex assemblies of different sizes. We plot them as gray lines in the phase diagram of Fig.~2b of the main text.

\subsection{Generalization of the defect engineering principle}
\label{supp:other_geo}

The defect engineering principle that underpins the simple vortex assembly discussed above is in fact much more general than this specific design. Figure~4 of the main text demonstrates this point by showing additional examples of its ability to control size in self-assembly. In each of these cases, the balance between crystalline and defect interactions defines an emergent length scale in the assembly that sets the external shape of the assembly and/or to its internal structure. Here we apply the approach of Sec.~\ref{supp:phase_diagram} to determine the most favorable assembly for each of the cases discussed in Fig. 4 of the main text, as well as the three-fold design mentioned in relation to it.

\paragraph*{Fiber-like assembly.}
Unlike the vortex design presented in Figs.~2 and 3 of the text, the fiber design shown in Fig.~4a-c yields assemblies that are size-limited in one direction and extended in another. The choice of the extended direction occurs through a spontaneous breaking of symmetry. The internal structure of the fiber displays a periodic organization that sets its width, and the associated size is an emergent length scale potentially much larger than that of the subunit. That length is set through a mechanism similar to the one at work in the vortex assembly; here we compute it by minimizing the energy per subunit of a fiber-like assembly with respect to its width. In this geometry, the triangular crystalline regions are in contact through defect interactions between subunits with opposite orientations. We derive the energy per subunit of an infinitely long fiber by determining the number of crystalline and defect interactions as a function of the crystalline domain size $r$, as shown in the inset of Fig~.~\ref{fig:EnergyCalcul}e, yielding
%\begin{subequations}
\begin{equation}\label{eq:fiber_energy}
    %N(r) &= r(r+1)/2 \\
    %N_c(r) &= 3r(r-1)/2+1\\
    %N_d(r) &= 2r-1\\
    e^{(\mathrm{fiber})}(r, \epsilon_c, \epsilon_d)
    %&= \frac{N_c(r)\epsilon_c +N_d(r)\epsilon_d}{N(r)}
    = \frac{3\epsilon_c r^2 + r(4\epsilon_d-3\epsilon_c)+2\epsilon_c-2\epsilon_d}{r(r+1)}.
\end{equation}
%\end{subequations}
Similar calculations to those of Sec.~\ref{supp:energy_minimization} allow to build the phase diagram of Fig. 4b of the main text from Eq.~\eqref{eq:fiber_energy}. Fibers of finite width are the most favorable assembly for $2\epsilon_d<3\epsilon_c<0$, and their size $r^*$ is given by
\begin{equation}
    \frac{\epsilon_c}{\epsilon_d} = \frac{2 (r^*)^2-2r^*-1}{3 (r^*)^2-2r^*-1}
\end{equation}
In the simulations in Fig. 4c of the main text, we use energy ratios corresponding to the integer values $r^*=2, 4 \text{ or }6$ as indicated in the figure.

\paragraph*{Patterned bulk.}
Here we investigate another design, presented in Fig.~4e of the main text. The resulting assembly is not limited in size, but it displays a periodic internal organization. Similar to the vortex assembly radius, the period of that lattice is a potentially large emergent length scale set by the ratio of the crystalline and defect energies. Here we compute it by minimizing the assembly's energy per subunit. The defect interactions leading to patterned bulks are illustrated in Fig.~4e of the main text and in Fig.~\ref{fig:EnergyCalcul}f. They include the defect interactions involved in the 6-fold vortex assembly design, with additional interactions to connect several vortices into an infinite two-dimensional assembly. Each triangular wedge now has three defect lines. Using the expressions shown in Fig.~\ref{fig:EnergyCalcul}f we find
\begin{equation}
    %N(r) &= r(r+1)/2 \\
    %N_c(r) &=3r(r-1)/2\\
    %N_d(r) &= 3(2r-1)/2\\
    e^{(\text{patterned})}(r, \epsilon_c, \epsilon_d)
    %&= \frac{N_c(r)\epsilon_c +N_d(r)\epsilon_d}{N(r)}
    = \frac{3\epsilon_c r^2 - r(3\epsilon_c-6\epsilon_d)-3\epsilon_d}{r(r+1)}.
     \label{eq:energy_square_pattern}
\end{equation} 
This energy has a minimum for a finite $r^*$ if and only if $\epsilon_d<\epsilon_c<0$, and the value of $r^*$ is given by
\begin{equation}
    \frac{\epsilon_c}{\epsilon_d} = \frac{2(r^*)^2-2r^*-1}{2(r^*)^2}.
\end{equation}
In the simulations in Fig. 4c of the main text, we use energy ratios corresponding to the integer values $r^*=2, 4 \text{ or }6$ as indicated in the figure.

\paragraph*{Four-fold vortex assembly.}
\label{sec:good_square} 

\begin{figure}[t]
    \centering
    \includegraphics[width=0.99\linewidth]{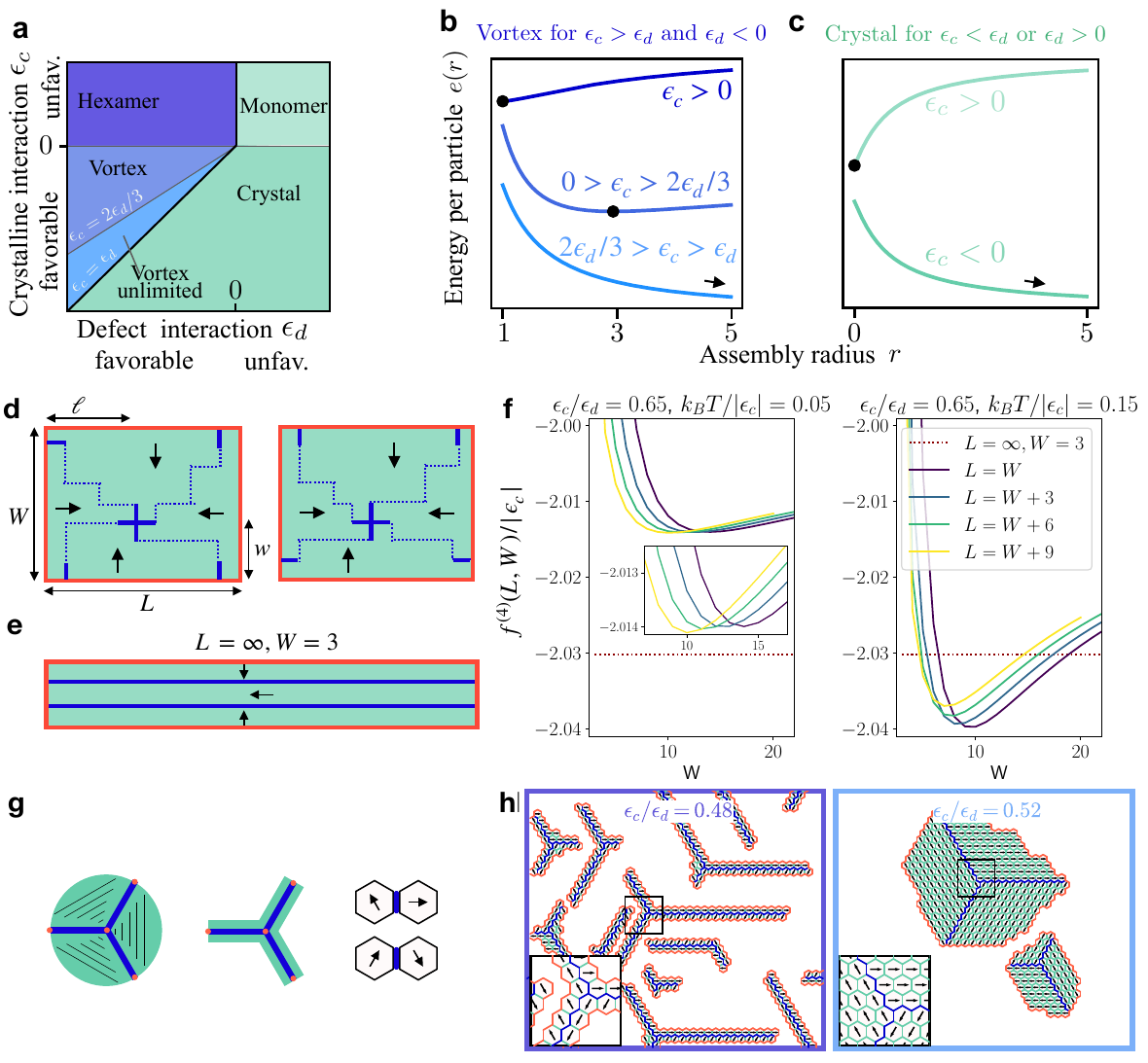}
    \caption{
    \textbf{Size selection in vortex assemblies}.\\
    (\textbf{a})~Favorable defects in hexagonal subunits result in the formation of vortex assemblies in the \emph{blue regions} of the phase diagram.
    (\textbf{b})~The vortex assembly energy has a minimum (\emph{black dot}) at finite $r^*$ if $2 \epsilon_d/3<\epsilon_c<0$.
    (\textbf{c})~There is no such non-trivial minimum for the crystal energy.
    (\textbf{d})~Rectangular vortex assemblies with length $L$ and width $W$ may differ by the conformation of their four grain boundaries, the position $(\ell, w$) of the central defect (\emph{blue cross}) and the chirality of the central defect with the corners (left-handed or right-handed). \emph{Black arrows} illustrates the orientation of the crystalline domains. The fluctuating defect lines are shown as \emph{dotted blue lines}.
    (\textbf{e})~Geometry of an infinitely long assembly of width $3$.
    (\textbf{f})~Free energies per subunit for rectangular vortex assemblies (\emph{solid lines}) and width-3 fibers (\emph{dotted line}) for two different temperatures. The legend for both plots is shown in the right panel. The magnification of the curves' minima in the inset of the left panel shows that symmetric assemblies are less favorable than assymetric ones at low temperature. The right panel shows that they are however more favorable at higher temperature.
    (\textbf{g})~A three-fold vortex design may form unlimited assemblies and thin defect lines. (\textbf{h})~Simulations show that thin defect lines dominate for $\epsilon_c/\epsilon_d<0.5$, and that unlimited vortices form for $\epsilon_c/\epsilon_d>0.5$. See details of the simulation parameters in Tab.~\ref{tab:simu_parameters}.
    }
    \label{fig:phaseD_square_threeL}
\end{figure}

As stated in the main text, understanding the stability of four-fold vortex assemblies requires a study of both their energy and entropy. Here we first compute the energy per subunit of square vortices, and show that it displays a minimum at a finite size $r$. This energy minimum however remains less favorable than that of very elongated assemblies of width $3$, implying that four-fold vortex assemblies should not form at $T=0$.
At higher temperatures, we show that the entropy associated with the many possible defect configurations stabilizes vortex assemblies and makes them the most favorable assemblies in terms of free energy.

We consider square subunits as in Fig. 4g of the main text. Using the schematic and expressions shown in Fig.~\ref{fig:EnergyCalcul}g, we compute the energy per subunit as:
\begin{equation}\label{eq:e_square}
    e^{(\text{4,vortex})}(r, \epsilon_c, \epsilon_d)
    %&= \frac{N_c(r)\epsilon_c +N_d(r)\epsilon_d}{N(r)}
    = 2\frac{\epsilon_c r^2 + r(2\epsilon_d-\epsilon_c)}{(r+1)^2}.
\end{equation}
For $\epsilon_d<3\epsilon_c/2<0$, this expression has a minimum $e^{(\text{4,vortex})}(r^*, \epsilon_c, \epsilon_d)=(4\epsilon_d^2-4\epsilon_d\epsilon_c+\epsilon_c^2)/[4(\epsilon_d-\epsilon_c)]$ for $r^*=(2\epsilon_d-\epsilon_c)/(2\epsilon_d-3\epsilon_c)$. This minimum is however always less favorable than the energy per subunit of the width-3 fiber illustrated in Fig.~\ref{fig:phaseD_square_threeL}d, which reads
\begin{equation}\label{eq:width-3_fiber_energy}
e^{(4,\text{3-fiber})}(\epsilon_c, \epsilon_d)=\epsilon_c+\frac{2}{3}\epsilon_d.
\end{equation}
This implies that square vortex assemblies are never stable at zero temperature. 

We now consider rectangular vortex assemblies with length $L\geq 4$ and width $W\geq 4$ (Fig.~\ref{fig:phaseD_square_threeL}d). The expression of Eq.~\eqref{eq:e_square} generalizes to 
\begin{equation}\label{eq:rectangle_energy}
    e^{(\text{4,vortex})}(L, W,\epsilon_c, \epsilon_d) = \frac{(2L+2W-4)\epsilon_d + (2LW - 3L - 3W +4)\epsilon_c}{LW},
\end{equation}
where Eq.~\eqref{eq:e_square} is recovered for $L=W=r+1$. To compute the conformational entropy associated with the grain boundaries of this assembly, we first consider the grain boundary that runs from the bottom-left corner of the assembly to the central defect as a discrete directed random walk. Each step in the walk is of unit length and proceeds either to the right or upwards. One such path is illustrated in Fig.~\ref{fig:phaseD_square_threeL}d as a dotted line, with the central defect marked by a cross. The number of paths from $(1,1)$ to $(\ell-1, w)$ respecting those conditions is given by the binomial coefficient $\binom{\ell-2+w-1}{\ell-2}$. We repeat this calculation for all four defect lines, accounting for the $2$ possible connectivities between the central defect and the defect lines, and the $2^4$ possible connectivites of the defects to the corners (illustrated in the two examples of Fig.~\ref{fig:phaseD_square_threeL}d). Finally we sum over all positions of the central defect and find
\begin{multline}\label{eq:rectangle_entropy}
    S^{(\text{4,vortex})}(L,W) =  k_B \ln \left[32 \sum_{\ell=2}^{L-1}\sum_{w=2}^{W-1} \binom{\ell+w-3}{\ell-2} \binom{L-\ell+w-3}{L-\ell-1} \binom{\ell+W-w-3}{\ell-1} \binom{L-\ell+W-3}{L-\ell-2} \right].
\end{multline}
Combining Eqs.~\eqref{eq:rectangle_energy} and \eqref{eq:rectangle_entropy}, we write the free energy per subunit of the square vortex assembly as
\begin{equation}
    f^{(\text{4,vortex})}(L,W,\epsilon_c, \epsilon_d) = e^{(\text{4,vortex})}(L, W,\epsilon_c, \epsilon_d) - T \frac{S^{(\text{4,vortex})}(L,W)}{LW}
\end{equation}

To determine the relative stability of the square vortex assembly and the width-$3$ fiber, we show the free energy of $f^\mathrm{(\text{4,vortex})}/|\epsilon_c|$ for two different values of $k_BT/|\epsilon_c|$ in Fig.~\ref{fig:phaseD_square_threeL}f. We compare it to the free energy per subunit of the width-3 fiber, which is equal to the expression of $e^\mathrm{(4,\text{3-fiber})}$ given in Eq.~\eqref{eq:width-3_fiber_energy} due to the fact that the fiber has no internal degrees of freedom and thus a vanishing entropy per subunit. This representation makes it apparent that the higher temperature stabilizes the vortex assemblies with respect to the fibers. Furthermore, among the possible vortex assemblies squares (\emph{i.e.}, assemblies with $L=W$) are favored at $k_BT/|\epsilon_c| = 0.15$, while more elongated morphologies have a lower free energy at $k_BT/|\epsilon_c| = 0.05$.

These findings are confirmed by the numerical simulation shown in Fig.~4f of the main text using the same parameters as in Fig.~\ref{fig:phaseD_square_threeL}f. %In 

\paragraph*{Non-size-controlled design.}
Here we show that the four-fold vortex assembly shown in Fig. 4g of the main text does not allow for size control.  
The schematic of Fig.~\ref{fig:EnergyCalcul}h indeed yields
\begin{equation}
    %N(r) &= r(r+1)/2 \label{eq:N_vortex}\\
    %N_c(r) &=r(r-1)\\
    %N_d(r) &= r\\
     e^{(\text{non-size-controlled})}(r, \epsilon_c, \epsilon_d)
     %&= 4\frac{N_c(r)\epsilon_c +N_d(r)\epsilon_d}{4N(r)}
     %= 2\frac{\epsilon_c r - (\epsilon_c-\epsilon_d)}{r+1}
     = \frac{2 \epsilon_c r - (\epsilon_d-2\epsilon_c)}{r}
     \label{eq:energy_square_invalid}
\end{equation}
This energy is a monotonic function of $r$ and thus cannot display a minimum for a finite and positive $r^*$. Additionally the defect lines associated with this design are straight and thus cannot give rise to the type of entropic effects discussed above.

\paragraph*{Three-fold vortex assembly.}

Here we study a type of assembly mentioned but not pictured in the main text, namely three-fold vortex assemblies. Similar to the non-size controlled square assemblies above, these do not give rise to size-controlled vortex assemblies. Instead they either form unlimited assemblies or a collection of thin defect lines.

The design considered here involves hexagonal subunits with defect interactions between subunits whose orientations differ by $2\pi/3$. These allow for the formation of ideal three-fold vortex assemblies, whose energy per subunit reads (Fig.~\ref{fig:EnergyCalcul}i):
\begin{equation}
    %N(r) &= r^2 \\
    %N_c(r) &=3r^2-4r+1\\
    %N_d(r) &= 2r-1\\
    e^{(\text{3,vortex})}(r, \epsilon_c, \epsilon_d)
    %= 3\frac{N_c(r)\epsilon_c +N_d(r)\epsilon_d}{3N(r)}
    = \frac{3\epsilon_c r^2 - r(2\epsilon_d-4\epsilon_c)+\epsilon_c-\epsilon_d}{r^2}
     \label{eq:energy_threefold}
\end{equation}
This energy has a minimum for a finite $r^*$ if and only if $\epsilon_d<2\epsilon_c<0$.
We however find that in this regime vortices with radii $r\geq 1$ are never as energetically favorable as the two-fiber illustrated in Fig.~\ref{fig:EnergyCalcul}j and whose energy per subunit reads 
\begin{equation}\label{eq:2-fiber}
    e^{(2\text{-fiber})}=\epsilon_d+\epsilon_c.
\end{equation}
By contrast, in the regime $2\epsilon_c<\epsilon_d<0$, the most favorable vortex has size $r^*\rightarrow\infty$ and is more stable than the width-2 fiber. Consistent with these estimates, the simulations of Fig.~\ref{fig:phaseD_square_threeL}g show the formation of width-2 fiber for $\epsilon_c/\epsilon_d<1/2$, and of large vortices for $\epsilon_c/\epsilon_d>1/2$

\paragraph*{Three-dimensional assemblies.}
\label{sec:supp_three_d}

\begin{figure}
    \centering
    \includegraphics[width=0.92\linewidth]{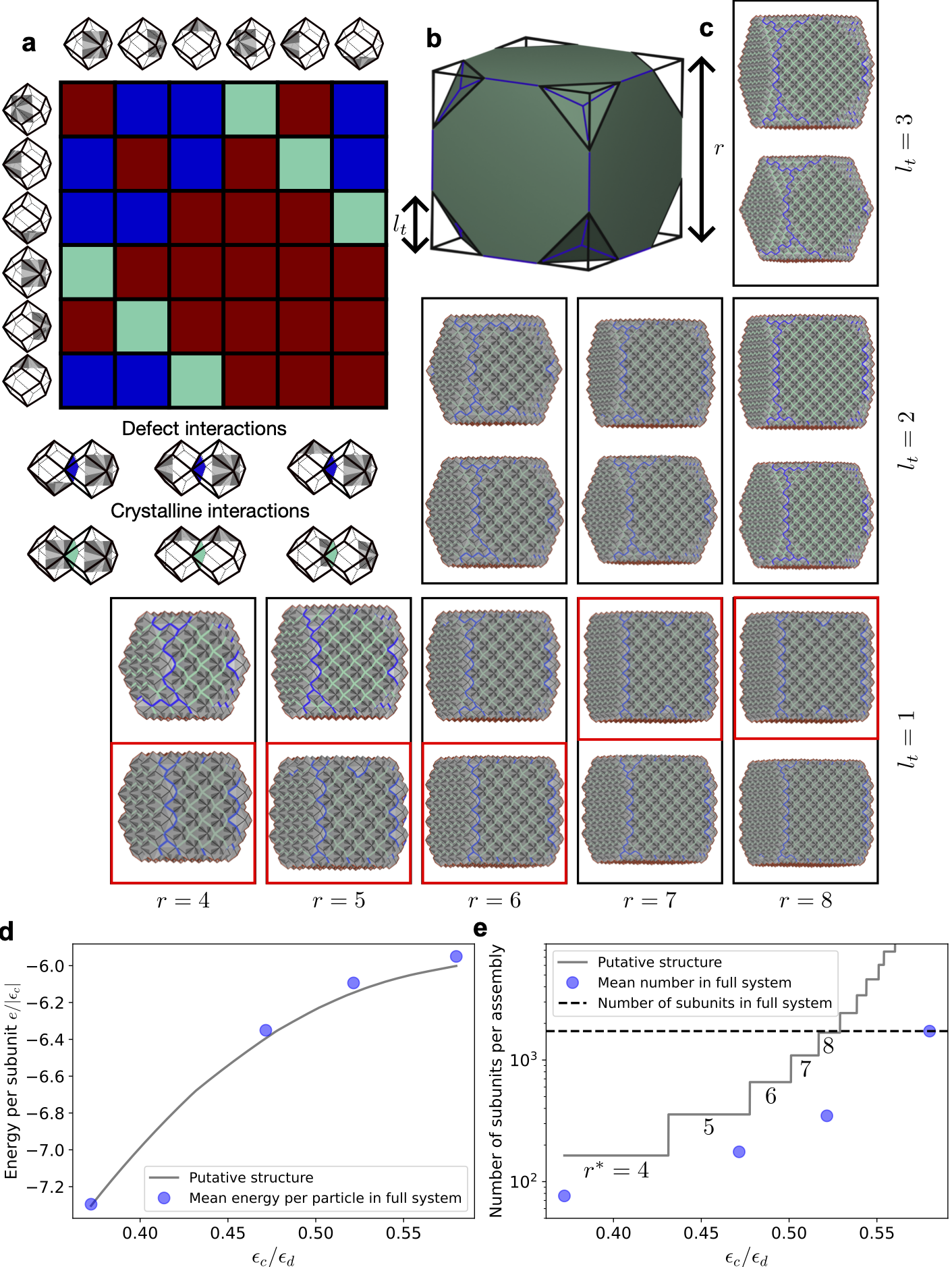}
    \caption{\textbf{Size limitation in the three-dimensional design.}\\
    (\textbf{a}) Full interaction map of our three-dimensional design (\emph{top}), accompanied by schematics highlighting the relative subunit orientations leading to defect and crystalline interactions (\emph{bottom}).
    (\textbf{b}) Putative truncated-cubic assembly structure.
    (\textbf{c}) Putative assemblies for radii $4 \leq r \leq 8$, and truncations $1 \leq l_{t} \leq \lfloor \frac{r}{2} - 1 \rfloor$. We show two structures for each $(r,l_t)$ as discussed in the text. The most favorable structures are outlined by \emph{red boxes}.
    (\textbf{d}) Comparison of the optimal energy per subunit (\emph{left}) and (mean) assembly size (\emph{right}) obtained on the one hand by comparing ideal putative structures, and on the other through the simulations of Fig.~4h. }
    \label{fig:supp_3d}
\end{figure}

Here we present energetic considerations to rationalize the observation of size control in the three-dimensional design shown in Fig. 4h. We show the full interaction map associated with this design in Fig.~\ref{fig:supp_3d}a. 
The orientations shown in this interaction map do not represent all 24 possible
orientations of a rhombic dodecahedron. Instead it shows the six non-redundant orientations for a subunit with $D_4$ symmetry about an axis that goes through two opposite vertices of the rhombic dodecahedron, as is the case in this design.
The geometry of the problem considered here is more complex than those of the two-dimensional systems studied above, and makes an analytical discussion more difficult. We thus instead base our discussion on numerical comparisons between the energies of a range of putative assembly structures.

The Monte-Carlo simulations of Fig. 4h typically produce assemblies with the structure of a truncated cube. We thus adopt this structure as our putative structure. We show our parametrization of the truncated cubes in Fig.~\ref{fig:supp_3d}b. The assembly radius $r$ denotes a length of the side of the cube, expressed in number of subunits.
At each corner of the cube, we remove a trirectangular tetrahedron whose right angle coincides with the cube corner, and whose legs of length $l_{t}$ run along the cube edges. These legs must be shorter than half of the edge length, imposing $1 \leq l_{t} \leq \lfloor \frac{r}{2} - 1 \rfloor$.
The assembly has six approximately pyramidal crystalline domains as illustrated in the left panel of Fig. 4h.

To generate all valid designs of radius $4 \leq r \leq 14$, we first compute the intersection of truncated cubes of given radii and truncation lengths
with a fully-packed FCC cell. We then orient the subunits in each crystalline domain towards the corresponding aggregate outer face. Due to the mismatch between the cubic geometry of the assembly and the rhombic-dodecahedral shape of the subunits that compose it, two different assemblies are possible for a given $(r, l_{t})$. These two possible morphologies differ by a translation of the truncated cube bounding the assembly by the length of one subunit in the direction of any of the cubic faces of the assembly. We take both possible morphologies associated with each $(r, l_{t})$ into account.

While this simple geometrical construction recapitulates the overall crystalline domain structure observed in our Monte-Carlo simulations, it does not optimize the specific orientations of its subunits close to the grain boundaries between these domains. To fine-tune this grain boundary structure, we conduct a simulation at $k_BT = 0$ where only subunit rotations are allowed. Our simulation consists in attempting $10^5$ Monte-Carlo moves par subunit. To illustrate the outcome, we show all post-annealing putative structures of radii $4 \leq r \leq 8$ in Fig.~\ref{fig:supp_3d}c.

For each annealed structure, we extract the total number of crystalline interactions, defect interactions and subunits. We then compute the energy per subunit $e(r)$ as a function of $\epsilon_c$ and $\epsilon_d$ for each structure as in the caption of Fig.~\ref{fig:EnergyCalcul}. This allows us to compare the energies per subunit of all our putative structures as a function of $\eced$
and to determine the most stable one for each value of this energy ratio. We find that the optimal value $r^*$ of the assembly radius depends on $\eced$ (Fig.~\ref{fig:supp_3d}d), while the optimal truncation length is always $\ell_t=1$ (Fig.~\ref{fig:supp_3d}c, red boxes).

The simulations shown in Fig. 4h of the main text are performed at $\eced$ values where we predict the formation of truncated cubes with assembly radii of 4, 5, and 8, in addition to an extra point at large $\eced$ where we expect an optimal assembly with more subunits than are present in the simulation (see Tab.~\ref{tab:simu_parameters}).
We find that the energy per subunit in the idealized and annealed structures are in close agreement (Fig.~\ref{fig:supp_3d}d). The latter systematically have larger energy per subunit than the former, probably due to imperfect annealing and/or finite size effects.
The corresponding assembly sizes are also in reasonable agreement (Fig.~\ref{fig:supp_3d}e).

\subsection{Effect of thermal fluctuations on vortex assemblies}
\label{supp:size_estimate}

The discussion of Sec.~\ref{supp:phase_diagram} and the resulting phase diagram of Fig.~2b relate to the ground state of our vortex assemblies. Here we assess the effect of thermal fluctuations, which are present in both our experiments and simulations, on the assembly structure. We find that they promote the formation of branches of the type depicted in Fig.~\ref{fig:EnergyCalcul}, but do not significantly affect the assembly radius $r$. The lines in Fig.~2d of the main text are computed using the results of this section.

\paragraph*{Energetic contribution of branches.}
The simulation results presented in Fig.~2c of the main text show that some vortex assemblies deviate from the perfectly hexagonal structures considered in Sec.~\ref{supp:phase_diagram}. Here we rationalize this observation by showing that forming branches is energetically favorable at low $\epsilon_c/\epsilon_d$.

We consider a hexagonal assembly with branches, and denote by $b$ the total number of subunits within these branches (shown in red in Fig.~\ref{fig:EnergyCalcul}b).
The total number of subunits, crystalline interactions and defect interactions in the whole assembly respectively read
\begin{subequations}
\begin{align}
N(r,b) &= 3r(r+1) + b\\
N_c(r,b) &= 6\times [3r(r-1)/2] + b\\
N_d (r,b) &= 6\times (2r-1) + b.
\end{align}
\end{subequations}
This yields an energy per subunit
\begin{equation}
\label{eq:e_cam_branch}
e^{(\text{branched vortex})}(r,b) = \frac{9r(r-1)\epsilon_c + b\epsilon_c + 6(2r-1)\epsilon_d + b\epsilon_d}{3r(r+1) + b}.
\end{equation}

To find the most energetically favorable values of $r$ and $b$ we first seek to minimize this expression with respect to $b$. We find 
\begin{equation}
    \partial_b e^{(\text{branched vortex})}(r,b) = - \epsilon_d(r-2)\left[r\left(1-2\frac{\epsilon_c}{\epsilon_d}\right)-1\right]
\end{equation}
This expression is independent of $b$, indicating that the branch length $b^*$ that minimizes $e^{(\text{branched vortex})}$ can only have either one of two values: $b^*=0$ if $\partial_b e^{(\text{branched vortex})}>0$ and $b^*=\infty$ if $\partial_b e^{(\text{branched vortex})}<0$. If $b^*=0$,   $e^{(\text{branched vortex})}(r,0) =  e^{(\text{vortex})}(r)$ and the energy minimum is found for $r=r^*$ [Eq.~\eqref{eq:rmin_camembert}]. If  $b^*=\infty$, the energy per subunit is independent of $r$ and equal to the energy $e^{(\text{2-fiber})}$ of an infinite fiber of width~2 [Eq.~\eqref{eq:2-fiber}]. Comparing $e^{(\text{vortex})}(r^*)$ and $e^{(\text{2-fiber})}$, we find that the former energy per subunit is lower than the latter as long as $\epsilon_c/\epsilon_d>1/4$. Therefore for $\epsilon_c/\epsilon_d>1/4$ branch formation is energetically suppressed.

In the case $\epsilon_c/\epsilon_d=1/4$, $e^{(\text{branched vortex})}$ does not have a single minimum but a line of degenerate minima with $r=2$ and arbitrary $b$. This implies that branch formation is not energetically suppressed, and thus rationalizes the presence of branches in the leftmost simulations panel of Fig.~2c of the main text. Smaller values of $\epsilon_c/\epsilon_d$ are not well described by our treatment of $r$ as a continuum variable and are not discussed here.

\paragraph*{Assembly size distribution.}
Thermal fluctuations can result in assemblies with radii other than $r^*$ or the formation of branches even in cases where they are not favored.
Here we determine the expected distribution of assembly and branch sizes.

We consider an ideal gas of non-interacting assemblies. We denote by $m(r,b)$ the number of assemblies of radius $r$ with branches containing a total of $b$ subunits. We denote the number of unassembled monomeric subunits as $m(0,0)$. Due to the lack of interactions between assemblies, the partition function of the total system $Z$ factorizes as
\begin{equation}
Z = \prod_{r,b}\frac{1}{m(r,b)!}\bigl[z(r,b)\bigr]^{m(r,b)},
\end{equation}
where the partition function $z(r,b)$ of an individual assembly reads
\begin{equation}
z(r,b)=N_\mathrm{sites} \times N_\text{config}(b)\times \exp\left[-\beta N(r,b)e^{(\text{branched vortex})}(r,b)\right].
\end{equation}
Here $N_\mathrm{sites}$ is the number of accessible sites in the whole system, $N_\text{config}(b)$ is the number of possible ways in which a vortex assembly can be decorated by a total of $b$ branch monomers, and $\beta=1/k_BT$ is the inverse temperature. The energy for a monomer is $e^{(\text{branched vortex})}(0,0)=0$.
Using the Stirling approximation and introducing the concentration of $(r,b)$-assemblies $c(r,b)=m(r,b)/N_\mathrm{sites}$ we obtain a free energy per unit volume 
\begin{equation}
f = -\frac{\ln Z}{N_\mathrm{sites}\beta} = \sum_{r,b} N(r,b) c(r,b) \left[ \frac{1}{N(r,b)\beta} (\ln c(r,b) -1) + N_\text{config}(b) e^{(\text{branched vortex})}(r,b)\right].
\end{equation}

We minimize $f$ with respect to the concentrations $c(r,b)$ while holding the total number of subunits constant by setting
\begin{subequations}
\begin{align}
\frac{\partial}{\partial c(r,b)} \left[f - \mu\sum_{r,b}N(r,b)c(r,b) \right]
&= 0\\
\sum_{r,b}N(r,b)c(r,b) &= \frac{N_{\mathrm{subunits}}}{N_{\mathrm{sites}}},\label{eq:r_b_distrib_sum}
\end{align}
\end{subequations}
where the thus introduced Lagrange multiplier $\mu$ is the subunit chemical potential. This yields
\begin{equation}\label{eq:concentration_distribution}
    c(r,b) = \left[c_0N_\text{config}(r,b)\text{e}^{-\beta e^{(\text{vortex branched})}(r,b)}\right]^{N(r,b)},
\end{equation}
where $c_0=c(0,0)=e^{-\beta\mu}$ denotes the concentration of the monomeric subunits.

To determine the number of configurations $N_\text{config}(b)$, we consider all possible ways in which the $b$ undistinguishable monomers can be partitioned into six groups representing the six distinguishable branches of a branchy vortex assembly. To illustrate this choice of a partition, consider a group of $b=5$ monomers. These can be distributed by assigning to the $i$th branch a number of monomer equal to $v_i$, where the vector $\mathbf{v}$ can for instance be given by $\mathbf{v}=(2,1,0,0,2,0$). We reason that choosing such a vector with $\sum_i v_i=b$ is equivalent to choosing $5$ sites out of a set of $b+5$. Indeed, consider a set of $5+5=10$ empty squares in our example:
$$\square\square\square\square\square\square\square\square\square\square$$
then choose 5 of them and color them black:
$$\square\square\blacksquare\square\blacksquare\blacksquare\blacksquare\square\square\blacksquare$$
In this example, the number of white squares between the beginning of the list and the first black square is $2$, implying $v_1=2$. Counting white squares between the first and the second black square yields $v_2=1$, \emph{etc}. until the whole vector $\mathbf{v}$ given above is recovered. There are $\binom{b+5}{5}$ ways of picking the black squares, and therefore
\begin{equation}
\label{eq:r_b_distribution}
N_\text{config}(b)=\binom{b+5}{5}.
\end{equation}
Finally, we determine $c_0$ numerically by solving Eq.~\eqref{eq:r_b_distrib_sum}.

\begin{figure}
    \centering
    \includegraphics[width=0.99\linewidth]{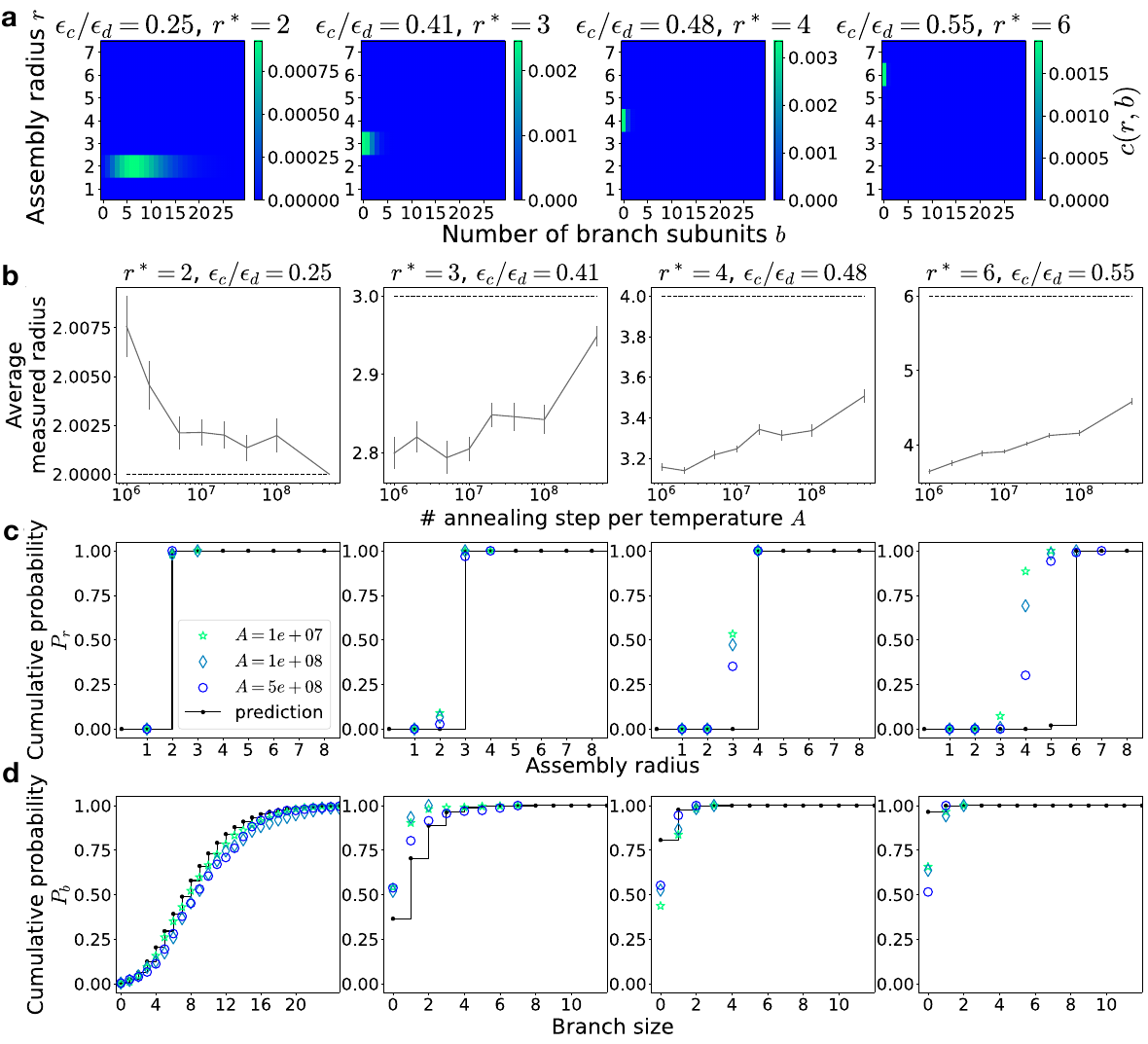}
    \caption{
    \textbf{Effect of entropy and annealing time on the assembly size and shape} \\ 
    (\textbf{a})~Theoretically predicted assembly radius and branch size distributions.
    Branches do not form for assemblies with equilibrium radii $r^*>2$, and their size distribution is determined by entropic effects for $r^*=2$.
    (\textbf{b})~The assembly radius increases with increasing annealing time in large assemblies, and tends to converge towards the theoretical prediction (\emph{dashed line}).
    (\textbf{c})~Upon sufficient equilibration, the assembly radius distribution defined in Eq.~(2) of the Methods is well predicted by the theory.
    (\textbf{d})~A similarly defined branch size distribution also agrees with theory.}
    \label{fig:branches_explanation}
\end{figure}

We plot the thus determined distributions in Fig.~\ref{fig:branches_explanation}a using the parameters of our main text simulations, and find that our formalism predicts very little branching except in the case $r^*=2$, as well as radii distributions that are highly peaked around the predicted ground-state $r^*$.
\begin{equation}
\label{eq:cumulative_distribution_theory}
P_r(r)=\text{Proba}(\text{radius}\leq r) =
\frac{N_\text{sites}}{N_\text{subunits}}\sum_{r'= 0}^{r}\sum_{b=0}^{+\infty}  c(r',b),
\end{equation}
and compare the result to our simulations by applying Eq.~2 of the Methods in Fig.~2d of the main text.

\paragraph*{Enhanced theory–simulation agreement via extended annealing.}

In Fig.~2d and 3h of the main text, we observe a small discrepancy between the predicted and simulated assembly sizes. Here, we show that increasing the annealing time steadily improves agreement with theory, highlighting the reliability of the prediction despite computational constraints. We vary the annealing time $A$ for the binding energy ratios $\epsilon_c/\epsilon_d$ discussed in the main text, and track how the average assembly radius and the distributions of radii and branch sizes evolve. As shown in Fig.~\ref{fig:branches_explanation}b, the assembly radius converges quickly toward the predicted equilibrium value $r^*$ for $r^*=2$ and $r^*=3$. For larger target sizes ($r^*=4$ and $r^*=6$), equilibration requires significantly more computational effort, explaining the small discrepancy observed in Fig.~3h. In panels c and d of Fig.~3, we display the distributions of assembly radii and branch sizes across different annealing protocols, compared to the theoretical predictions derived from Eq.~\eqref{eq:cumulative_distribution_theory}, and confirm that assemblies are not fully equilibrated for $r^*=4$ and $r^*=6$. Longer annealing times lead to closer agreement between simulated and theoretical assembly properties. The qualitative agreement of our simulation results with our predictions even in cases of imperfect equilibration highlights the robustness of our defect engineering concept.

\section{\label{sec:robustness}Robustness of self-limiting assembly}

The theoretical model and lattice simulations presented in the main text involve several idealizations of the self-assembly process. Here we assess the robustness of the formation of vortex assemblies as we relax some of these assumptions. In Sec.~\ref{sec:constant_T}, we show that their formation does not require simulated annealing, and also occurs in constant-temperature simulations. Sec.~\ref{sec:randomized} demonstrates they are moreover robust to errors in the interactions between subunits, as well as to the introduction of faulty subunits. The simulations in these two subsections are conducted according to the protocol described in the main text and methods unless otherwise specified. Finally in Sec.~\ref{sec:off_lattice} we recapitulate vortex assembly in off-lattice simulations, implying that our main text results are not mere artifacts of our lattice formalism.

\subsection{\label{sec:constant_T}Comparing annealing with constant-temperature simulations}

Throughout the main text we present lattice-based simulations that are meant to approximate the equilibrium state of our system, as motivated by the observation that our DNA origami vortex assemblies appear to equilibrate over the course of our experiments (Fig.~3f of the main text). To speed up the equilibration process, we implement a standard simulated annealing scheme whereby we slowly decrease the temperature of the system over the course of the simulation (see Methods). This procedure may not be a good model for the experiment due to the fact that self-assembly is often strongly influenced by kinetic traps. Such traps could indeed dramatically impact the outcome of a constant-temperature assembly process similar to the one at work in our experiments, while our annealed simulation might bypass them. This would make our simulation a poor model of the experiment.

\begin{figure}
    \centering
    \includegraphics[width=\linewidth]{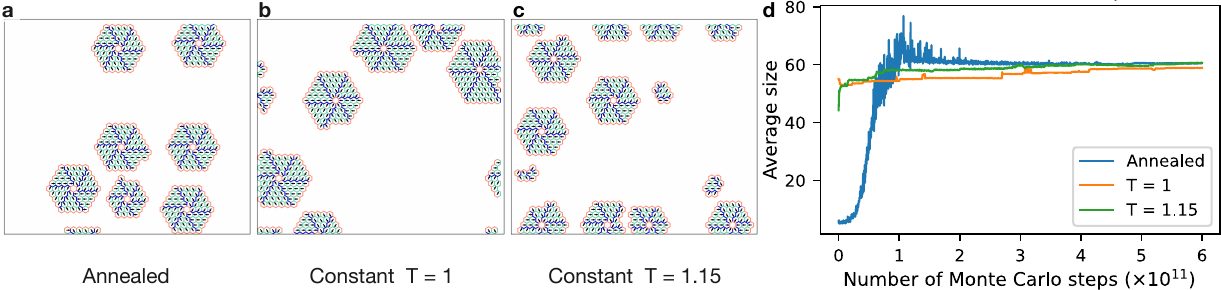}
    \caption{
    \textbf{Simulated annealing makes equilibration more effective without noticeably biasing the outcome of self-assembly.}\\
    (\textbf{a})~Outcome of a vortex assembly simulation for $\epsilon_c/\epsilon_d=0.52$ (target radius $r^*=5$) using simulated annealing. The simulation parameters are similar to the ones used in the corresponding panel of Fig.~2c of the main text. The simulation performs $3\times 10^8$ Monte-Carlo steps for each of the $2000$ temperature steps, hence a total of $6\times 10^{11}$ Monte-Carlo steps. See Movie~SM2 for the full simulation run.
    (\textbf{b})~Outcome of a constant temperature simulation for the same parameter values and a temperature equal to $1$. The simulation comprises $6\times 10^{11}$ Monte-Carlo steps. See Movie~SM3 for the full simulation run.
    (\textbf{c})~Outcome of a constant temperature simulation for the same parameter values and a temperature equal to $1.15$. The simulation comprises $6\times 10^{11}$ Monte-Carlo steps.
    (\textbf{d})~Mean assembly size over the course of the three simulation schemes described in the previous panels. The statistics of each constant temperature curve are collected over eight simulation runs, while the annealing statistics are collected over four runs.
    }
    \label{fig:anneal_vs_constant_T}
\end{figure}

To verify that our annealing procedure does not introduce any such strong bias in our simulations, we compare its outcome to that of a constant-temperature simulation. As shown in Fig.~\ref{fig:anneal_vs_constant_T}a-b, the simulations yield similar final configurations. To demonstrate that this result does not strongly depend on the value of the temperature of the finite-temperature simulation, we additionally conducting another at slightly higher constant temperature (Fig.~\ref{fig:anneal_vs_constant_T}c). We furthermore monitor the evolution of the average assembly size over the course of equilibration in all three simulations schemes in Fig.~\ref{fig:anneal_vs_constant_T}d. In all cases the assembly sizes appear to converge towards the same long-time limit. This convergence is slower in the constant-temperature equilibration however, and the implied added computational cost motivates our use of simulated annealing to equilibrate our system in all lattice simulations presented in the main text.

Finally, while the similarity between the outcomes of the annealed and the constant-temperature simulations indicates that our simulation scheme is robust, it does not guarantee that it actually corresponds to the system's equilibrium state. It indeed leaves open the possibility that the system enters a kinetic trap that affects the annealed and constant-temperature simulations alike. The existence and nature of such a trap is further discussed in Sec.~\ref{sec:circumvent_nucleation}.

\subsection{\label{sec:randomized}Randomized interaction maps}

Interaction energies in experiments are unlikely to be exactly equal to the ones prescribed by theory. Here we demonstrate that such deviations do not compromise our designs. We do so by conducting simulations of two variants of the $\epsilon_c/\epsilon_d=0.4792$ setup of Fig.~2c of the main text, which corresponds to a target radius $r^*=4$.

\begin{figure}
    \centering
    \includegraphics[width=0.9\linewidth]{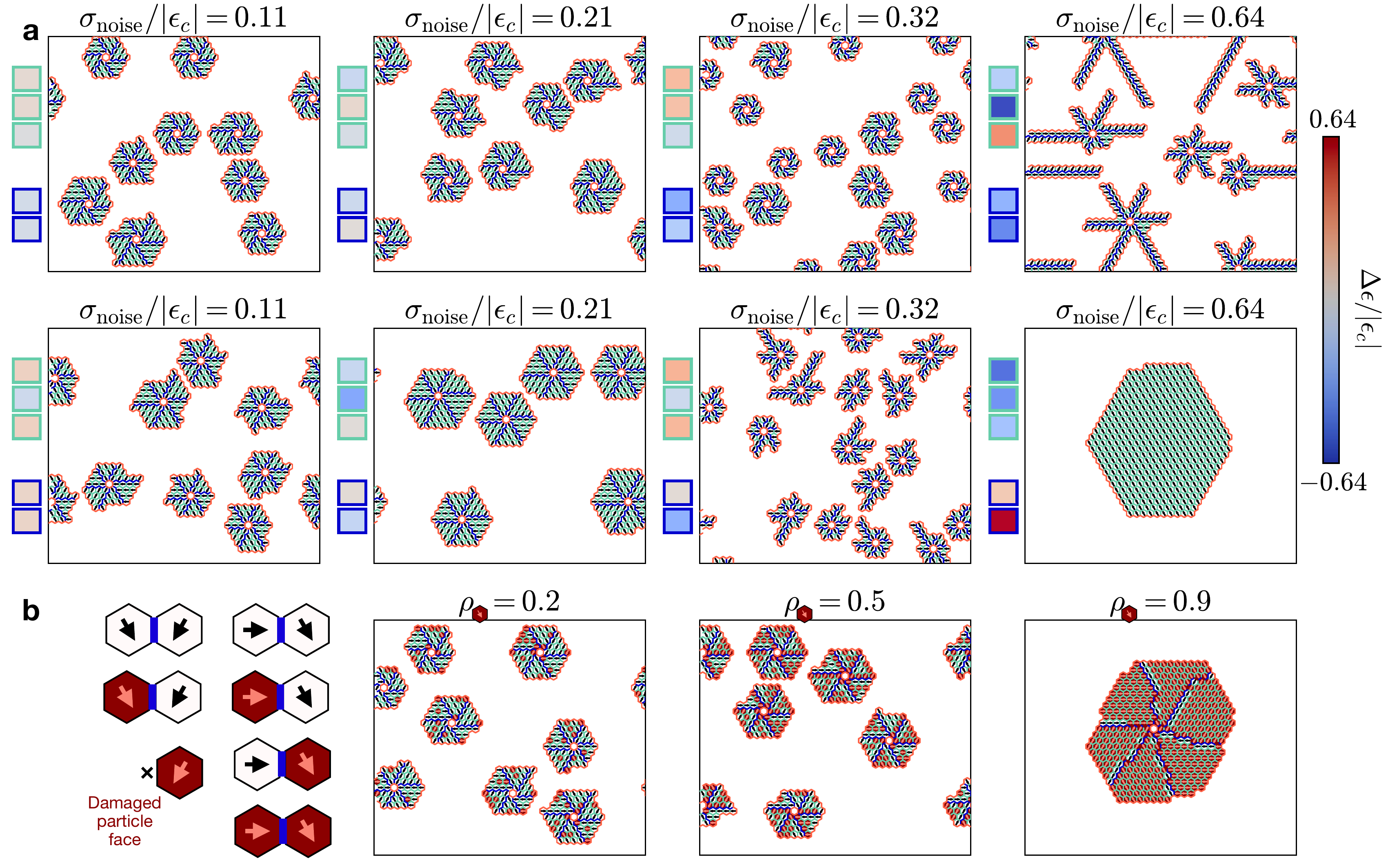}
    \caption{
    \textbf{Vortex assemblies are robust to noisy interactions.}\\
    (\textbf{a})~Effect of a zero-mean Gaussian noise of variance $\sigma_\text{noise}^2$ in the crystalline and defect interaction. The colors in the boxes on the left of each snapshot show the magnitude $\Delta \epsilon$ of the excess (\emph{red}) or deficit (\emph{blue}) associated with each energy. \emph{Green (blue) box outlines} denote crystalline (defect) interactions. The sizes of the vortex assemblies change only at high noise levels, while their shapes remain consistent with those observed in the ideal case discussed in the main text.
    (\textbf{b})~Assembly of a mixture of normal subunits and faulty subunits (shown in \emph{red}) missing one defect interaction. Assembly sizes and shapes remain unchanged for faulty subunit fractions up to $\rho=0.5$.
    }    \label{fig:randomized_maps}
\end{figure}

We first investigate a possible systematic deviation of the interactions from the prescribed values $\epsilon_c = -18.7 k_BT$ and $\epsilon_d = -39.03 k_BT$. We draw each of the three crystalline interactions and each of the two defect interactions from a Gaussian distribution centered on those ideal values, using the standard deviation of the distribution as a parameter. We present our results in Fig.~\ref{fig:randomized_maps}a, where the subunits within each snapshot are identical and correspond to a specific set of interactions. We see that shifts of the interaction energies by an amount of the order of $10\%$ of the crystalline interaction hardly modify the final assemblies, and that a perturbation of the order $20\%$ only induces a slight shift in the vortex size, while preserving their overall morphology.

Another possible imperfection consists in the presence of faulty subunits, for instance due to a failure to incorporate some DNA staples over the course of the assembly of the DNA nanocylinders. To illustrate this case, in Fig.~\ref{fig:randomized_maps}b we consider a situation where a fraction $\rho$ of the subunits in the system are faulty. In this example, one of the faces of the faulty subunits is damaged and is thus incapable of forming defect interactions. This could model a situation where some DNA nanocylinders miss the interaction strands labeled $\text{D}_21$ and $\text{D}_22$ in Fig.~3a of the main text. The schematic on the left of Fig.~\ref{fig:randomized_maps}b illustrates the remaining possible defect interactions. The simulation results show little change in the vortex morphologies when up to $50\%$ of the subunits are faulty. The faulty subunits are indeed incorporated into the crystalline domains of the vortex assembly, where their impaired defect interaction are not used anyway. By contrast, using a very large fraction of faulty subunits shifts the equilibrium in favor of the formation of very large crystalline domains, and thus distorts the final shape of the vortex assembly. Although such large faulty populations are unlikely to occur using the experimental protocol presented in the main text, from a design perspective a similar diversity in subunit binding profiles could be induced on purpose and harnessed as an additional lever to control the aggregate morphology. A discussion of defect engineering in the presence of multiple subunit types is however beyond the scope of the present study.

\subsection{\label{sec:off_lattice} Off-lattice simulations}

\begin{figure}
    \centering
    \includegraphics[width=\linewidth]{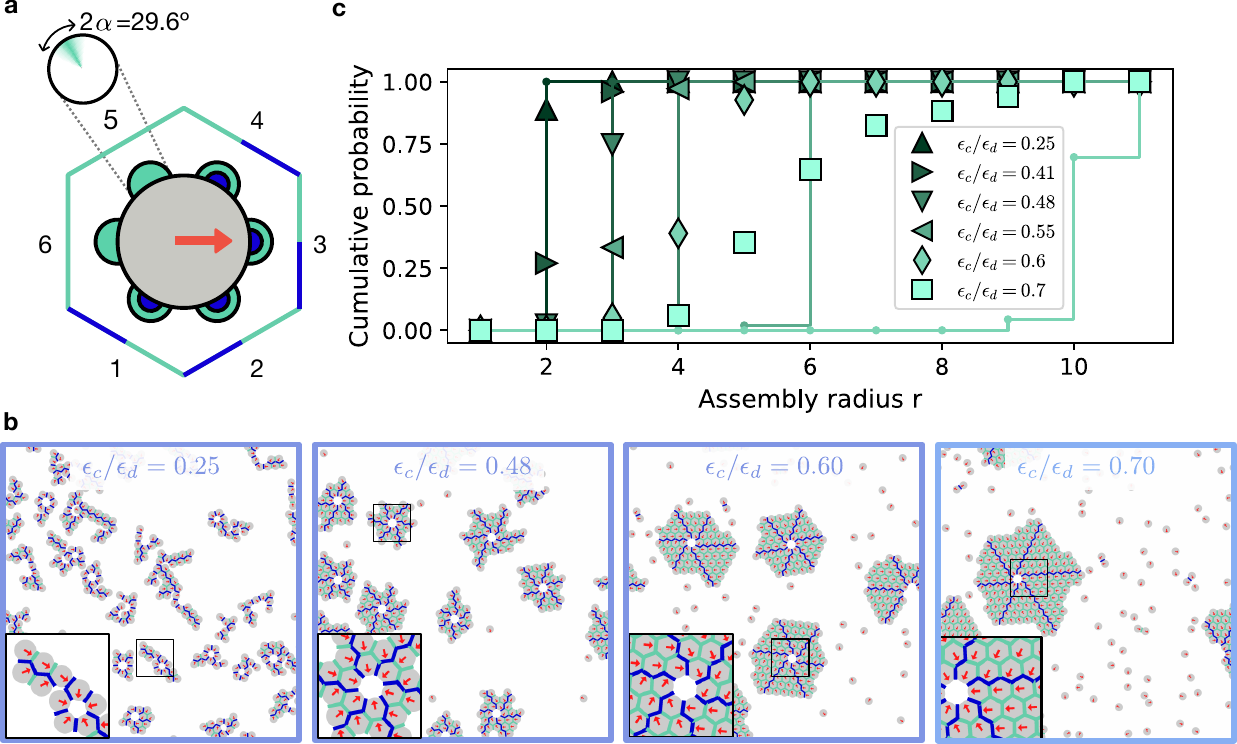}
    \caption{
    \textbf{Off-lattice simulations reproduce the size control of on-lattice simulations and experiments.}\\
    (\textbf{a})~Subunit design shown with a representation similar to Fig.~3a of the main text. The inset represents the angular segment $2\alpha$ over which each patch interacts with other patches.
    (\textbf{b})~Final structures of the assemblies for selected values of $\epsilon_c / \epsilon_d$. The subunits are initially uniformly spread over the periodic box and the simulations run for a time $\approx 10^4\tau_D$, where $\tau_D$ indicates the time that it takes for a subunit to diffuse over a distance equal to its diameter. Here $\epsilon_c/k_BT=5.5$, which provides a good compromise between assembly stability and fast equilibration. The first three panels correspond to values of $\epsilon_c/\epsilon_d$ theoretically predicted to yield size-limited vortex assembly, while the fourth panel is in the size-unlimited theoretical regime. Movies~SM4-SM7 show the full simulation runs associated with these final snapshots.
    (\textbf{c})~Distributions of final assembly sizes at the end of the simulations (\emph{symbols}) and comparison with the on-lattice ideal-gas-of-clusters theory of Sec.~\ref{supp:size_estimate}. Each set of markers corresponds to an ensemble average over eight simulation runs.\\
    }
    \label{fig:off-lattice}
\end{figure}

To verify that the on-lattice nature of the simulations presented in the main text do not introduce significant artifacts in our self-assembly predictions, here we perform off-lattice Brownian dynamics simulations of vortex self-assembly. The technical details are described in Sec.~I\,D of the Methods. Briefly, we use hard-disk-like subunits covered with six smaller, regularly spaced patches as shown in Fig.~\ref{fig:off-lattice}a. Patches belonging to two different subunits interact through directional, short-range interactions corresponding to the scheme of Fig.~2a of the main text.

Figure~\ref{fig:off-lattice}b shows the result of the constant-temperature assembly of 400 subunits for different values of the energy ratio $\epsilon_c/\epsilon_d$. At the smallest value of $\epsilon_c/\epsilon_d$ shown there, we see assemblies whose structures deviate from the ideal lattice. This includes the formation of five or seven-fold rings as well as the presence of small gaps between subunits belonging to the same assembly. For higher values of this ratio however, crystalline domains build up and constrain the subunits into adopting a triangular lattice-like structure. This suggests that our lattice approach is relevant in this parameter regime, which is the main regime of interest for our study. Indeed, the assembly formed there are strikingly similar to those observed both in our lattice simulations and in our experiments.

We report the size statistics of these vortex assemblies in Fig.~\ref{fig:off-lattice}c as well as Fig.~3h of the main text. We find that the results of the off-lattice simulations are shifted to larger $\epsilon_c/\epsilon_d$ ratios compared to our lattice predictions, similar to our experimental results. We speculate that this resemblance could be due to the off-lattice nature of both systems. Indeed, while in lattice simulations small assemblies do not have any conformational freedom, in continuum space they can deform and fluctuate. This endows them with a larger entropy than big vortex assemblies, which have a more rigid lattice structure that suppresses these fluctuations. This could explain why for moderate values of $\epsilon_c/\epsilon_d$, smaller vortices appear more favored off-lattice than on-lattice. Correspondingly, larger values of $\epsilon_c/\epsilon_d$ appear to be required to assemble large off-lattice vortices. Finally, in Fig.~\ref{fig:off-lattice}c the discrepancy between actual and targeted vortex size for large targeted sizes is consistent with our discussion of Sec.~\ref{sec:circumvent_nucleation}. This confirms the relevance of lattice simulations, including in addressing questions related to nucleation.

\section{\label{sec:number_of_designs}Estimate of the number of possible defect-engineered designs}
Here we provide a rationale for the assertion made in the main text that our framework can generate a number of designs of order $10^{67}$. Specifically we show that this number is larger than $8\times 10^{66}$. Section~\ref{sec:hex_2D} introduces our reasoning in the simpler case of hexagonal subunits in two dimensions, and Sec.~\ref{sec:rhodec_3D} derives the estimate for rhombic dodecahedral subunits in three dimensions. We discuss the relevance of these estimates in Sec.~\ref{sec:bound_relevance}.

\subsection{\label{sec:hex_2D}Hexagonal subunits in two dimensions}
\paragraph*{Size of the interaction map.} Our reasoning is most easily visualized in the case of two-dimensional hexagonal subunits. Consider the interaction maps shown in Fig.~\ref{fig:EnergyCalcul}. Since each subunit has $6$ faces, these maps are $6\times 6$ matrices whose entries represent the energies associated with each possible face-to-face interaction. In the language of group theory, the set of all orientations of the subunits is obtained by letting an element of the cyclic group of order 6, denoted by $\textsf{C}_6$ in Sch\"onflies notation, act on a reference subunit. This group is of order $n=|\textsf{C}_6|=6$, resulting in an interaction map of size $n\times n$.

\begin{figure}
    \centering
    \includegraphics[width=0.8\textwidth]{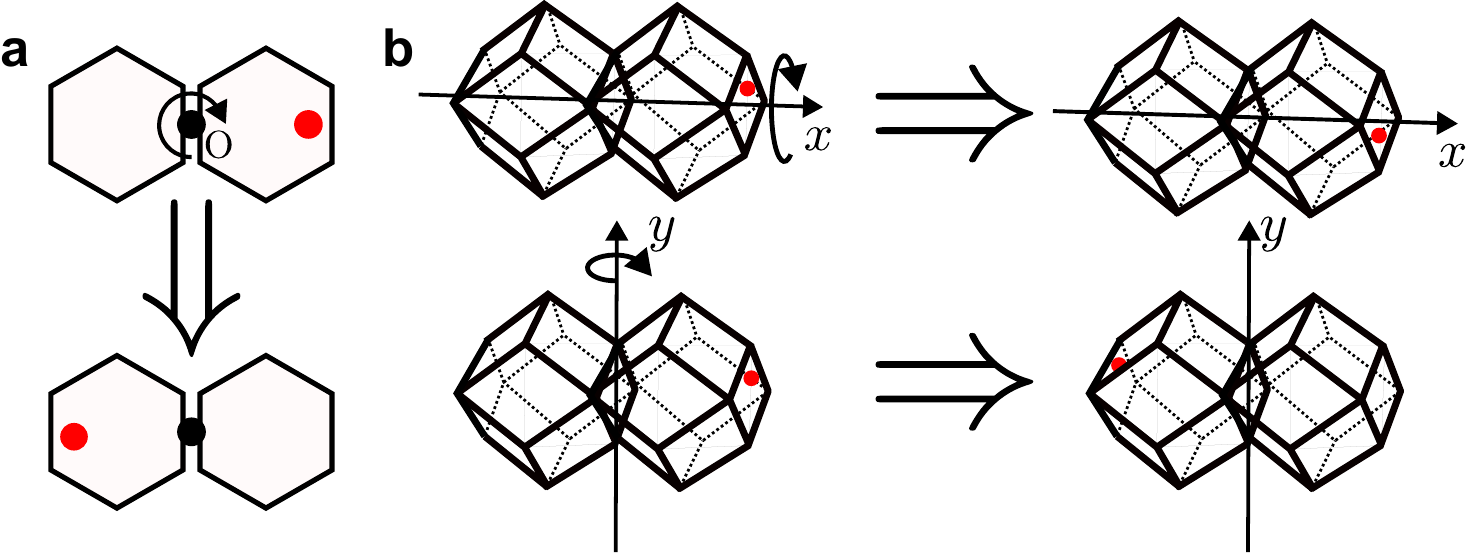}
    \caption{\textbf{Symmetries in subunits interactions.}\\
    (\textbf{a})~The energy of an interacting pair of two-dimensional hexagons must be invariant under a rotation by $180^\circ$ around point O, which exchanges the contacting faces and modifies the subunits' orientations. The red marking is a cue in visualizing the rotation.
    (\textbf{b})~The energy of an interacting pair of three-dimensional rhombic dodecahedra must be invariant under rotations by $180^\circ$ around both around the $x$ and $y$ axes. The former flips the orientations of the two contacting faces, while the latter exchanges the faces themselves while modifying the subunits' orientations. The red marking provide visual help in visualizing the rotations.}
    \label{fig:counting_designs}
\end{figure}

\paragraph*{Number of independent interactions.} The interaction between two subunits is unchanged upon an exchange of the two subunits followed by a $180^\circ$-rotation of each of the subunits, which is equivalent to subjecting the whole pair of subunits to a $180^\circ$ solid rotation without unbinding them [Fig.~\ref{fig:counting_designs}a]. As a result, the interaction map is a symmetric matrix and thus has only $n(n+1)/2=21$ independent entries. The set of these entries may be visualized as the upper triangular submatrix of the interaction map. Out of these 21 entries, $c=3$ are dedicated crystalline interactions (green squares in Fig.~\ref{fig:EnergyCalcul}d), leaving $n(n+1)/2-c=18$ other interactions. We reason that designing a subunit is equivalent to choosing which of these remaining interactions are repulsive (red squares), and which are defect interactions with energy $\epsilon_d$ (blue squares). Since this boils down to one binary choice per remaining interaction, this reasoning implies that there are $\mathcal{D}_0=2^{n(n+1)/2-c}$ possible hexagonal subunit designs.

\paragraph*{Number of non-redundant designs.} Some of the designs enumerated above are redundant. For instance, the hexagonal subunit presented in Fig.~\ref{fig:EnergyCalcul}d where each side $i$ has a crystalline interaction with side $(i+3 \text{ mod }3)$ and where side~1 has a defect interaction with side~3 and side~2 has a defect interaction with side~6, is equivalent to a subunit with the same crystalline interactions and where side~2 has a defect interaction with side~4 while side~3 has a defect interaction with side~1. Indeed, the latter is identical to the former up to a relabeling of its sides; or equivalently the latter can be obtained from the latter through a rotation by $60^\circ$. This multiple counting does not apply to all subunits however: for instance, the rotationally symmetric design of Fig.~\ref{fig:EnergyCalcul}c is counted only once in $\mathcal{D}_0$. A similar reasoning applies to reflections: since two subunits that are mirror images of each other give rise to the same self-assembly behavior (up to a reflection of the whole system), we deduce that chiral designs are counted twice in $\mathcal{D}_0$. By contrast, achiral designs are counted only once. Overall, whether of not a specific subunit design gives rise to multiple counting or not depends on the symmetry group of this subunit.

To provide an estimate of the number of independent subunit designs, we reason that the largest number of such multiple countings is the order of the symmetry group of the hexagon. This group is the six-fold dihedral group $\textsf{D}_6$ and has $|\textsf{D}_6|= 12$. This upper bound implies a lower bound of the number $\mathcal{D}$ of non-redundant design through
\begin{equation}
    \mathcal{D}^{\text{(hexagons)}} \geq \frac{\mathcal{D}_0^{\text{(hexagons)}}}{|\textsf{D}_6|} = \frac{2^{n(n+1)/2-c}}{|\textsf{D}_6|} = \frac{262\,144}{12}>21\,845.
\end{equation}
 Therefore there are more than $21\,845$ non-redundant defect engineering designs for hexagonal subunits with crystalline interactions and one defect energy $\epsilon_d$.

\subsection{\label{sec:rhodec_3D}Rhombic dodecahedral subunits in three dimensions}
\paragraph*{Size of the interaction map.} We now adapt this reasoning to the case of rhombic dodecahedral subunits in three dimensions. The orientations of the rhombic dodecahera are generated by the action of the chiral octahedral symmetry group $\textsf{O}$ with $n=|\textsf{O}|=24$, implying a most general interaction map with size $n\times n$. We note in passing that the interaction map of the specific rhombic dodecahedral subunits considered in the main text affords a much more compact representation than this general case. Indeed, these special four-fold symmetric rhombic dodecahedral subunits have a rotational symmetry group $\textsf{C}_4(d)$, where $d$ denotes their axis of symmetry, and an associated $n'=|\textsf{O}/\textsf{C}_4(d)|=6$, consistent with the $6\times 6$ interaction map in Fig.~\ref{fig:supp_3d}.

\paragraph*{Number of independent interactions.} Going back to the case of generic dodecahedral subunits, we note that the interaction between two subunits lined up on the $x$ axis is invariant both under a rotation of $180^\circ$ about the $x$-axis, and under a rotation of $180^\circ$ about the $y$-axis [Fig.~\ref{fig:counting_designs}b]. The former rotation maps each of the two interacting faces on itself, but with a reversed orientation. The latter rotation exchanges the contacting faces of the two subunits without modifying their orientation. The invariance of the interaction under these two transformations dictates the form of the interaction map. We write the $24\times 24$ interaction map in such a way that its first 12 coordinates refer to 12 different faces, and that the next 12 recapitulate the same faces with opposite orientations
\begin{equation}
    \text{Interaction map} = \left(\begin{array}{cc}
         \mathbf{I}_{(11)} & \mathbf{I}_{(12)} \\
         \mathbf{I}_{(21)} & \mathbf{I}_{(22)}
    \end{array}\right)
\end{equation}

The symmetry under rotation around $x$ implies that the interaction map must be invariant under an exchanges of the faces and so that $\mathbf{I}_{(11)}=\mathbf{I}_{(22)}$ and $\mathbf{I}_{(12)}=\mathbf{I}_{(21)}$. The symmetry upon rotation around $y$ functions in a similar fashion as the one discussed in Sec.~\ref{sec:hex_2D} and implies a symmetric interaction map: $\mathbf{I}_{(11)}=\mathbf{I}^T_{(11)}$, $\mathbf{I}_{(12)}=\mathbf{I}^T_{(21)}$ and $\mathbf{I}_{(22)}=\mathbf{I}^T_{(22)}$. Taken together, these constraints imply that there are only $2(2+1)/2\times (n/2)(n/2+1)/2=3n(n+2)/8=234$ independent entries in the interaction map. Substracting $c=6$ crystalline interactions then leaves $3n(n+2)/8-c=228$ potential defect interactions, implying $\mathcal{D}_0=2^{3n(n+2)/8-c}$ possible rhombic dodecaheral subunit designs.

\paragraph*{Number of non-redundant designs.} As in Sec.~\ref{sec:hex_2D}, this number $\mathcal{D}_0$ counts equivalent subunit designs multiple times. The largest possible number of multiple countings is given by the order of the full octahedral group $\textsf{O}_h$. As in the previous section we obtain a lower bound of the number $\mathcal{D}$ of possible designs by dividing $\mathcal{D}_0$ by the largest possible number of multiple counts $|\textsf{O}_h|=48$. This yields
\begin{equation}
\mathcal{D}^{\text{(rhombic dodecahera)}}
\geq \frac{\mathcal{D}_0^{\text{(rhombic dodecahera)}}}{|\textsf{O}_h|}
= \frac{2^{3n(n+2)/8-c}}{|\textsf{O}_h|}
> 8\times 10^{66}.
\end{equation}
Therefore there are more than $8\times 10^{66}$ non-redundant defect engineering designs for rhombic dodecahedral subunits with crystalline interactions and one defect energy $\epsilon_d$.

\subsection{\label{sec:bound_relevance}Relevance of these Bounds}
The lower bounds computed above overestimate the number of possible designs due to the fact that some of them have additional symmetries (\emph{e.g.}, reflection symmetry) compared to the most generic case. Given the complexity of our interaction maps, we however expect that such cases are a small minority of all possible designs and thus that our lower bounds are quite tight.

We suspect that an overwhelming majority of possible defect engineered designs discussed do not lead to assembly size control or any other useful feature. Nevertheless, the enormous magnitude of $\mathcal{D}$ suggests that setting them aside might still leave a large number of interesting designs to be exploited. We believe that this is enough to motivate a search for a deeper understanding of the principles underlying defect engineering.

\section{Enhancing the performance of defect engineered designs}

As discussed in Sec.~\ref{sec:number_of_designs}, defect engineering opens a colossally large design space. A systematic exploration of this space is beyond the scope of the present work, and our main text instead focuses on the presentation of a small number of simple, easily understandable designs. This emphasis on intelligibility does not necessarily produce completely optimized designs from a functional point of view. Indeed our simplest, most pedagogical example, the vortex assembly, leaves room for improvement in several respects.

Here we demonstrate that these vortex-specific limitations are not fundamental features of defect engineering, and can be improved upon within its scope. We first discuss the relevance of uncertainties in the values of the crystalline and defect energies, and show that the vortex assembly can be made more robust to them through a modification of its interaction map (Sec.~\ref{sec:enlarge_domain}). We then discuss the relatively large shape fluctuations in our vortices in Sec.~\ref{sec:suppress_fluctuations} alongside strategies to suppress them as needed. In Sec.~\ref{sec:circumvent_nucleation}, we tackle the issue of vortex nucleation and the resulting diminished control over the sizes of large assemblies; again we find that this effect can be mitigated using defect engineering. Sec.~\ref{sec:rules_of_thumb} then discusses some heuristic considerations in identifying new designs that lead to self-limitation.

Beyond these defect-engineering-specific considerations, we note that while our approach is distinct from existing strategies such as self-closing, deformable and addressable assembly~\cite{hagan2021equilibrium}, it can be combined with them. This will enable synergies and offer additional avenues to improve upon technical issues, similar to existing schemes combining self-closing and addressable assembly~\cite{Videbaek:2022,Duque:2024,Tyukodi:2025}.

\subsection{\label{sec:enlarge_domain}Experimental precision and broadening of the range of self-limited assembly}

In the phase diagram of Fig.~2b of the main text, large vortex assemblies are associated with narrow parameter regimes. This implies that large assemblies can only be produced reliably if the ratio $\epsilon_c/\epsilon_d$ is controlled with a sufficient precision. This precision requirement is common in size-limited assembly. It also arises when strategies different from ours are used, and existing work indicates that it does not compromise their ability to achieve self-limiting assembly.

In self-closing assembly, where curved subunits assemble until they form a closed structure, fluctuations in the angles between consecutive subunits result in a fairly broad distribution of assembly sizes~\cite{hayakawa2022geometrically}. Electron microscopy suggests that the standard deviation of these angular fluctuations is of the order of $10^\circ$ to $20^\circ$ at ambient temperature~\cite{Saha:2025}. This is comparable to the precision required to ensure that a cylinder with a circumference of 4 subunits (which requires inter-subunit angles of $52^\circ$) forms more readily than a cylinder with 5 (which requires inter-subunit angles of $42^\circ$; here we use values that are twice the ``bevel angle'' of Fig.~S2a of Ref.~\cite{hayakawa2022geometrically} for achiral cylinders). In self-assembly strategies based on strain accumulation, the magnitude of the unintended modes of deformation of the individual subunits is similar to that of the intended mode (see, \emph{e.g.}, the quantification in Fig.~4b of Ref.~\cite{berengut2020self}). In other words, the uncertainty over the deformation is of the same order as the deformation itself. These complex subunit distortions result in a broad dispersion of assembly sizes, typically of the order of the mean size itself.

Here we argue that the main control parameter of our strategy, namely the ratio $\epsilon_c/\epsilon_d$, is easier to precisely control than the subunit angles or deformations used as control parameters in these other approaches. This implies that precision should be less of an issue in defect engineering than there. We compute a nominal experimental uncertainty over that ratio, and find it to be small enough that it would in principle allow us to control the sizes of our vortex assemblies up to target radii $r^*=9$ (\emph{i.e.}, assemblies comprising up to 270 subunits). While in practice the formation of such large vortex assemblies is complicated by nucleation issues, these can be mitigated using other related defect-engineering designs as discussed in Sec.~\ref{sec:circumvent_nucleation}). We then demonstrate that defect engineering additionally provides resources to loosen the precision requirements associated with the parameter $\epsilon_c/\epsilon_d$. We do this by introducing a variant of our vortex design where the parameter regimes associated with size-controlled vortices are systematically broader than those shown in the phase diagram of Fig.~2b of the main text.

\paragraph*{Experimental precision over $\epsilon_c/\epsilon_d$.} In our experiments, the ratio $\epsilon_c/\epsilon_d$ is set by the binding free energies of the hybridizing DNA strands illustrated in Fig.~3a-b of the main text. We estimate that the main experimental imprecisions leading to fluctuations in these quantities are fluctuations in the temperature $T$ and the $\text{MgCl}_2$ concentration of our buffer. We conservatively estimate the associated standard deviations to be $ \delta T \simeq5\,\degreeCelsius$ and $\delta[\mathrm{MgCl_2}]\simeq1\,\mathrm{mM}$. Using first-order (linear) error propagation, we find

\begin{equation}
    \delta\!\left(\frac{\epsilon_c}{\epsilon_d}\right) =
    \sqrt{
    \left[
        \frac{1}{\epsilon_d}\frac{\partial\epsilon_c}{\partial T}
        -
        \frac{\epsilon_c}{\epsilon_d^2}\frac{\partial\epsilon_d}{\partial T}
    \right]^2 \delta T^2
    +
    \left[
        \frac{1}{\epsilon_d}\frac{\partial\epsilon_c}{\partial[\mathrm{MgCl_2}]}
        - 
        \frac{\epsilon_c}{\epsilon_d^2}\frac{\partial\epsilon_d}{\partial[\mathrm{MgCl_2}]}
    \right]^2 \delta[\mathrm{MgCl_2}]^2
    } \simeq 0.0048,
    \label{eq:error_ratio}
\end{equation}
where we used the following values for the binding free energies, and computed their susceptibilities to changes in the temperature and magnesium ion concentration using the NUPACK software suite: 
%\begin{subequations}
    \begin{align*}
        \epsilon_c &\simeq -13.1\,\mathrm{kcal/mol}\\
        \epsilon_d &\simeq -42.7\,\mathrm{kcal/mol}\\
        \frac{\partial\epsilon_c}{\partial T}&\simeq0.19\,\mathrm{kcal/mol/}\degreeCelsius\\
        \frac{\partial\epsilon_d}{\partial T}&\simeq0.50\,\mathrm{kcal/mol/}\degreeCelsius\\
        \frac{\partial\epsilon_c}{\partial[\mathrm{MgCl_2}]}&\simeq-0.09\,\mathrm{kcal/mol/mM}\\
        \frac{\partial\epsilon_d}{\partial[\mathrm{MgCl_2}]}&\simeq-0.20\,\mathrm{kcal/mol/mM}.
    \end{align*}
%\end{subequations}
The uncertainty resulting from the calculation of Eq.~\eqref{eq:error_ratio} is quite small. This is partly due to the fact that changes in temperature or ion concentration modify $\epsilon_c$ and $\epsilon_d$ in similar proportions, implying only small changes in the ratio of these quantities.

The theoretical uncertainty computed in Eq.~\eqref{eq:error_ratio} is to be put in perspective with the width of the domains of the phase diagram of Fig.~2b of the main text. In Table~\ref{tab:eced_intervals} we list the widths $\Delta(\epsilon_c/\epsilon_d)$ of the intervals of $\epsilon_c/\epsilon_d$ corresponding to various vortex sizes. We reason that fluctuations in this parameter may compromise our control over self-assembly if their standard deviation $\delta(\epsilon_c/\epsilon_d)$ is larger than $\frac{1}{2}\Delta(\epsilon_c/\epsilon_d)$. By that rationale, the value computed in Eq.~\eqref{eq:error_ratio} should allow us to assemble vortices with sizes or the order of $r^*=9$. This estimate suggests that our imperfect control over our experimental temperature and ion concentrations are not substantial obstacles to assembling large monodisperse vortices, or in principle even larger polydisperse assemblies. This stands in stark contrast with our earlier discussion of assembly strategies based on self-closing or strain accumulation, which display substantial relative fluctuations in the subunit-wise interactions responsible for size control. Despite these challenges, investigators using these strategies have mitigated the impact of fluctuations through the use of favorable geometries, \emph{e.g.}, highly geometrically constrained icosahedral shells instead of cylinders~\cite{Sigl:2021}. As discussed in Sec.~\ref{sec:suppress_fluctuations}, defect engineering also offers opportunities to harness similar geometrical effects.

\begin{table}[t]
    \centering
    \begin{tabular}{c|c|c|c|c|c|c|c|c|c|c|c|c|c|c}
        $r^*$ & 2 & 3 & 4 & 5 & 6 & 7 & 8 & 9 & 10 & 11 & 12 & 13 & 14 & 15 \\
        \hline
        \# of subunits & 18 & 36 & 60 & 90 & 126 & 168 & 216 & 270 & 330 & 396 & 468 & 546 & 630 & 720\\
        $\Delta(\epsilon_c/\epsilon_d)$ & 0.3333 & 0.1111 & 0.0556 & 0.0333 & 0.0222 & 0.0159 & 0.0119 & 0.0093 & 0.0074 & 0.0061 & 0.0051 & 0.0043 & 0.0037 & 0.0032 \\
    \end{tabular}
    \caption{Width $\Delta(\epsilon_c/\epsilon_d)$ of the energy ratio intervals over which vortex assemblies of radii $r^*$ dominate the phase diagram of Fig.~2b of the main text. These values are computed from Eq.~\eqref{eq:ratio_star} through $\Delta(\epsilon_c/\epsilon_d) (r^*)=(\epsilon_c/\epsilon_d)_\text{lim}(r^*+1)-(\epsilon_c/\epsilon_d)_\text{lim}(r^*)$. The table also specifies the number of subunits in the target assemblies.}
    \label{tab:eced_intervals}
\end{table}

Our reasoning suggests that because of their small magnitude, parameter fluctuations may be less of an issue than other limitations discussed in the remainder of this section. We note that the discussion above focuses on the issue of the precision of our control of $\epsilon_c/\epsilon_d$ and leaves aside the question of its accuracy. We indeed reason that any systematic, reproducible bias in our estimation of $\epsilon_c/\epsilon_d$ can be compensated for through trial and error and/or mitigated through the use of more detailed models of DNA hybridization between our nanocylinders.

\paragraph*{Increasing the robustness of vortex assemblies by modifying their interaction map.}
To increase the area of the size-limited domains of the phase diagram of Fig.~2b of the main text, we modify the interaction map of the vortex design so as to change the magnitude of the point-like energy $\tilde{\epsilon}_p$ introduced in Fig.~1 of the main text. This energy is defined as the sum of all contributions to the vortex energy that scale like $r^0$. In the basic vortex design and as discussed in the main text, this term is dictated by the crystalline and defect energies through $\tilde{\epsilon}_p=6(\epsilon_c-\epsilon_d)$.

\begin{figure}[t!]
    \centering
    \includegraphics[width=\linewidth]{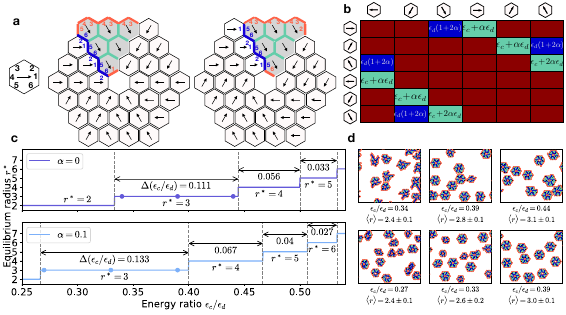}
    \caption{\textbf{Penalizing the corners of the vortex assembly broadens the size-limited domains of the phase diagram.}\\
    (\textbf{a})~Vortex assemblies exist in two polymorphs with equal interaction energies. The polymorph on the \emph{left} (\emph{right}) has a small number of orange interfaces involving sides $1$ and $5$ ($2$ and $6$) and an excess of blue defect interactions between sides $2$ and $6$ ($1$ and $5$).
    (\textbf{b})~We modify the interaction map through a dimensionless parameter $\alpha\geq 0$ to promote these specific blue interfaces and penalize the orange ones. This specific combination modifies $\tilde{\epsilon}_p$ without changing $\tilde{\epsilon}_c$ and $\tilde{\epsilon}_d$ in the notation of Fig.~1 of the main text.
    (\textbf{c})~Increasing $\alpha$ increases the range $\Delta(\epsilon_c/\epsilon_d)$ over which a given vortex size $r^*$ is favored.
    (\textbf{d})~Simulation snapshots at the $\epsilon_c/\epsilon_d$ values shown by dots in the previous panel confirm the formation of $r^*=3$ vortices over an increased range (simulation panels at the \emph{top} and \emph{bottom} correspond to simulations with $\alpha=0$ and $\alpha=0.1$, respectively).
    }
    \label{fig:increase_range}
\end{figure}

To be able to tune $\tilde{\epsilon}_p$ independent of the values of $\epsilon_c$ and $\epsilon_d$, we first note that the vortex design favors the formation of two closely related polymorphs which have the same size but slightly different grain boundary positions (Fig.~\ref{fig:increase_range}a). The outer and inner rims of these polymorphs are comprised of interfaces between empty sites and specific subunit faces. The interfaces with faces 3 and 4 are the most abundant, and their number scales linearly with the radius of the assembly. Each polymorph additionally displays two more types of interfaces, respectively with faces $\lbrace 1,5\rbrace$ and $\lbrace 2,6\rbrace$. These interfaces are only present at the vortex outer corners and at its center, and there always are exactly six interfaces of each type per completed vortex. Modifying their energies thus affords us some additional control over $\tilde{\epsilon}_p$. In practice, we modify the energy of the interface between an empty site and face $i$, by acting on the interactions involving $i$ and all other faces in our interaction map. Specifically, we increase or decrease the energy of every crystalline or defect interaction that involves face $i$. Indeed, doing so respectively decreases or increases the relative cost of forming the corresponding interface. The second component of our strategy is to note that the two different defect interactions involved in the vortex assembly (between faces 1 and 5 on the one hand and between 2 and 6 on the other) are not equally represented in each polymorph. Indeed, each polymorph comprises six more of one type than the other. Assigning different energies to each type of defect again provides us a way to control $\tilde{\epsilon}_p$.

We use these two levers to design a modified interaction map where the crystalline and defect interactions are corrected through a dimensionless parameter $\alpha$ as shown in Fig.~\ref{fig:increase_range}b. The case $\alpha=0$ corresponds to the original design. The two polymorphs still have identical energies in our modified design, and their energy per subunit reads:
\begin{equation}
    e^{(\mathrm{vortex})}(r,\alpha)=\frac{3r^2 \epsilon_c + r(-3\epsilon_c+4\epsilon_d) - 2 (1+2\alpha) \epsilon_d}{r(r+1)}.
\end{equation}
As in Eq.~\eqref{eq:ratio_star}, we compute the energy ratio at which a vortex of size $r^*$ becomes more stable than a vortex of size $r^*-1$. This provides the positions of the boundaries of the new phase diagram as
\begin{equation}
    (\epsilon_c/\epsilon_d)_\mathrm{lim}(r^*) = \frac{2}{3} \frac{r^*-2}{r^*-1} - \frac{4\alpha}{3(r^*-1)}.
\end{equation}

We show in Fig.~\ref{fig:increase_range}c that using $\alpha>0$ broadens the range $\Delta(\epsilon_c/\epsilon_d)$ over which each assembly size is stable, which addresses the narrow-range limitation of the original vortex design. Figure~\ref{fig:increase_range}d shows corresponding simulation results, and demonstrates that the indented vortex sizes form over the whole intended range. This example illustrates the potential to improve upon a wide range of practical issues using the vast design freedom associated with defect engineering.

\subsection{\label{sec:suppress_fluctuations}Suppression of the assembly shape fluctuations}

\begin{figure}[t]
    \centering
    \includegraphics[width=1.0\linewidth]{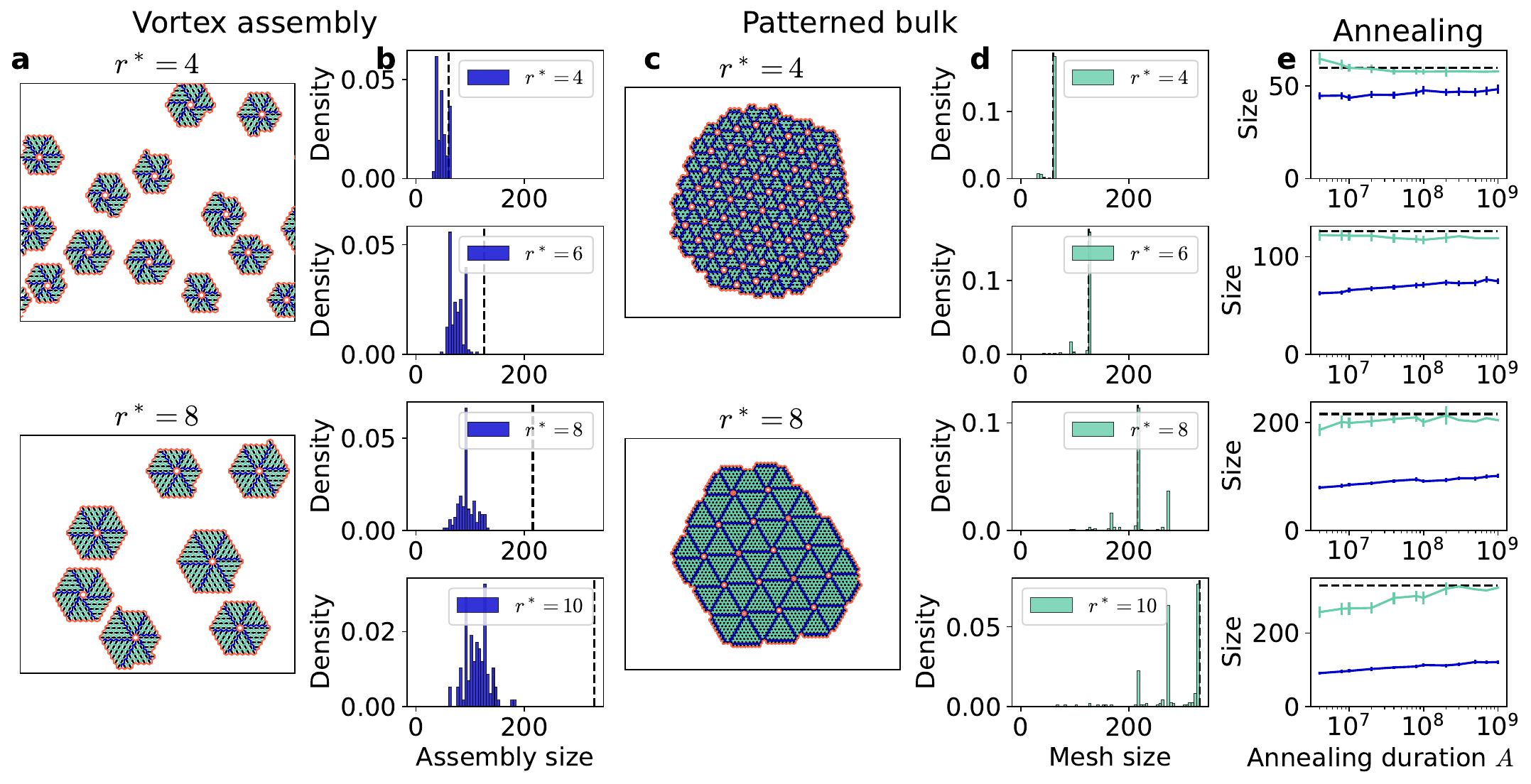}
    \caption{\textbf{Shape fluctuations and slow equilibration are much less pronounced in patterned bulks than vortex assemblies.}\\
    (\textbf{a})~Simulations of vortex assembly using $648$ subunits for two different target sizes $r^*$. 
    (\textbf{b})~Size histograms for different vortex assembly target sizes (\emph{dashed lines}), expressed in the number of subunits per assembly.
    (\textbf{c})~Simulations of patterned bulk assembly for two different target sizes using respectively $1620$ and $2088$ subunits.
    (\textbf{d})~Size histograms for different vortex assembly target sizes (\emph{dashed lines}), expressed in number of subunits per unit cell.
    The annealing time for these first four panels is $A=2\times 10^8$ Monte-Carlo steps per value of the temperature (see Methods).
    (\textbf{e})~Comparison between the vortex (\emph{blue}) and patterned bulk (\emph{green}) sizes obtained in simulations with four different target sizes (indicated by the \emph{dashed lines} in each panel) as a function of the annealing time $A$ expressed in number of Monte-Carlo steps per value of the temperature (see Methods).
    }
    \label{fig:annealing_and_fluctuation}
\end{figure}

The vortex assemblies seen in simulations and experiments do not perfectly conform to the highly symmetric hexagonal assemblies illustrated in the inset of Fig.~2b of the main text. Instead, the sizes of each of the six triangular crystalline domains that surround the vortex core often fluctuate in on-lattice simulations (main text Fig.~2c and Fig.~\ref{fig:annealing_and_fluctuation}a), experiments (main text Fig.~3 and Fig.~\ref{fig:vortex_fov_0.57_gallery_2x2.jpg}) as well as off-lattice simulations (Fig.~\ref{fig:off-lattice}). By contrast, the core itself typically does largely conform to the ideal model. We quantify the size fluctuations of the vortex design in Fig.~\ref{fig:annealing_and_fluctuation}b, which shows that it is associated with relatively broad histograms of assembly sizes.

This relatively loose control of the size of these crystalline domains is a specificity of the vortex design. It stems from the weak coupling between the domains due to the fact that each of them is flanked by only two other domains. As a result, the energetic cost of a fluctuation in the domain's size is relatively low. Consider for instance an increase of the radius of one of the domains by one unit. To accommodate this change, the two neighboring domains need only add a step of height 1 on top of their structure, which implies an energetic cost of the order of one unit of interaction energy.

Other defect engineered designs have dramatically smaller fluctuations. One example is the defect-engineered patterned bulks presented in Fig.~4e of the main text. As discussed in Sec.~\ref{supp:other_geo}, this patterned bulk can be viewed as a juxtaposition of unit cells each in the shape of a vortex assembly. Figure~\ref{fig:annealing_and_fluctuation}c shows simulations of two large assemblies of this type. They display crystalline domains with extremely well-defined sizes, in stark contrast with the vortices. This regularity is further quantified in Fig.~\ref{fig:annealing_and_fluctuation}d, which indeed shows very peaked histograms for the structures' unit cell sizes (each unit cell is comprised of six triangular crystalline domains). This difference in behavior is the result of the strong geometrical constraints exerted on each crystalline domain by its neighbors. Indeed, not only is each domain surrounded by other domains on all three sides, but these domains are themselves constrained by other domains in all directions. It is thus impossible to change the size of a triangular domain without either changing those of many others, or inducing significant deformations throughout the lattice. Although we have not investigated the matter in detail, we suggest that this delocalized control presents many of the hallmark of a two-dimensional phase transition, where fluctuations are strongly suppressed on large length scales due to the cooperativity of local interactions.

On a more local scale, we note that the three-dimensional assemblies presented in Fig.~4h of the main text also display smaller fluctuations than vortex assemblies and a peaked distribution of assembly sizes similar to that of Fig.~\ref{fig:annealing_and_fluctuation}d. Indeed, in this design each crystalline domain is constrained in four different directions by four neighbors instead of the vortices' two. This effect is reinforced by the three-dimensional nature of the assembly. Indeed, similar to the vortex case, an increase of the radius of one of the domains requires the creation of a step in all neighboring domains. These steps however differ from those discussed in the vortex case: while each step involves a single subunit in the two-dimensional vortex, three-dimensional steps have a lateral extension proportional to the assembly radius $r$. This implies that their energetic cost is much larger than in two dimensions, which dramatically suppresses domain size fluctuations in large assemblies. This qualitative difference between the cost of creating terraces in a one-dimensional interface (the outer boundary of the vortex) \emph{vs.} a two-dimensional interface (the outer surface of the 3D assembly) is reminiscent of the physics of the roughening transition, and of the fact that equilibrium two-dimensional interfaces are macroscopically smooth at low temperature while one-dimensional interfaces are not.

Overall, we suggest that shape fluctuations, while pronounced in the vortex design case, can be avoided through a suitable choice of defect-engineered design in the more general case. In addition to those discussed above, this could include size-limited designs obtained by folding the motif of Fig.~\ref{fig:annealing_and_fluctuation}c into a closed three-dimensional hollow shell.

\subsection{\label{sec:circumvent_nucleation}Circumventing nucleation-induced size limitations}

In simulations, our control over the size of large vortex assemblies is hampered by nucleation issues. This is apparent in Supplementary movies  SM2 and SM3, which show that the nucleation of vortex assemblies is a relatively irreversible process. This is problematic in the relatively dense systems featured in our simulations, where many new vortices tend to quickly nucleate early in the dynamics. Following this first stage, the remaining isolated monomers condense onto the nucleated structures. At this stage they are thus shared relatively evenly between many assemblies, and the assembly sizes tend to be smaller than they should be at equilibrium. This transient issue obviously gets corrected as the system relaxes to equilibrium, but the mechanism to do so is slow. Indeed, the system undergoes Ostwald ripening, whereby individual monomers are stochastically removed from suboptimal assemblies to allow others to reach their optimal size. This ripening process is especially slow for large assemblies, which explains why the actual mean vortex assembly sizes in Fig.~\ref{fig:annealing_and_fluctuation}b tend to undershoot their target for intended radii equal to or larger than $r^*=6$. This slow ripening dynamics is illustrated in Fig.~\ref{fig:annealing_and_fluctuation}e, which shows that a longer annealing time $A$ (defined in Sec.~I\,A of the Methods) results in the formation of larger vortex assemblies, although the cases with $r^*\geq 6$ still remain far from target even with slow annealing.

Unlike vortices, patterned bulks very reliably achieve their target sizes up to $r^*=8$ as shown in Fig.~\ref{fig:annealing_and_fluctuation}c-d. The case $r^*=10$ display a mixture of $r=9$ and $r=10$ assemblies for moderate annealing times, and can be brought to its target with a longer annealing (Fig.~\ref{fig:annealing_and_fluctuation}e). Overall, the patterned bulk case demonstrates that defect engineering can be used to achieve very large size-limited objects with a single subunit type. As discussed in Sec.~\ref{sec:suppress_fluctuations}, the advantages of the patterned bulk can in principle be exploited to assemble finite-size, non-periodic objects. To put our results into perspective, in Table~\ref{tab:SLA_litterature} we compare the self-limited sizes achieved within the present first attempt to implement defect engineering in self-assembly with the sizes achieved using much more established techniques optimized over more than a decade. We believe that these results provide a very encouraging indication that defect engineering could significantly outperform current strategies at the cost of modest optimization steps.

\begin{table}[t]
    \centering
        \begin{tabular}{c|c|c|c|c|c}
         Reference & $\#$ units & Relative std %& Yield 
         & Mechanism & Subunit & Assembly \\
         \hline
         \cite{stradner2004equilibrium} & 5-20 &Exp. distrib.  %& -- 
         & Interaction competition%\footnotemark[1]
         & Colloids & 0D assemblies \\
         \cite{hong2008clusters} & 9 & Exp. distrib. %& -- 
         & Self-closing & Janus particles & 0D assemblies \\
         %Sutter 2008 \cite{sutter2008structural} & $60$ & $\pm 0$%\footnotemark[2]
         %& Self-closing &     Protein & Polyhedron \\
         \cite{king2014accurate} & $24$ %& --
         & -- & Self-closing & Synth. proteins & Polyhedron \\
         \cite{wagenbauer2017gigadalton}& $12$ %& --
         & -- & Self-closing & DNA origami & Dodecahedron \\
         \cite{wagenbauer2017gigadalton}& $10$ & $10\%$ %& $90\%$ 
         & Self-closing & DNA origami & Planar ring 22/55\\
         \cite{wagenbauer2017gigadalton}& $15$ & $13\%$ %& $90\%$ 
         & Self-closing & DNA origami & Planar ring 14/37\\
         \cite{wagenbauer2017gigadalton}& $28$ & $7\%$ %& -- 
         & Self-closing & DNA origami & Planar ring 5/18\\
         \cite{berengut2020self} & $5$ & $40\%$ %& -- 
         & Strain accumulation & DNA origami & 1D-chain \\
         \cite{sigl2021programmable} & $20$ & --
         & Self-closing & DNA origami & $T=1$ Icosahedron \\
         \cite{sigl2021programmable} & $60$ & --
         & Self-closing & DNA origami & $T=3$ Icosahedron \\
         \cite{hayakawa2022geometrically} & $10$% \footnotemark[3]
         & $13\%$ %& - 
         & Self-closing & DNA origami & Tubule \\
         \cite{videbaek2024economical} & $10 $%\footnotemark[3]
         & $20\%$ %& $15\%$ 
         & Self-closing & DNA origami & Tubule \\
         this study & $\mathbf{96}$ & $22\%$ %$\pm21$ 
         & Defect engineering & DNA origami & Vortex assembly \\
         this study & $\mathbf{68}$ & $23\%$ %$\pm16$ 
         & Defect engineering & Off-lattice simulation & Vortex assembly\\
         this study & $\mathbf{112}$ & $20\%$ %$\pm14$
         & Defect engineering & On-lattice simulation & Vortex assembly\\
         this study & $\mathbf{324}$ & $4\%$ %$\pm14$
         & Defect engineering & On-lattice simulation & Patterned bulk \\
         this study & $ \mathbf{429} $ & $18\%$ & Defect engineering & On-lattice simulation & 3D assembly 
     \end{tabular}
    \caption{
    Comparison of the largest self-limited assembly sizes (expressed in number of subunits) obtained using different self-assembly techniques involving a single subunit type. The sizes reported for tubules and our patterned bulks are respectively the tubule perimeter and the size of the lattice unit cell. The distributions observed in our patterned bulks and 3D assemblies correspond to a polydisperse collection of assemblies with well-defined sizes and shapes, in contrast with the more continuous distributions obtained in the other examples.
    Most existing approaches consist in self-closing assemblies that reproduce a discrete lattice pattern \cite{hong2008clusters,king2014accurate,wagenbauer2017gigadalton,hayakawa2022geometrically}. Experiments that go beyond $60$ subunits in artificial systems based on this scheme use of multiple subunit types~\cite{sigl2021programmable}. A more recent strategy relies instead on geometrical frustration arising from the accumulation of strain in the subunits.
    Reference~\cite{stradner2004equilibrium} uses a competition between long-range and short-range interactions, which can be controlled by the salt concentration. The number of subunits in our size-controlled assemblies exceeds those achieved with all three of these approaches.
    }
    \label{tab:SLA_litterature}
\end{table}

\subsection{\label{sec:rules_of_thumb}Heuristic rules for the design of self-limited assemblies}
% In response to Ref.~2 Qn.~6

While our defect engineering strategy allows for many possible designs (Sec.~\ref{sec:number_of_designs}), not all lead to self-limitation. While we do not yet have a comprehensive understanding of the rules at play, here we make two observations that constrain the production of self-limited assemblies.

Our first comment regards the form of the large-$r$ expansion of the energy per subunit discussed in Fig.~1 of the main text, namely
\begin{equation}\label{eq:e_large_r_2d}
    e(r)=-|\tilde{\epsilon}_c|-\frac{|\tilde{\epsilon}_d|}{r}+\frac{\tilde{\epsilon}_p}{r^2}.
\end{equation}
This expression can only have a minimum if $\tilde{\epsilon}_p>0$, \emph{i.e.}, if singular points in the assembly carry an energy penalty. This is the case in the vortex design. Indeed, due to the details of the structure of its central hole and its outer corners, each of its defect lines comprises $2r-1$ defect interactions. In other words, the length of each defect line is not quite proportional to $r$; it ``misses'' one defect interaction for that to be the case, and that missing interaction contributes to making $\tilde{\epsilon}_p$ positive. A similar reasoning applies to the square vortices of Fig.~4f of the main text, where the grain boundaries do not reach all the way to the corners of the assemblies but stop one lattice spacing away from them. By contrast, the designs of main text Fig.~4g (also discussed in Fig.~\ref{fig:EnergyCalcul}h) and of Fig.~\ref{fig:EnergyCalcul}i do not have any such missing interaction. They have $\tilde{\epsilon}_p=0$, and as a result do not allow for self-limitation. This observation thus suggests that defect interactions that induce complex point defects at the ends of the defect lines favor self-limitation. 

This comment has interesting implications for assembly in 3D, where the energy per subunit may be expanded as
\begin{equation}\label{eq:e_large_r_3d}
    e(r)=-|\tilde{\epsilon}_c|+\frac{\tilde{\epsilon}_{d2}}{r}+\frac{\tilde{\epsilon}_{d1}}{r^2}+\frac{\tilde{\epsilon}_p}{r^3},
\end{equation}
where the coefficients $\tilde{\epsilon}_{d2}$ and $\tilde{\epsilon}_{d1}$ respectively denote the interaction energies associated with two-dimensional defect planes and one-dimensional defect lines. There are many more ways to induce a minimum in the function of Eq.~\eqref{eq:e_large_r_3d} than in Eq.~\eqref{eq:e_large_r_2d}. Some of those involve the same kind of energetically costly point defects discussed above ($\tilde{\epsilon}_p>0$) in combination with energetically favorable defect planes ($\tilde{\epsilon}_{d2}<0$) or lines ($\tilde{\epsilon}_{d1}<0$). Another strategy could be to induce favorable defect planes ($\tilde{\epsilon}_{d2}<0$) alongside unfavorable defect lines ($\tilde{\epsilon}_{d1}>0$). This would potentially constitute a more robust design. Indeed it would not depend on the system's behavior at a single point, which can potentially be disrupted by the presence of a single impurity. We do not yet have a systematic understanding of the ways in which the entries on the interaction maps between 3D subunits (of which there are 234 in the case of our rhombic dodecahedral subunits) favor or penalize defect planes and defect lines, but we believe this to be a promising area for further investigations.

\begin{figure}[t!]
    \centering
    \includegraphics[width=\linewidth]{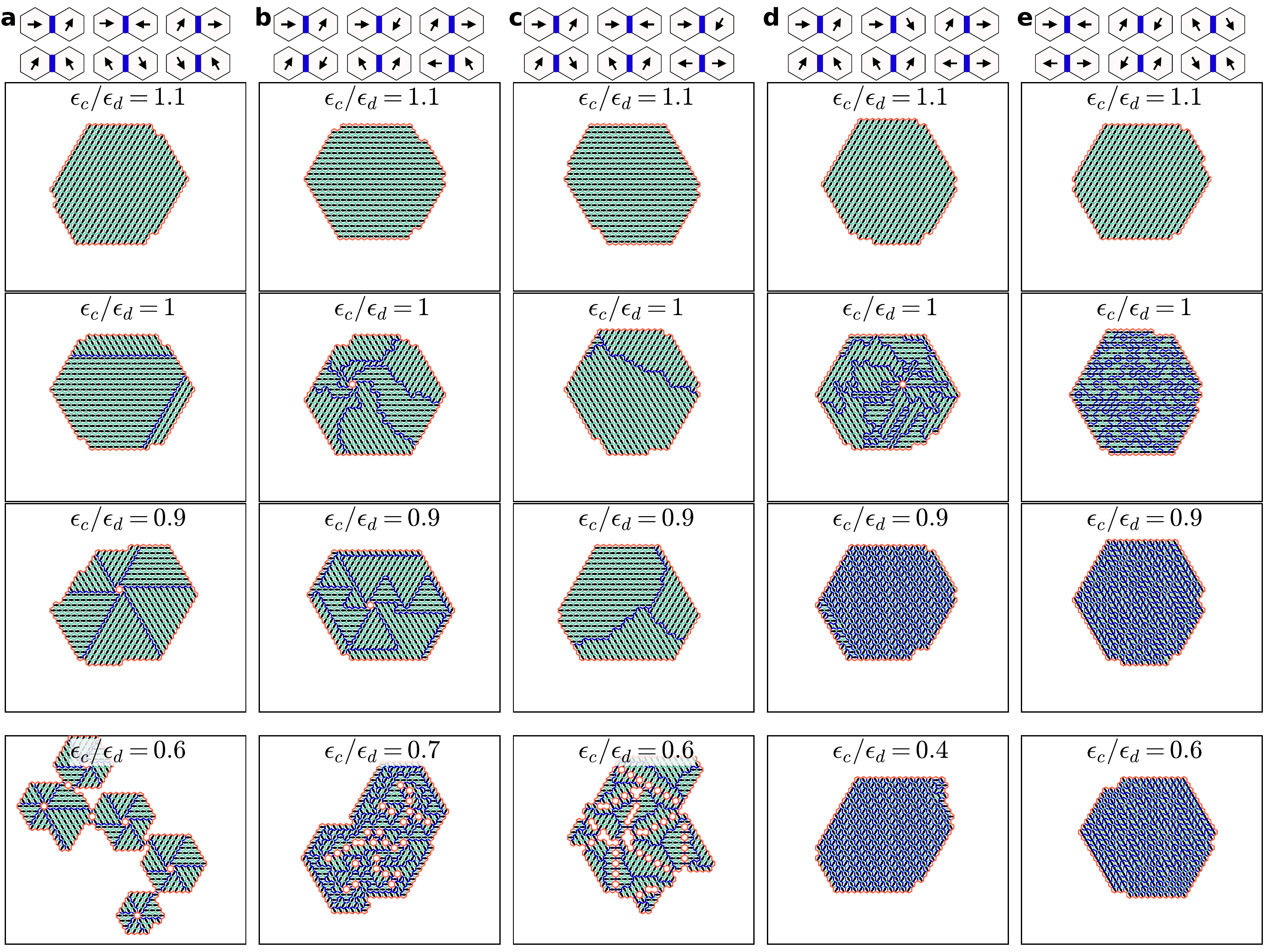}
    \caption{
    \textbf{Proliferating defect interactions prevent self-limitation.}\\
    Here we show the assemblies formed by six different types of subunits, each with its own combination of defect interactions (pictured at the \emph{top}). In the first row of simulations, crystalline interactions are more favorable than defect interactions, implying that only the former are present. We increase the strength of the defect interactions in the next two rows and thus reveal two different types of behaviors. In columns (\textbf{a}-\textbf{c}), only a limited number of defect interactions enter the system, forming the same type of geometrically constrained grain boundaries discussed in the main text and in Sec.~\ref{supp:geometrical_considerations}. By contrast, in columns (\textbf{d}-\textbf{e}) defect interactions proliferate throughout the assemblies, implying that our size-control mechanism cannot operate. The last row shows simulation snapshots in the range of defect interaction strength that allows for self-limitation, and reveals some new morphologies not discussed in the text.
    }
    \label{fig:heuristics}
\end{figure}

Our second comment is a warning against the inclusion of an excessive number of defect interactions in our designs. Indeed the very premise of our defect engineering strategy is that topological constraints prevent defect interactions from proliferating throughout the bulk of a large assembly. Excessively generous defect interactions appear to abolish these constraints, as in the last two columns of Fig.~\ref{fig:heuristics}. In hexagonal subunits, our anecdotal experience suggests that  proliferation is rare when fewer than six defect interactions are present, and predominant when there are more. Figure~\ref{fig:heuristics} illustrates the marginal case of six defect interactions. This proliferation is more likely to occur when the individual subunits have fewer sides; for instance we have not managed to design a self-limited assembly based on triangular subunits.

%\newpage
\section{Geometrical considerations}
\label{supp:geometrical_considerations}

\begin{figure}
    \centering
    \includegraphics[width=0.99\linewidth]{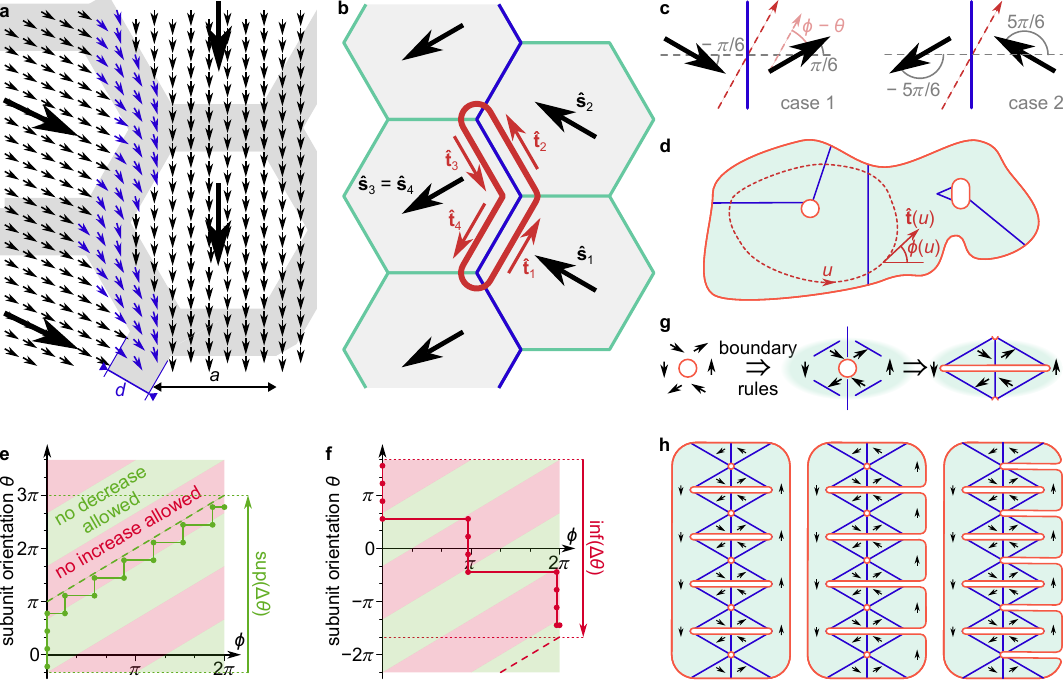}
    \caption{
    \textbf{Geometrical constraints on the presence of defect interactions in an assembly.}\\
    (\textbf{a})~Interpolation of a smooth orientation field $\tilde{\theta}(\mathbf{x})$ (\emph{small arrows}) from the subunit orientations (\emph{large black arrows}, with the orientation convention of the main text). The arrows within a neighborhood $d$ of a defect interaction interpolate between the subunit orientations on either side of it (\emph{dark blue}).
    (\textbf{b})~Integration domain $\omega$ (\emph{dark red}) used to relate the discrete and continuum expressions for the defect energy. Here the subunit orientation is pictured with the convention introduced in Sec.~\ref{sec:field}.
    (\textbf{c})~Coarse-graining the corrugated defect boundary of panel~B to a straight line produces two possible cases for the crossing of this boundary (\emph{dark blue}) by a locally straight, oriented curve (\emph{dashed red arrow}). Cases~1 and 2 respectively correspond to crossing that boundary from right to left, and from left to right. The \emph{dashed gray line} is the normal to the boundary, and we show the angle $\phi-\theta$ between the orientation of the tangent to the curve and the orientation of the subunits in one of the crystalline domains.
    (\textbf{d})~Illustration of a convex loop (\emph{dashed red line}) used to evaluate the winding number of the subunit orientation. \emph{Blue lines} show the coarse-grained straight grain boundaries.
    (\textbf{e})~Illustration of the maximum possible increase of the subunit orientation $\theta$ upon one turn around the loop of panel~D.
    (\textbf{f})~Example of maximum possible decrease of the subunit orientation $\theta$ upon one turn around the loop of panel~D.
    (\textbf{g})~Construction of the grain boundary geometry surrounding a topological defect with winding number $-1$. \emph{Inset:} local structure of the inner boundary of the $-1$ defect. We observe that the tangent vector of every fourth edge of the \emph{orange boundary} is orthogonal to the subunit orientation, while the others make an angle $\pm\pi/6$ with it.
    (\textbf{h})~Examples of assemblies with many topological defects. We recall that a contour that encloses several defects has a winding number equal to the sum of the winding numbers of the individual defects~\cite{Mermin:1979}. From \emph{left} to \emph{right}: convex assembly with winding number $+1$ (the outer boundary encloses five $+1$ defects and four $-1$ defects), assembly whose outer boundary is associated with a large positive winding number (the outer boundary encloses five $+1$ defects and no $-1$ defects), and assembly with a large negative winding number (the outer boundary encloses no $+1$ defects and four $-1$ defects).
    }
    \label{fig:geometry}
\end{figure}

Here we use theory to demonstrate constraints on the spatial arrangement of defects in our assemblies. These constraints are a new feature compared to several classical statistical mechanics models. To make this comparison, we first note that our defect engineering setup requires energetically favorable domain boundaries. In the present work we implement these boundaries using subunits with discrete orientations. This setup is reminiscent of, \emph{e.g.}, the well-known Potts model, as well as the so-called clock model~\cite{Wu:1982}.

In the Potts model, the sites of a lattice each adopt a discrete state often referred to as a ``color''. In contrast with our defect interactions, the interaction between neighboring sites with different colors typically consists in an energetic penalty. This favors single-color domains. To mimic our favorable-boundary scheme within this framework, one may consider making the interaction between different colors favorable instead. This would however result in a drastically different outcome. Indeed, the ground state of such a modified Potts model is a random mosaic of colors. The domain boundaries would thus proliferate throughout the system similar to the two bottom rows of Fig.~\ref{fig:heuristics}d-e, and a macroscopic fraction of the bonds of the lattice would be part of such domain boundaries. A similar conclusion holds for the clock model. In this variant of the Potts model, the colors are replaced by explicit discrete orientations. The interaction between two neighboring sites is computed from their relative orientations to favor either alignment or anti-alignment. The latter choice yields an antiferromagnetic ground state, and thus again a proliferation of the domain boundaries.

By contrast, in our vortex design the fraction of defect interactions in a large, compact assembly is small and proportional to its diameter. This behavior is enabled by a key difference between our model and the Potts/clock models: while there the interactions between two neighboring sites is a function only of their internal states (color or orientation), in our model the interaction also depends on their relative positions on the lattice. For instance, in Fig.~\ref{fig:EnergyCalcul}d a subunit with a $0^\circ$ orientation located to the left of a subunit with a $-60^\circ$ orientation has a defect interaction with its neighbor [matrix element $(1,3)$ of the interaction map]. The situation is completely different if the positions of the two subunits are reversed, which yields a forbidden interaction [matrix element $(6,4)$]. In other words, our model is a variant of the Potts model where each subunit displays a different color to each of its neighbors, hence the added complexity.
 % Addresses Referee comment 1.3

To highlight the geometric considerations that specifically arise from our model, here we prove two key implications of the interaction rules pictured in Fig.~2a of the main text. First, that a large compact assembly must contain a vanishingly small fraction of defect interactions. Second, that alternatives to our six-fold vortex assembly topology are either geometrically forbidden or energetically unstable.

We build our formalism over the first two subsections below. In Sec.~\ref{sec:field}, we map the discrete description of the orientations of our subunits to an equivalent field formalism. We then express the assemblies' defect energy in this formalism in Sec.~\ref{sec:continuum_defect}. In Sec.~\ref{sec:bound}, we use our formalism to derive an upper bound for the number of defect interactions in an assembly, which proves our first point. This upper bound forms the basis of the scaling reasoning presented in the caption of Fig.~1 of the main text. Section~\ref{sec:topological} then proves our second point by establishing that only some assembly topologies are allowed, and by discussing why vortex assemblies are energetically favored among the permissible topologies. Finally, Sec.~\ref{sec:more_topological} discusses a possible extension of this topological point of view to the other designs presented in Fig.~4 of the main text.

For the purposes of this section it is convenient to use a different convention for our interaction energies than that described in the main text. The energy used in the main text reads
\begin{equation}\label{eq:energy}
    E = N_c\epsilon_c+N_d\epsilon_d,
\end{equation}
where $N_c$ and $N_d$ respectively denote the number of crystalline and defect interfaces, and $\epsilon_c$ and $\epsilon_d$ their energies. We also introduce the number $S$ of interfaces between a subunit and an empty site (represented in orange in all figures). Denoting the total number of subunits in our system as $N$, and recalling that each subunit has six sides, we write the total number of sides in our system as $6N$. That number is moreover also equal to $2N_c+2N_d+S$. Subtracting the constant $3N\epsilon_c$ from Eq.~\eqref{eq:energy} and using this equality, we obtain the physically equivalent form
\begin{equation}\label{eq:alt_energy}
    E = N_d(\epsilon_d-\epsilon_c)-S\frac{\epsilon_c}{2},
\end{equation}
which boils down to assigning an energy 0 to crystalline interfaces, an energy $\epsilon_d-\epsilon_c$ to defect interfaces and an energy $-\epsilon_c/2$ to interfaces at the surface of the assembly. We adopt this point of view throughout this section.

\subsection{\label{sec:field}Interpolation of a field theory}
{Here we map the discrete description of our assemblies used in the main text to a continuous field theory. This enable our subsequent use of some convenient tools of vector calculus.}
Consider a connected assembly of oriented hexagonal subunits on a triangular lattice. In this assembly, interfaces between neighboring subunits are either of the crystalline type or of the defect type illustrated in Fig.~2a of the main text. We denote the length of a subunit edge by $a$.

We define an orientation field $\tilde{\theta}(\mathbf{x})$ at every point $\mathbf{x}$ within the assembly through the procedure illustrated in Fig.~\ref{fig:geometry}a. In the procedure, points within a subunit are assigned the orientation of the subunit unless they are within a small boundary region of width $d$ centered around a defect interface. Within this boundary region, the field instead interpolates between the orientations on either side by rotating by an angle $\pi/3$. We consider the $d\ll a$ limit, and within it the detailed form of the interpolating function does not matter for our discussion. We however assume that it is designed in such a way that (i)~the function $\tilde{\theta}(\mathbf{x})$ is differentiable everywhere within the assembly and (ii)~That the field monotonically interpolates between the orientation of two neighboring subunits while choosing the shortest path on the unit circle to do so. For instance, we assume that the interpolation between $\tilde{\theta}=0$ and $\tilde{\theta}=\pi/3$ proceeds through a monotonic increase over the interval $0\leqslant\tilde{\theta}\leqslant\pi/3$ as opposed to a decrease over $2\pi\geqslant\tilde{\theta}\geqslant\pi/3$.

Within this continuous formalism, we denote the set of points belonging to the assembly (\emph{i.e.}, located within any of the hexagonal subunits of the assembly) as $\Omega$. The orientation within $\Omega$ can be equivalently parametrized by any field $\theta(\mathbf{x})=\tilde{\theta}(\mathbf{x})+\theta_0$, where $\theta_0$ is an arbitrary constant. In the following we choose an offset $\theta_0=-2\pi/3$ compared to the convention of the main text. We define the unitary orientation vector field as
\begin{equation}
    \hat{\mathbf{s}}(\mathbf{x})=\left(\begin{array}{c}\cos\theta(\mathbf{x})\\\sin\theta(\mathbf{x})\end{array}\right).
\end{equation}
In this new convention, the unit vector field $\hat{\mathbf{s}}$
is tangent to the overall direction of the outer boundary of the ideal vortex assembly shown in the inset of Fig.~2b of the main text.

\subsection{\label{sec:continuum_defect}Continuum Expression for the Defect Energy}
{Here we show that the assembly defect energy takes a convenient form in our field theory.}
Using our orientation field formalism, we {thus} propose to write the defect energy of the assembly as
\begin{equation}\label{eq:field_ed}
    E_d = \frac{2(\epsilon_d-\epsilon_c)}{\sqrt{3}a} \int\!\!\!\!\int_\Omega \nabla\times\hat{\mathbf{s}}\,\text{d}A,
\end{equation}
where $\text{d}A$ denotes the area element. To verify that this expression matches the discrete definition of the main text to lowest order in $d/a$, we first note that the value of its integrand vanishes everywhere except within our boundary regions of thickness $d$. We then compute its value for the representative portion of grain boundary represented in Fig.~\ref{fig:geometry}b.  We denote the thus-defined domain by $\omega$ and use Stokes' theorem:
\begin{equation}\label{eq:boundary_Stokes}
    E_d = \frac{2(\epsilon_d-\epsilon_c)}{\sqrt{3}a} \int\!\!\!\!\int_\omega \nabla\times\hat{\mathbf{s}}\,\text{d}A = \frac{2(\epsilon_d-\epsilon_c)}{\sqrt{3}} \oint_{\partial\omega} \hat{\mathbf{s}}\cdot \hat{\mathbf{t}} \,\text{d}\ell,
\end{equation}
where $\hat{\mathbf{t}}$ is the tangent vector to domain $\omega$ and $\text{d}\ell$ is the associated length element. To lowest order in $d$, the line integral of Eq.~\eqref{eq:boundary_Stokes} can be approximated by a sum of straight segments which we each label by $i$, yielding
\begin{equation}\label{eq:segment_ed}
    E_d =\frac{2(\epsilon_d-\epsilon_c)}{\sqrt{3}a} \sum_i a\, \hat{\mathbf{s}}_i\cdot \hat{\mathbf{t}}_i
\end{equation}
The example of Fig.~\ref{fig:geometry}b displays the two possible types of defect interfaces associated with our design. It yields $\hat{\mathbf{s}}_1\cdot \hat{\mathbf{t}}_1 = \hat{\mathbf{s}}_3\cdot \hat{\mathbf{t}}_3 = 0$ and $\hat{\mathbf{s}}_2\cdot \hat{\mathbf{t}}_2 = \hat{\mathbf{s}}_4\cdot \hat{\mathbf{t}}_4 = \sqrt{3}/2$. Combining this result with Eq.~\eqref{eq:segment_ed} implies that each interface brings a contribution $\epsilon_d-\epsilon_c$ to the defect energy, thus validating the expression proposed in Eq.~\eqref{eq:field_ed}.

\subsection{\label{sec:bound}Upper bound on the number of defect interactions}
We now use the expression of Eq.~\eqref{eq:field_ed} to establish a bound on the assembly defect energy. Since as discussed in Eq.~\eqref{eq:alt_energy} this energy can alternatively be written as $E_d=N_d(\epsilon_d-\epsilon_c)$, we write 
\begin{equation}\label{eq:Nd}
    N_d=\frac{2}{\sqrt{3}a} \int\!\!\!\!\int_\Omega \nabla\times\hat{\mathbf{s}}\,\text{d}A = \frac{2}{\sqrt{3}a}\oint_{\partial\Omega} \hat{\mathbf{s}}\cdot \hat{\mathbf{t}} \,\text{d}\ell,
\end{equation}
where the second equality again uses Stokes' theorem. In Eq.~\eqref{eq:Nd}, the line integral runs along the outer perimeter of the whole assembly in the counterclockwise direction, and in the clockwise direction along the surface of any internal holes.

The perimeter of our assembly is comprised of a collection of subunit edges, and the value of the tangent vector $\hat{\mathbf{t}}$ thus always matches the discrete orientations of these edges. Let us consider an individual subunit, \emph{e.g.}, in Fig.~\ref{fig:geometry}b. It is clear that relative to the orientation vector $\hat{\mathbf{s}}$ of this subunit, the orientation of the edge vector $\hat{\mathbf{t}}$ can only be $\pm \pi/6$, $\pm \pi/2$ or $\pm 5\pi/6$. We thus always have $\hat{\mathbf{s}}\cdot \hat{\mathbf{t}}\leq (\hat{\mathbf{s}}\cdot \hat{\mathbf{t}})_\text{max}=\sqrt{3}/2$. Using Eq.~\eqref{eq:Nd}, this implies
\begin{equation}\label{eq:Nd_bounds}
    0\leq N_d \leq \frac{2}{\sqrt{3}a}\oint_{\partial\Omega} (\hat{\mathbf{s}}\cdot \hat{\mathbf{t}})_\text{max}\, \text{d}\ell = P.
\end{equation}
Here $P$ is the perimeter of the assembly counted in units of the subunit edge length. This perimeter includes the length of the surface of any internal holes.
In an assembly where this upper bound would be reached, the defect energy would play the same role as a negative surface tension (assuming $\epsilon_d<\epsilon_c$).

In the system considered in Figs.~2 and Fig.~3 of the main text, the case that comes closest to the upper bound of Eq.~\eqref{eq:Nd_bounds} is that of the ideal vortex assembly. There the dot product $\hat{\mathbf{s}}\cdot \hat{\mathbf{t}}$ is maximized at every site of the internal hole of the assembly; it is also maximized along the outer perimeter except at the corners, resulting in 
\begin{equation}\label{eq:vortex_Nd}
N_d=P-18.    
\end{equation}
This offset between the actual maximal value of $N_d$ and the upper bound of Eq.~\eqref{eq:Nd_bounds} results from the microscopic details of our model, and turns out to be crucial in allowing size-controlled assembly. This can be seen in the example of the non-sized-controlled design of Fig. 4h of the main text. There, the maximum number of defect interfaces is exactly equal to $P/2$, and therefore proportional to $P$. The defect energy thus acts exactly as a correction to the assembly surface tension. As a result, the only transition in the assembly ground state morphology is between an absence of growth (for a positive effective surface tension) to unbounded growth (for a negative effective surface tension).

Equation~\eqref{eq:Nd_bounds} implies a severe restriction on the number of defect interactions allowed in a large assembly. In an infinite bulk without an extensive number of holes this result imposes a vanishingly small fraction of defect interactions, as stated in the main text. In addition, in assemblies characterized by a single length scale such as the one of Fig.~1a of the main text, the number of defect interactions must at most scale like the lateral size of the assembly rather than its area. This statement forms the basis of the scaling reasoning presented in the caption of Fig.~1a of the main text. 

\subsection{\label{sec:topological}Topological and energetic constraints resulting from the interaction rules}
As implied by its name, a vortex assembly is centered around a single topological defect with winding number $+1$. One might speculate that the vectorial nature of our order parameter could allow for defects with any other integer winding number. However we do not observe them in simulations, nor do we reproducibly observe single assemblies including more than one vortex. Here we discuss the geometrical and energetic constraints that account for these observations. This additionally allow us to rationalize why our assemblies never take the form of a bulk with an extensive numbers of holes.

\paragraph*{Convex real-space loops can have winding numbers $-1$, $0$ or $1$.}
We first demonstrate a geometrical limitation imposed by our choice of interactions between subunits. For simplicity we present a coarse-grained argument where the defective grain boundaries between domains with different orientations $\theta$ are straight lines instead of straight-on-average zigzagging lines. As illustrated in Fig.~\ref{fig:geometry}c, in that setting the orientations $\theta$ of the subunits on either side of the boundary make angles $\pm\pi/6$ with the normal to the boundary, or $\pm 5\pi/6$ depending on the conventional choice of the direction of this normal.

We consider a differentiable closed loop fully comprised within our assembly (Fig.~\ref{fig:geometry}d). We parametrize it by its curvilinear coordinate $u\in[0,\ell]$ where $\ell$ is the total length of the loop. We choose $u$ in such a way that it increases when traveling along the loop in the counterclockwise direction, and denote the angle of its tangent vector $\hat{\textbf{t}}(u)$ with the horizontal axis by $\phi(u)$. We consider a convex curve with a turning number of one, \emph{i.e.}, one whose tangent has a winding number of one (qualitatively, a closed curve that loops only once). This implies
\begin{equation}
    \forall u\in[0,\ell]\quad \frac{\text{d} \phi}{\text{d} u}(u)\geq 0
    ~~\textrm{and}~~
    \phi(u)\in[0,2\pi].
\end{equation}
We denote by $\theta(u)$ the map that gives the subunit orientation of a point with curvilinear coordinate $u$.

Close to a grain boundary our closed loop is locally a straight line, which we represent by a dashed line in Fig.~\ref{fig:geometry}c. Thus any point where the loop intersects the boundary realizes either one of the following scenarios corresponding to the two panels of Fig.~\ref{fig:geometry}c:
\begin{itemize}
    \item \textbf{Case 1:} The tangent to the loop is about to poke through the boundary in the same direction as the orientation vectors $\hat{\textbf{s}}$, which is equivalent to stating that $-\frac{\pi}{3}<\phi-\theta<\frac{2\pi}{3}$. Then $\theta(u)$ increases by an amount $\pi/3$ as the loop crosses the boundary.
    \item \textbf{Case 2:} The tangent to the loop is about to poke through the boundary in the opposite direction compared to the orientation vectors $\hat{\textbf{s}}$, which is equivalent to stating that $\frac{2\pi}{3}<\phi-\theta<\frac{5\pi}{3}$. Then $\theta(u)$ decreases by an amount $\pi/3$ as the loop crosses the boundary.
\end{itemize}
This implies that the angle $\theta$ cannot increase if $\frac{2\pi}{3}\leq\phi-\theta\leq\frac{5\pi}{3}$, and that it cannot decrease if $-\frac{\pi}{3}\leq\phi-\theta\leq\frac{2\pi}{3}$. As shown in  Fig.~\ref{fig:geometry}e-f, these conditions divide the ($\phi,\theta$) plane in three types of regions: those where only an increase of $\theta$ by increments of $\pi/3$ is possible; those where only an decrease of $\theta$ by decrements of $\pi/3$ is possible; and the boundaries between those two types of regions, where the loop runs parallel to the grain boundary and $\theta$ can neither increase nor decrease.

We point out two trajectories on these schematics. Since $\phi$ is an increasing function of $u$, we parametrize the trajectory along our convex loop by $\phi$ instead of $u$:
\begin{itemize}
\item Figure~\ref{fig:geometry}e shows a trajectory of maximum possible increase of $\theta$. We start with $\theta\in(-\pi/3,0)$, which places us in a region where increases are allowed. We immediately increase $\theta$ by three increments of $\pi/3$. The trajectory is then stuck in a no-increase region. We must then wait for $\phi$ to increase to again find ourselves in the region that allows $\theta$ to increase. When that happens, we increase $\theta$ by $\pi/3$. We renew the operation whenever possible until $\phi$ reaches $2\pi$. Clearly this trajectory is constrained to lie strictly below the $\theta=\phi+\pi$ green dashed line. Thus its end point $\theta(\ell)$ lies strictly below the end point of the dashed line, whose vertical coordinate is $3\pi$, \emph{i.e.}, $\theta(\ell)<3\pi$. Recalling our assumption that $\theta(0)>-\pi/3$, this implies $\theta(\ell)-\theta(0)<10\pi/3$. In other words, the total increase of $\theta$ over the trajectory is bounded by above by $10\pi/3$.
\item Figure~\ref{fig:geometry}f shows a trajectory of maximum possible decrease of $\theta$. We start with $\theta\in(4\pi/3,5\pi/3)$, which immediately allows for three consecutive decreases by $\pi/3$ each. The trajectory is then stuck in a no-decrease region, and must wait until $\phi$ increases by a little less than $\pi$ to undergo three additional consecutive decreases. This scenario then repeats itself once. The whole trajectory must remain above the $\theta=\phi-11\pi/3$ red dashed line. A similar reasoning to the one detailed above in our discussed of Figure~\ref{fig:geometry}e thus implies that the total decrease of $\theta$ over the course of the trajectory is bounded by below by $-10\pi/3$.
\end{itemize}

This reasoning thus implies
\begin{equation}\label{eq:angle_constraint}
    -\frac{10\pi}{3} < \theta(\ell)-\theta(0) < \frac{10\pi}{3}.
\end{equation}
Since the continuity of the orientation field additionally imposes that $\theta(\ell)=\theta(0)\mod 2\pi$, we conclude $\theta$ can only change by $-2\pi$, $0$ or $2\pi$. The geometry of our defect interactions thus only allows winding numbers of the order parameter over our convex closed loop with values $-1$, $0$ or $+1$.

\paragraph*{Isolated point-like defects can only have winding number $+1$.}
Consider a large assembly with one or more isolated topological defects, which we take to mean that we can individually enclose each of them in a finite-size convex loop as in Fig.~\ref{fig:geometry}d. As a result of the property proved above, the winding number associated with such an isolated defect is either $+1$ or $-1$. Here we show that the geometry of our interactions forces the $-1$ defects to have a finite core size (\emph{i.e.}, prevents them from being point-like), which compromises its local stability.

We recall that the winding number of a defect is defined by considering the change in the angle $\theta(u)$ as $u$ goes around a loop that encloses it. In the case of a counterclockwise loop, a $+1$ defect is characterized by a $2\pi$-increase of $\theta$, \emph{i.e.}, the vector $\hat{\mathbf{s}}$ undergoes one counterclockwise rotation as we go around the loop. Conversely, a $-1$ defect corresponds to a decrease by $-2\pi$ or equivalently a clockwise rotation of $\hat{\mathbf{s}}$.
A $-1$ defect must thus be surrounded by crystalline domains with all six possible subunit orientations in the opposite order to that of the orientations found around a vortex assembly, which we illustrate in Fig.~\ref{fig:geometry}g. As shown there, the relative orientations of these domains fixes the orientation of the grain boundaries between them according to our interaction rules. This construction makes it evident that lines cannot be extended indefinitely without intersecting their counterparts. Since our interaction rules forbid an interface between two domains whose orientations differ by $2\pi/3$, these intersections must either give rise to a defect or occur at the outer boundary of the assembly. As shown on the right of Fig.~\ref{fig:geometry}g, the orientation of the grain boundaries imposes that the horizontal size of the $-1$ defect is of the order of the vertical distance between these two defects or outer boundaries; therefore isolated $-1$ defects are not point-like.

To assess the cost of opening such a defect, we note in the inset of Fig.~\ref{fig:geometry}g that a quarter of the edges that constitutes its inner boundary has $\hat{\mathbf{t}}\cdot\hat{\mathbf{s}}=0$ while the other three quarters have $\hat{\mathbf{t}}\cdot\hat{\mathbf{s}}=\sqrt{3}/2$. Equation~\eqref{eq:Nd} thus implies that the total energy per unit length associated with the defect's inner boundary is $3/4(\epsilon_d-\epsilon_c)-\epsilon_c/2=3/4\epsilon_d-5/4\epsilon_c$. In cases where this energy is negative, \emph{i.e.}, whenever $\epsilon_c>3\epsilon_d/5$, such defects tend to gain energy by extending laterally. As demonstrated in Sec.~\ref{supp:phase_diagram}, this condition is fulfilled in all regimes of the phase diagram of Fig.~2b of the main text where finite vortices form. As a result, an extended isolated $-1$ defect that forms at the center of a large assembly locally tends to grow until it spans across the whole assembly and splits it in half, thereby deleting itself.

\paragraph*{Convex assemblies with a single vortex are favored over other defective assemblies.}
Beyond assemblies containing a single isolated topological defect, we now discuss assemblies with multiple defects. This discussion provides some assembly-wide insights on their stability which complement the local arguments discussed in the previous paragraph. For the purpose of this discussion we distinguish between convex and non-convex assemblies, \emph{i.e.}, between assemblies whose outer boundary (colored orange in Fig.~\ref{fig:geometry}) is a convex curve and those for which it is not.

We first consider a convex assembly and draw a loop along its outer boundary. Our previous result implies that the net winding number of all defects contained within this loop is $-1$, $0$ or $+1$. They thus contain as many $+1$ defects as $-1$ defects plus or minus one unit. We illustrate one such assembly in Fig.~\ref{fig:geometry}h. Let us discuss its defect energy as well as its surface energy while approximating the $+1$ topological defects as point-like, implying that their contribution to the line integral of Eq.~\eqref{eq:Nd} is negligible. The finite-size $-1$ defects, which cannot be point-like as previously discussed, contribute to this integral to the rate of $3/4$ per unit length. To leading order in the assembly perimeter $P$, the number of defects in the assembly is thus
\begin{equation}
    N_d = \left[(1-f) + \frac{3}{4}f\right]P + \mathcal{O}\left(P^0\right)=\left(1-\frac{f}{4}\right)P+ \mathcal{O}\left(P^0\right),
\end{equation}
where $f\in(0,1)$ is the fraction of the total assembly perimeter associated with an inner boundary. By contrast, breaking up the same number of subunits into a set of vortex assemblies with the same perimeter would result in a larger number of defects $N_d = P + \mathcal{O}\left(P^0\right)$ [see Eq.~\eqref{eq:vortex_Nd}]. The latter option has the same surface energy as the large, multiple-defect assembly, but a more favorable defect energy, and is thus the energetically preferred outcome.

As illustrated in Fig.~\ref{fig:geometry}h, non-convex assemblies may display arbitrary large positive or negative winding numbers. Both types of assemblies display the same types of inner boundary which contribute to the integral of Eq.~\eqref{eq:Nd} at an average rate of $3/4$, plus additional boundaries which are perpendicular to the subunit orientations and thus do not contribute at all. The basic energetic argument made above for a convex assembly therefore equally applies here, and is compounded by an additional surface energy cost due to the creation of new assembly boundaries that run perpendicular to the local subunit orientation.

Overall, these arguments indicate that multiple-defect assemblies are less energetically favorable than vortex assemblies. This rationalizes the absence of multiple-defect as well as non-convex assemblies in the results presented in the main text.

\subsection{\label{sec:more_topological}Possible topological nature of other two-dimensional designs}

In our vortex assemblies, defect lines mark the boundaries between crystalline domains. These domain boundaries join at the center of the assembly and form a vortex there, \emph{i.e.}, a point-like defect with a nontrivial topological character in the sense of homotopy theory~\cite{Mermin:1979}. While it is clear from Secs.~\ref{supp:phase_diagram} and \ref{supp:other_geo} that domain boundaries are essential to all our defect-engineered designs, it is less obvious whether a topological defect is also required. To begin to address this question, here we discuss the presence of topological defects in the designs presented in Fig.~4 of the main text.

We first consider the vortex-like assemblies presented in Fig.~4e-f. These examples clearly include a topological defect. Indeed we can unambiguously apply the interpolation procedure described in Sec.~\ref{sec:field} to convert the subunit orientations into a continuous orientation field. This field has a winding number of $+1$ at the center of the assembly, which puts these objects in the same category as the original vortex assembly. Figure~4f, which does not have a self-limitation, furthermore demonstrates that having a topological defect does not guarantee the existence of a such a mechanism.

Another simple case is the three-dimensional assembly of Fig.~4h. As discussed in Sec.~\ref{supp:other_geo}, we intentionally designed its subunits to be invariant through a rotation around their main axis. This makes them vector-like, and their orientations can be mapped to a vector field in three dimensions through a simple generalization of the procedure of Sec.~\ref{sec:field}. This field has a ``hedgehog'' topological defect at the center of the assembly by design.

\begin{figure}
    \centering
    \includegraphics[width=100mm]{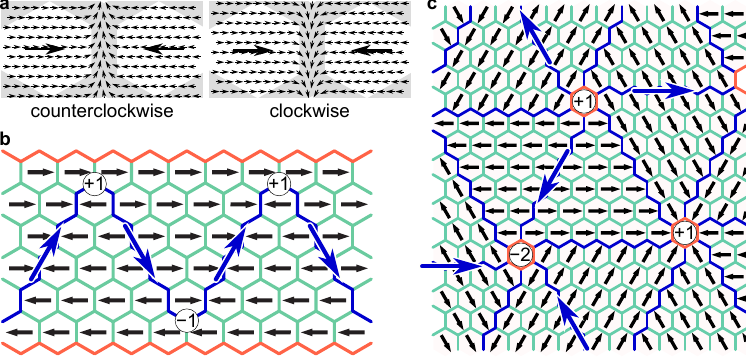}
    \caption{
    \textbf{Assignment of a winding number to two-dimensional point defects in two ambiguous cases.}\\
    (\textbf{a})~There are two equally valid procedures to interpolate the orientation field across a boundary where the subunit orientation flips by $180^\circ$. The graphical conventions in this panel are the same as in Fig.~\ref{fig:geometry}a.
    (\textbf{b})~Interpolation in a fiber using the counterclockwise convention. The \emph{large blue arrows} on the defect lines highlight the direction of the interpolation. The resulting field has $+1$ topological defects along the assembly's upper boundary and $-1$ defects along the lower boundary (\emph{circled numbers}). Using the clockwise convention would exchange the location of these defects, as it the whole fiber had been rotated by $180^\circ$.
    (\textbf{c})~Interpolation of an orientation field and defect winding numbers on the scale of a unit cell of the periodic bulk.
    }
    \label{fig:topological_interpolation}
\end{figure}

The case of the fiber of Fig.~4a-c is less straightforward. Unlike in the examples above, the interpolation of its subunit orientations into a vector field is ambiguous. The cause of this ambiguity is that the subunit orientations on either side of a domain boundary point in exactly opposite directions. This implies that there are two equally natural ways to interpolate between their orientations, namely by rotating the arrow anticlockwise or clockwise as we go from left to right in Fig.~\ref{fig:topological_interpolation}a. The implications of this arbitrary choice are limited however. To understand why, first consider Fig.~\ref{fig:topological_interpolation}b where we made the former choice. We see that one side of the fiber is decorated by vortices (topological defects with a winding number $+1$), while the other has antivortices (topological defects with a winding number $+1$). This conclusion would remain unchanged had we made the opposite, clockwise choice, although the vortices and antivortices would exchange their positions. 

The patterned bulk of Fig.~4e of the main text is a periodic structure whose unit cell comprises six triangular domains, each with a different orientation. This unit cell comprises three geometrically inequivalent point defects. Assigning them a winding number requires a similar arbitrary choice as in the fiber example, with different choices resulting in a reshuffling of the winding numbers between the three point defects. We display one possible convention in Fig.~\ref{fig:topological_interpolation}c, and find two vortices with winding number $+1$ and one point defect with winding number $-2$. We recall that the group structure of the homotopy group imposes that the winding number computed along the edge of an assembly is equal to the sum of the winding numbers of the point defects within. In the case of an infinite periodic bulk, the former is a surface term, which scales with the first power of the assembly radius, while the latter is a bulk term that scales with its square. Thus this equality can only be realized in the limit of large assemblies if the mean winding number per unit area goes to zero for large assemblies. In other words, the winding numbers of the point defects must cancel out on average, as is the case here. 

Overall, the simple designs considered here all appear to be interpretable in terms of topological defects of the subunit orientation field. We however note that such a description requires stretching the continuum concepts that underlie homotopy theory to accommodate our discrete subunits, suggesting that the problem may be better approached using other mathematical tools. We also want to emphasize that due to the enormous number of possible defect engineered designs (Sec.~\ref{sec:number_of_designs}), our present presentation is limited to those interactions that we could build from intuition. This likely biases the sample presented here, and leaves the possibility that a more systematic exploration of all possible size-limited defect engineered designs may not display as clear a connection with topological defects.

\newpage

%%%%%%%%%%%%%% SUPPLEMENTARY FIGURES %%%%%%%%%%%%%%%
\section{Supplementary experimental figures and simulations parameters}

\begin{figure}[h]
    \centering
     \includegraphics[width=.8\linewidth]{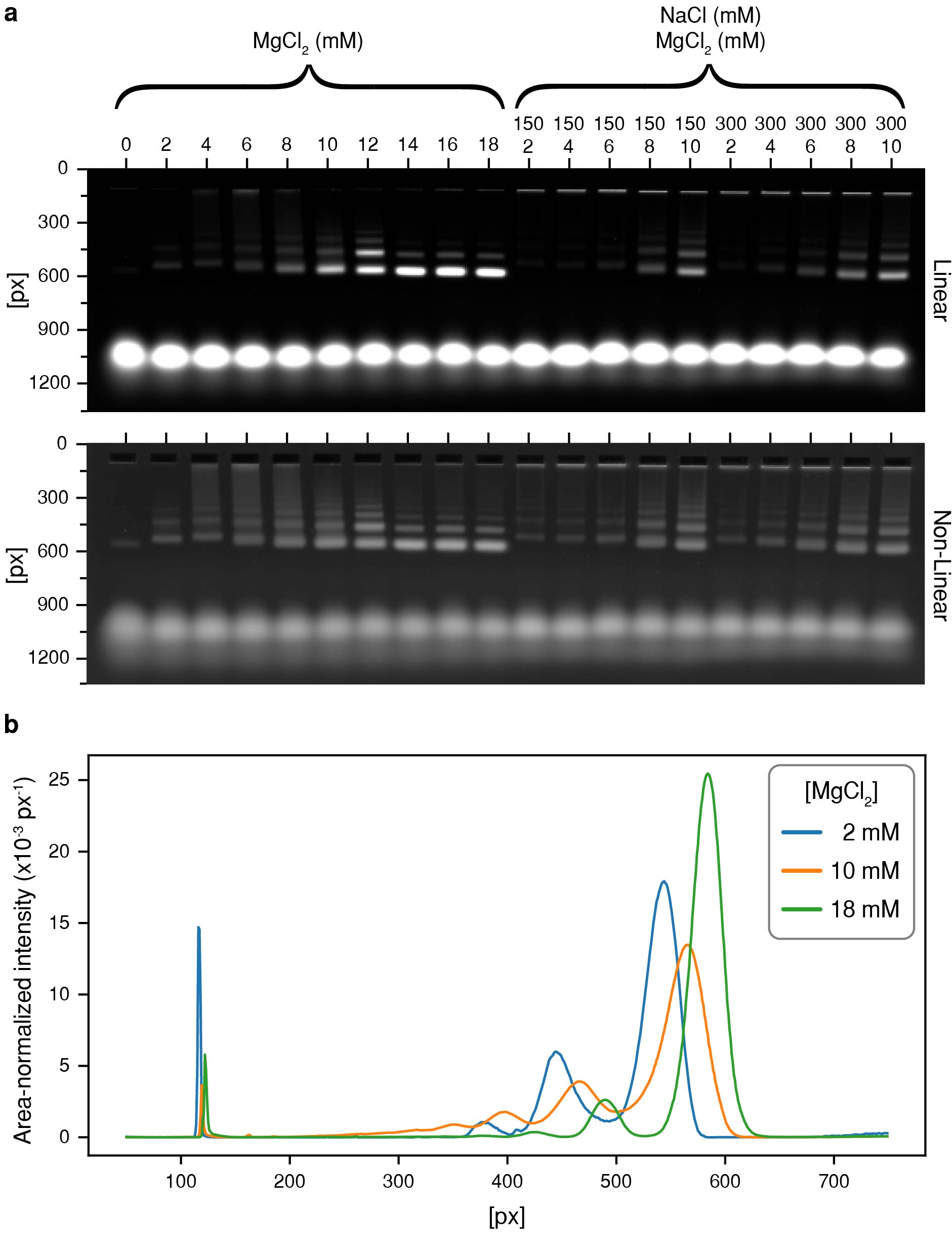}
    \caption{\textbf{Influence of buffer composition on nanocylinder yield.}\\
    (\textbf{a})~Agarose gel electrophoresis scans (top: linear, bottom: non-linear intensity adjustment) of nanocylinders folded in FoB buffer containing varying NaCl and MgCl$_2$ concentrations. Samples contained \qty{50}{\nano\molar} p2873 scaffold and \qty{200}{\nano\molar} staples. Monomer bands appear above 600~px, followed by multimer/aggregate bands; excess staples migrate below 600~px.
    (\textbf{b})~Area-normalized band profiles for \qty{2}{\milli\molar}, \qty{10}{\milli\molar}, and \qty{18}{\milli\molar} MgCl$_2$, analyzed via \textit{gelpy}~\cite{Gelpy:2024}. Peak monomer yield is observed at \qty{18}{\milli\molar} MgCl$_2$.}
    \label{fig:nanocylinder_folding_screen}
\end{figure}

\begin{figure}[p]
    \centering
     \includegraphics[width=\linewidth]{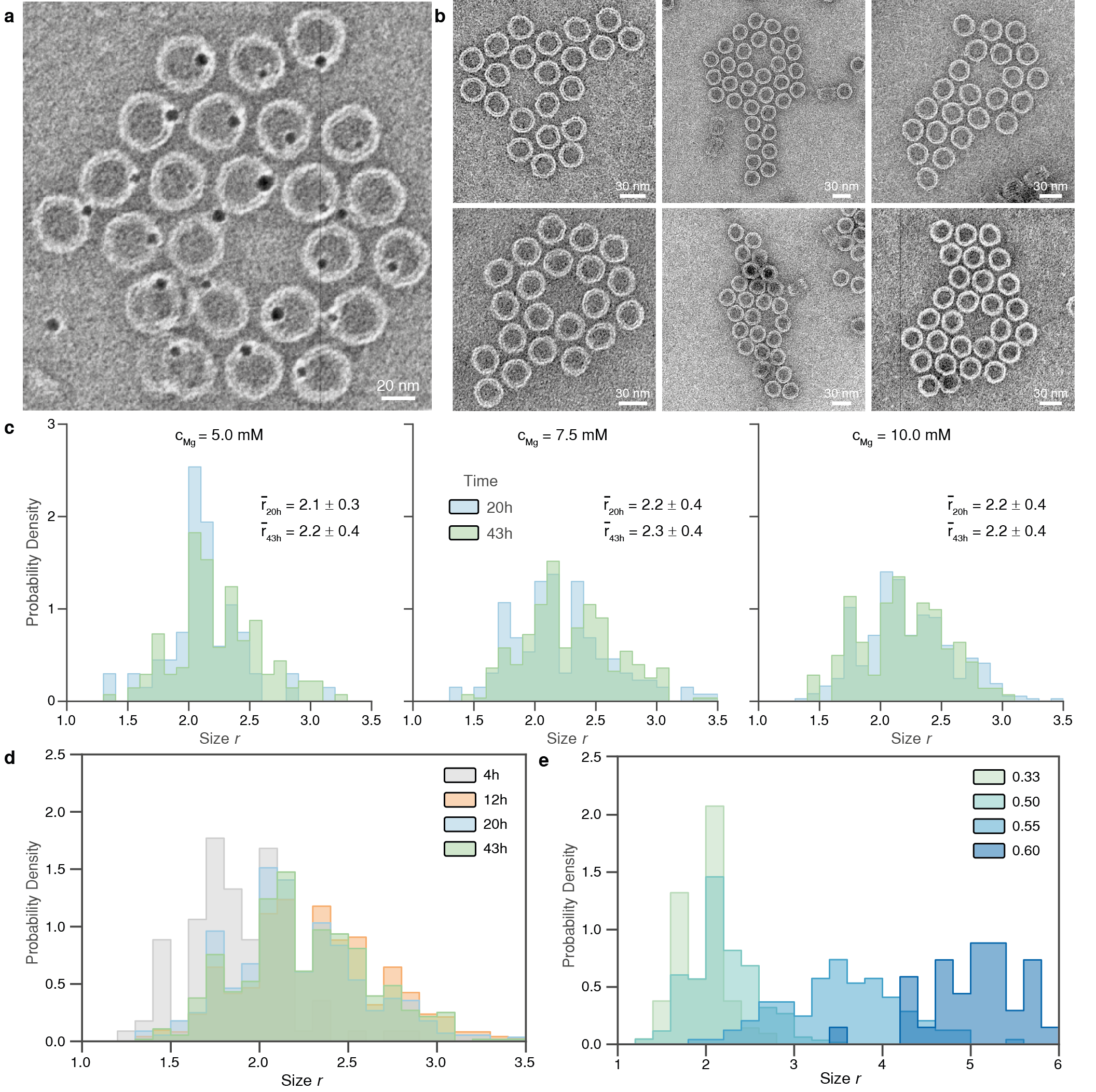}
    \caption{\textbf{TEM micrographs and histograms showing various aspects of spatial and temporal vortex assembly behaviour.}\\
    (\textbf{a})~This TEM micrograph shows the same assembly as Fig.~3d with annotations removed for visibility. Here $\epsilon_c/\epsilon_d=0.52$.
    (\textbf{b})~TEM micrographs of defect-interaction-induced appendages reminiscent of those observed in the numerical simulation of Fig.~2c of the main text. Another instance of these appendages is visible on the micrograph of Fig.~3c of the main text.
    (\textbf{c})~Size distributions for isothermally self-assembled samples ($\epsilon_c/\epsilon_d=0.52$) for \qty{20}{\hour} and \qty{43}{\hour} respectively. MgCl$_2$ concentrations of \qty{5}{\milli\molar} ($N_{20\text{h}}=67$, $N_{43\text{h}}=137$), \qty{7.5}{\milli\molar} ($N_{20\text{h}}=131$, $N_{43\text{h}}=278$) and \qty{10}{\milli\molar} ($N_{20\text{h}}=364$, $N_{43\text{h}}=141$) were screened. Errors of mean radii represent Bessel corrected sample standard deviations.
    (\textbf{d})~Time series of size distribution of isothermally self-assembled samples ($\epsilon_c/\epsilon_d=0.52$, $[\text{MgCl}_2]=\qty{5}{\milli\molar}$, $N_{4\text{h}}=113$, $N_{12\text{h}}=1223$, $N_{20\text{h}}=562$, $N_{43\text{h}}=693$). A Kolmogorov–Smirnov p-value test reveals the similarity between histograms at different timepoints with $p_{12\text{h}\rightarrow20\text{h}}=\qty{0.12}{}$ and $p_{20\text{h}\rightarrow43\text{h}}=\qty{0.07}{}$, indicating the convergence of the underlying distribution towards a steady state.
    (\textbf{e})~Size distributions for isothermally self-assembled samples ($t=20\,\text{h}$, $[\text{MgCl}_2]=\qty{5}{\milli\molar}$) for $\epsilon_c/\epsilon_d=0.31$, 0.52, 0.57 and 0.61 respectively. ($N_{0.31}=106$, $N_{0.52}=562$, $N_{0.57}=122$, $N_{0.61}=38)$.}
    \label{fig:cmbassembly}
\end{figure}

\begin{figure}[p]
    \centering
     \includegraphics[width=\linewidth]{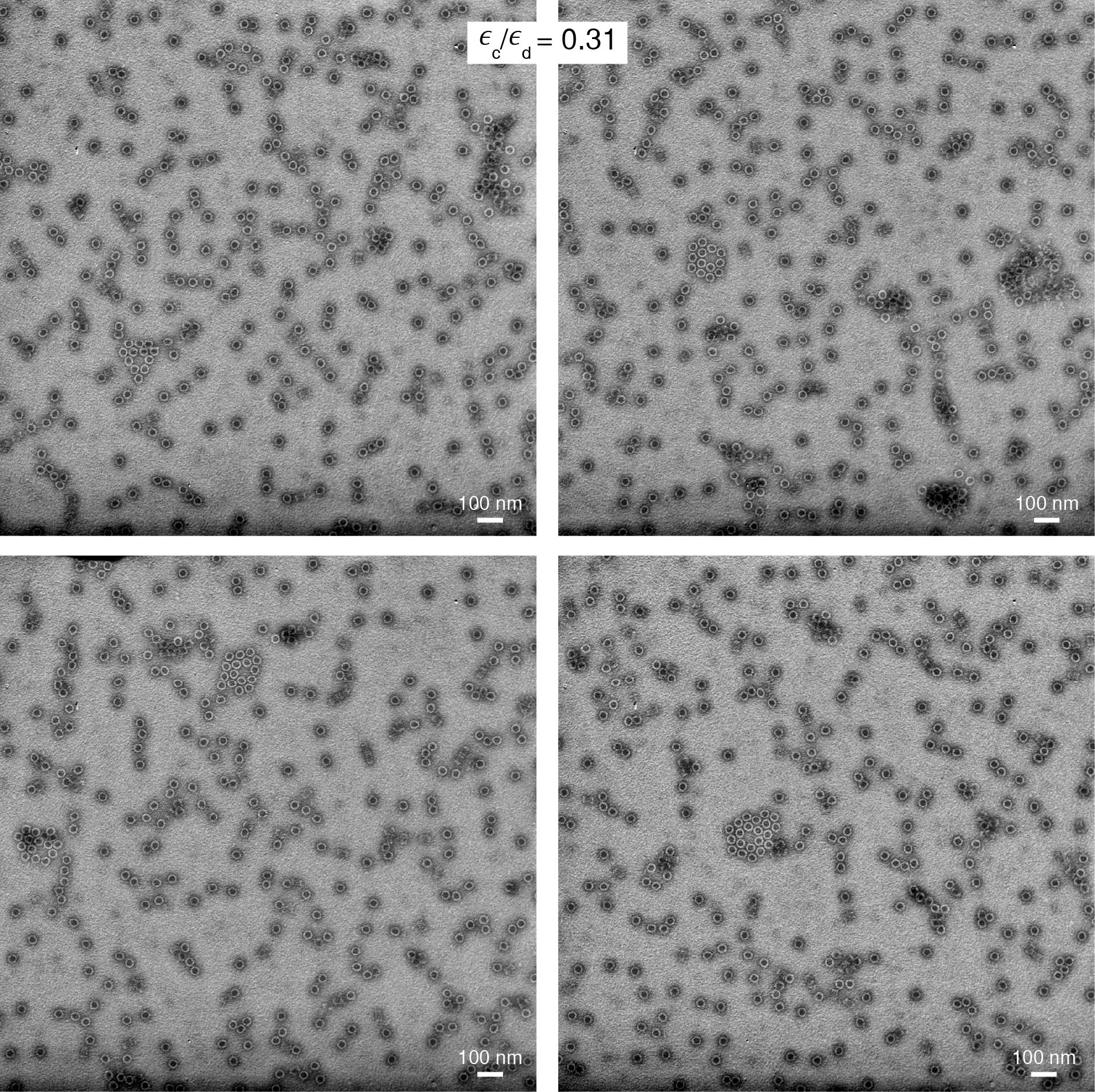}
    \caption{\textbf{TEM micrographs of vortex assemblies at $\epsilon_{\mathrm{c}}/\epsilon_{\mathrm{d}} = 0.31$.}\\
    Isothermally self-assembled samples (\qty{36}{\degreeCelsius}, \qty{5}{\milli\molar} MgCl$_2$) reveal vortex structures at various intermediate growth stages. Observed transitional and final vortex sizes range within $1 < r \lesssim 2$. The conspicuous absence of hexameric seed structures or their corresponding precursors indicates that nucleation constitutes the primary kinetic bottleneck in the early phase of vortex assembly.}
    \label{fig:vortex_fov_0.31_gallery_2x2.jpg}
\end{figure}

\begin{figure}[p]
    \centering
     \includegraphics[width=\linewidth]{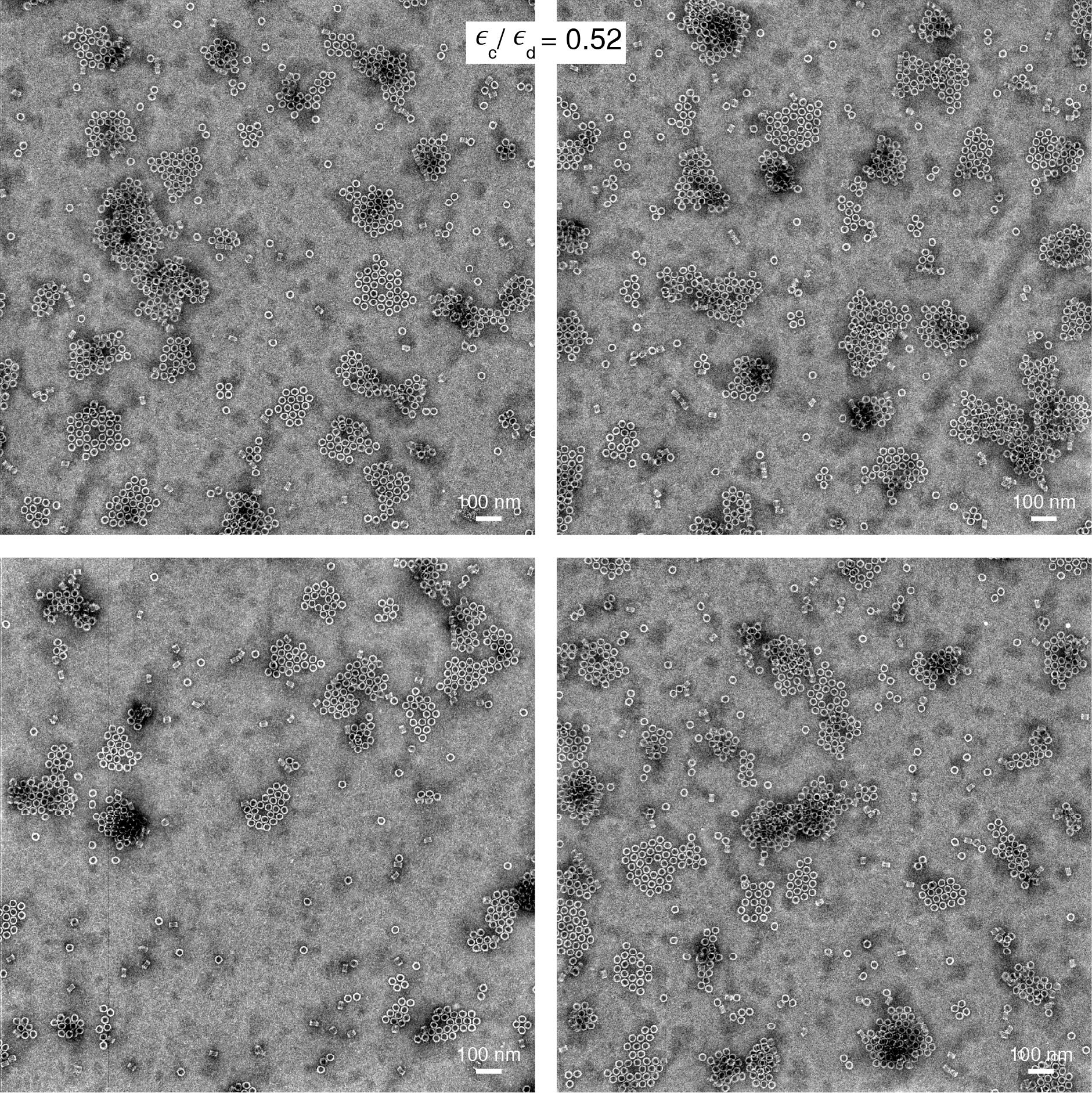}
    \caption{\textbf{TEM micrographs of vortex assemblies at $\epsilon_{\mathrm{c}}/\epsilon_{\mathrm{d}} = 0.52$.}\\
    Isothermally self-assembled samples (\qty{36}{\degreeCelsius}, \qty{5}{\milli\molar} MgCl$_2$) display vortex structures across multiple developmental steps, with characteristic sizes ranging within $2 \lesssim r \lesssim 4$. In addition to target vortex architectures, smaller assemblies possessing pentameric pseudo-seeds are observed. However, their subsequent growth is structurally impeded by geometric incompatibilities between the seed layout and global topological constraints.}
    \label{fig:vortex_fov_0.52_gallery_2x2.jpg}
\end{figure}

\begin{figure}[p]
    \centering
     \includegraphics[width=\linewidth]{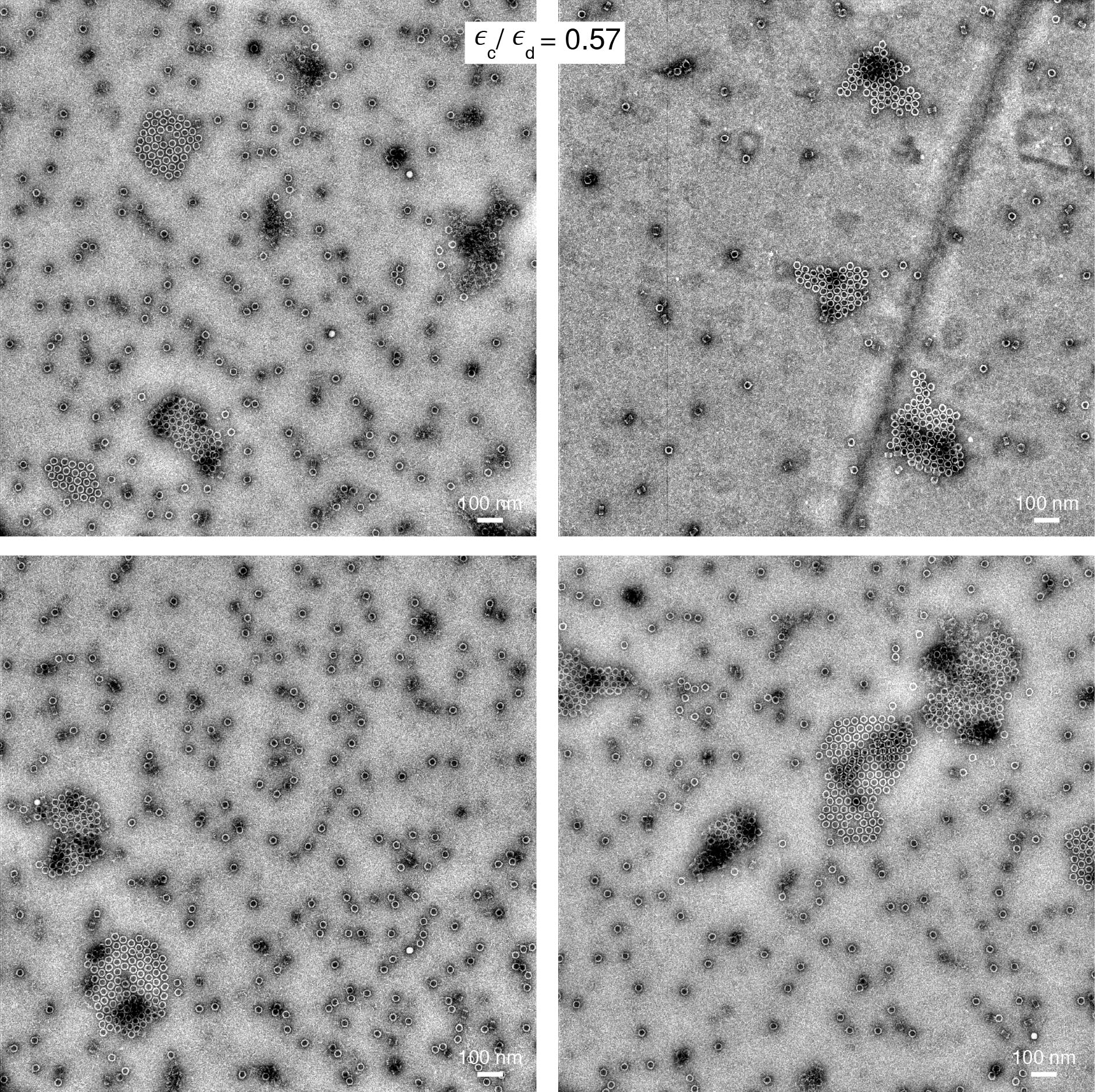}
    \caption{\textbf{TEM micrographs of vortex assemblies at $\epsilon_{\mathrm{c}}/\epsilon_{\mathrm{d}} = 0.57$.}\\
    Isothermally self-assembled samples (\qty{36}{\degreeCelsius}, \qty{5}{\milli\molar} MgCl$_2$) reveal vortex structures at distinct growth increments, with intermediate and final sizes ranging within $3 \lesssim r \lesssim 5$. The total number of observed stable vortices is greatly diminished relative to samples assembled at lower ratios ($\epsilon_{\mathrm{c}}/\epsilon_{\mathrm{d}} = 0.31$ and $0.52$). This behavior is attributed to rapid monomer consumption by rapidly growing seeds, which depletes the free subunit pool below the critical concentration necessary for nucleation and halts further seed formation on experimentally accessible timescales.}
    \label{fig:vortex_fov_0.57_gallery_2x2.jpg}
\end{figure}

\begin{figure}[p]
    \centering
     \includegraphics[width=\linewidth]{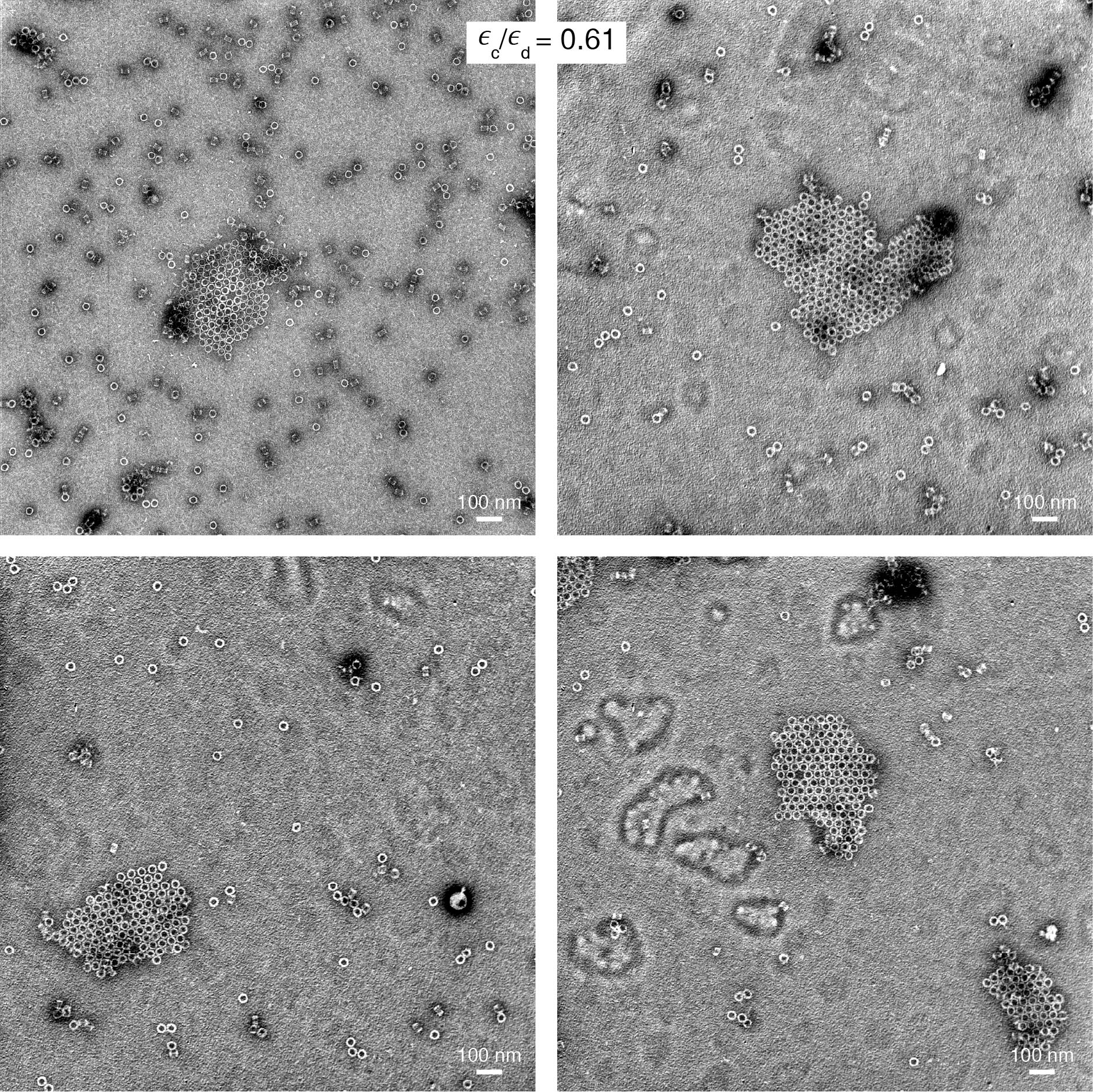}
    \caption{\textbf{TEM micrographs of vortex assemblies at $\epsilon_{\mathrm{c}}/\epsilon_{\mathrm{d}} = 0.61$.}\\
    Isothermally self-assembled samples (\qty{36}{\degreeCelsius}, \qty{5}{\milli\molar} MgCl$_2$) resolve architectures at various growth stages with dimensions of $4 \lesssim r \lesssim 6$. At this energy ratio, vortices predominantly exhibit open morphologies with partially completed boundaries that indicate the terminal size limit $r^*$.}
    \label{fig:vortex_fov_0.61_gallery_2x2.jpg}
\end{figure}

\begin{figure}[p]
    \centering
     \includegraphics[width=\linewidth]{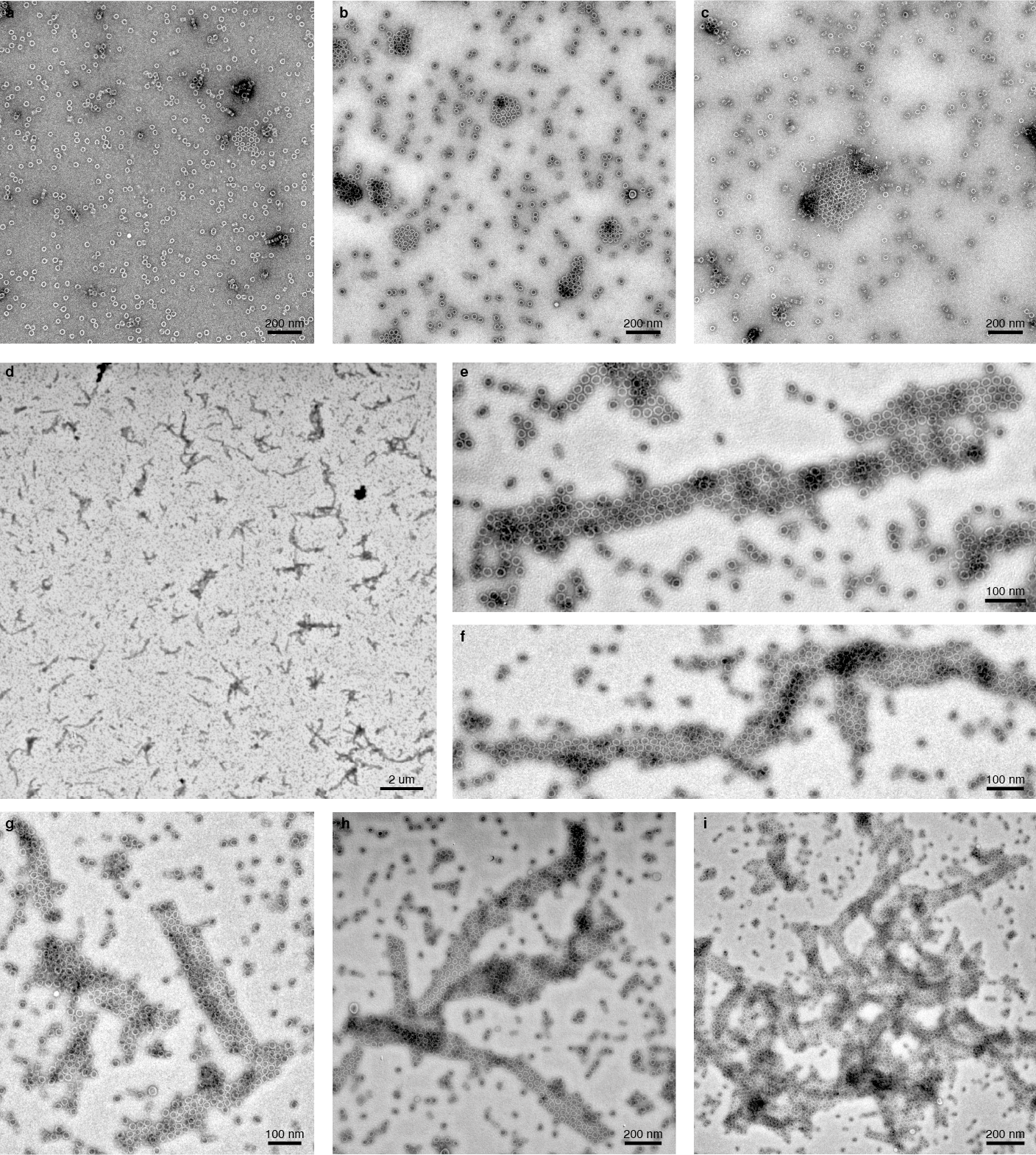}
    \caption{\textbf{TEM micrographs of size-controlled vortex and fiber assemblies with DNA origami.}\\ 
    (\textbf{a})~Typical field of view of vortex assemblies at $\epsilon_c/\epsilon_d=0.31$.
    (\textbf{b})~Typical field of view of vortex assemblies at $\epsilon_c/\epsilon_d=0.52$.
    (\textbf{c})~Typical field of view of vortex assemblies at $\epsilon_c/\epsilon_d=0.61$.
    (\textbf{d})~Typical field of view of fiber assemblies at $\epsilon_c/\epsilon_d=0.50$.
    (\textbf{e})-(\textbf{f})~Close-ups on standalone fibers under the same conditions.
    (\textbf{g})-(\textbf{i})~Close-ups on overlapping and branching fibers under the same conditions.
    }
    \label{fig:gallery}
\end{figure}

\begin{figure}[p]
    \centering
     \includegraphics[width=\linewidth]{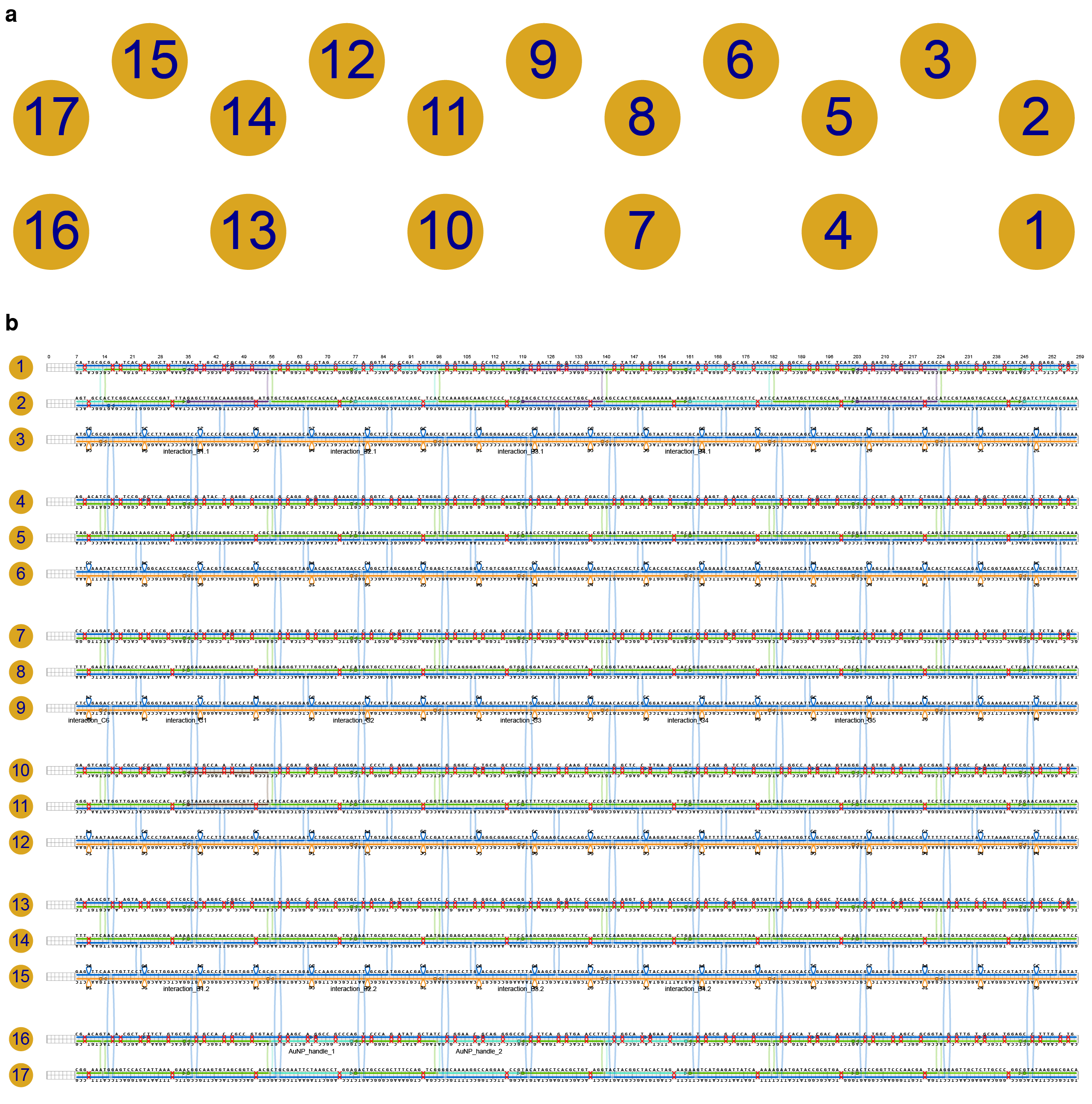}
    \caption{\textbf{Design file of a DNA origami subunit as extracted from the scadnano software.}\\
    (\textbf{a})~Spatial arrangement of DNA helices (numbered circles) that form the side wall of the DNA origami nanocylinder.
    (\textbf{b})~Main view of the design file showing scaffold routing, staple breaking and DNA sequences. Positions of single-stranded extensions for defect/crystalline interaction strands and AuNP linker strands are labeled. The corresponding sequence information can be found in Data S3 and Data S4.}.
    \label{fig:scadnano}
\end{figure}

\begin{figure}[p]
    \centering
     \includegraphics[width=\linewidth]{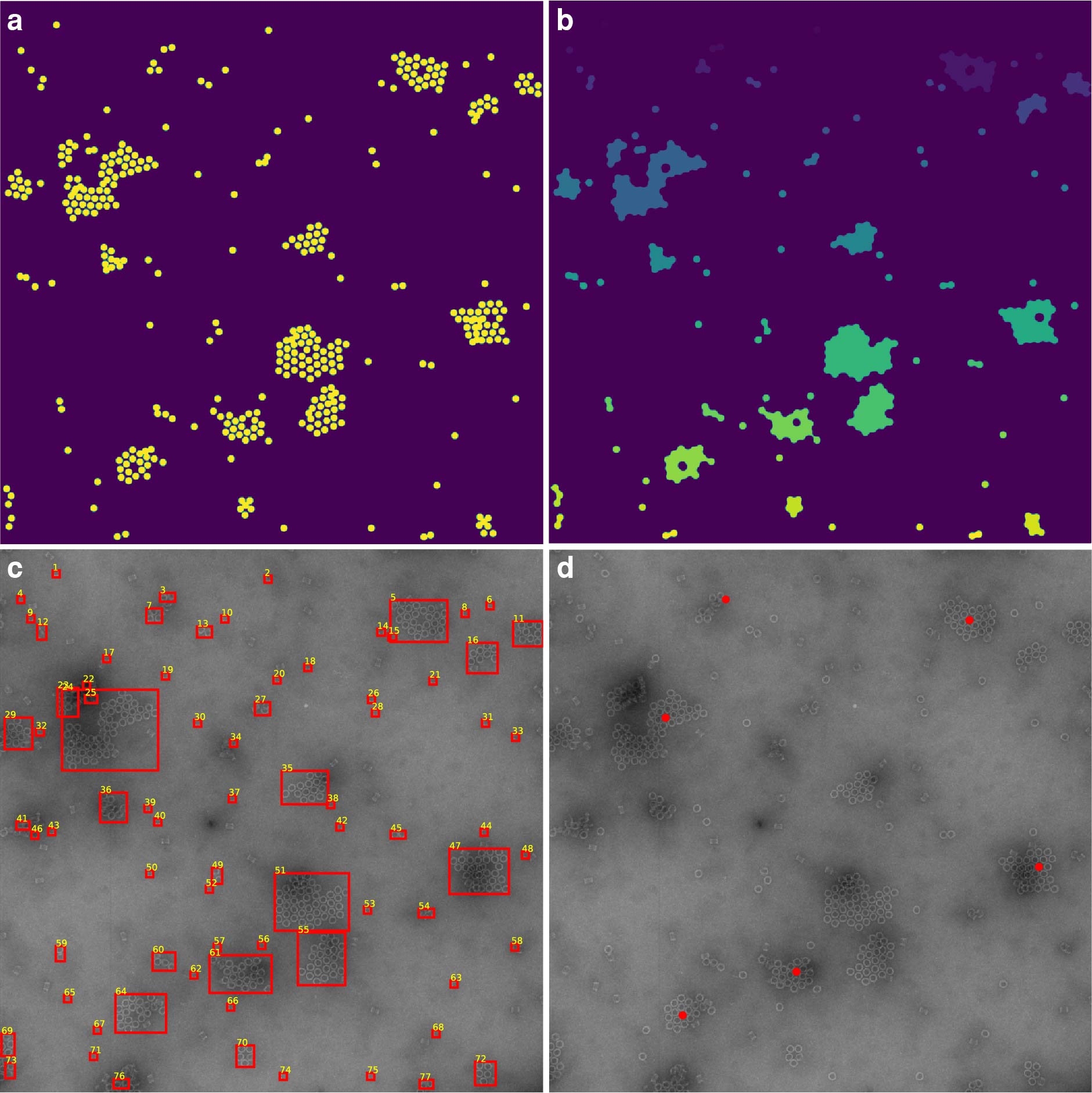}
    \caption{\textbf{Automated assembly detection ensures reproducible data analysis.}\\
    (\textbf{a})~A template representing the mean subunit structure is matched to TEM micrographs to identify individual subunits and their positions (\textit{yellow dots}).
    (\textbf{b})~A gap-filling algorithm merges neighboring subunits into connected regions corresponding to assemblies.
    (\textbf{c})~Detected assemblies are labeled, and relevant features are extracted for downstream analysis.
    (\textbf{d})~Labeled assemblies are filtered for vortex structures by detecting the characteristic central hole. Filtering criteria are applied to assembly area, aspect ratio, and ellipticity. Assemblies classified as vortices are highlighted by a \textit{red dot} marking the verified central hole.}
    \label{fig:automated_detection}
\end{figure}

%%%%%%%%%%%%%%%% SUPPLEMENTARY TABLES %%%%%%%%%%%%%%%

\begin{table}[p]% Do not use \begin{table*}
    \centering
    \begin{tabular}{|c|c|c|c|c|c|c|c|}
    %\begin{tabular}{|C|C|C|C|C|C|C|}
    \hline
        Figure & Assembly type &  $A$ & $\#$ subunits & $\epsilon_c$ & $\epsilon_d$ & $\epsilon_c/\epsilon_d$ & $\epsilon_f$  \\ \hline
        Fig. 2c & 6-fold vortex &  $ 5\times 10^{8}$ & 400 & $-18.7$ & $-74.8$ & $0.25$ &  10  \\
        Fig. 2c & 6-fold vortex &  $ 5\times 10^{8}$ & 400 & $-18.7$ & $-45.9$ & $0.407$&  10  \\
        Fig. 2c & 6-fold vortex &  $ 5\times 10^{8}$ & 400 & $-18.7$ & $-39.0261$ & $0.479$&  10 \\
        Fig. 2c & 6-fold vortex & $ 5\times 10^{8}$ & 400 & $-18.7$ & $-34.2305$ & $0.546$&  10 \\
        Fig. 2c & 6-fold vortex & $4 \times 10^{7}$ & 400 & $-18.7$ & $-22.44$ & $5/6$&  10 \\
        Fig. 2c & 6-fold vortex & $4 \times 10^{7}$ & 400 & $1$ & $-74.8$ & &  10 \\
        Fig. 2c & 6-fold vortex & $4 \times 10^{7}$ & 400 & $-18.7$ & $1$ & &  10 \\
        Fig. 2c & 6-fold vortex & $4 \times 10^{7}$ & 400 & $1$ & $1$ & &  10\\
        Fig. 4c & Fiber  &  $4 \times 10^{7}$ & 300 & $-18.7$ & $-43.6333$ & $0.43$ & 50 \\
        Fig. 4c & Fiber  &  $4 \times 10^{7}$ & 300 & $-18.7$ & $-31.7087$ & $0.59$ & 50 \\
        Fig. 4c & Fiber  &  $4 \times 10^{7}$ & 300 & $-18.7$ & $-30.1102$ & $0.62$ & 50 \\
        Fig. 4c&Fiber  &  $4 \times 10^{7}$ & 300 & $-18.7$ & $-22.44$ & $ 5/6$ & 50 \\
        Fig. 4c & Fiber  &  $4 \times 10^{7}$ & 300 & $-18.7$ & $1$ &  & 50 \\
        Fig. 4c & Fiber  &  $4 \times 10^{7}$ & 300 & $1$ & $-43.6$ &  & 50 \\
        Fig. S2h & 3-fold vortex &  $4 \times 10^{7}$ & 400 & $-18.7$ & $-38.9583$ & $0.48$ & 50 \\
        Fig. S2h & 3-fold vortex  &  $4 \times 10^{7}$ & 400 & $-18.7$ & $-35.9615$ & $0.52$ & 50 \\
        Fig. 4f &4-fold vortex  &  $4 \times 10^{7}$ & 400 & $-18.7$ & $-28.9$ &$0.647$ & 10\\
        Fig. 4f & 4-fold vortex &  $4 \times 10^{7}$ & 400 & $-6.7$ & $ -10.3545$ & $0.647$ & 10 \\
        Fig. 4g &Non-size-controlled design  &  $4 \times 10^{7}$ & 400 & $-18.7$ & $-34.63$ & $0.54$ & 10 \\
        Fig. 4g &Non-size-controlled design  &  $4 \times 10^{7}$ & 400 & $-18.7$ & $-31.16$ & $0.6$ & 10 \\
        Fig. 4e &Patterned bulk &  $4 \times 10^{7}$ & 400 & $-18.7$ & $-49.8667$ & $0.375$ & 10 \\
        Fig. 4e &Patterned bulk  &  $4 \times 10^{7}$ & 400 & $-18.7$ & $-26.0174$ & $0.719$ & 10 \\
        Fig. 4e &Patterned bulk  &  $4 \times 10^{7}$ & 400 & $-18.7$ & $-22.8203$ & $0.819$ & 10 \\
        Fig. 4e & Patterned bulk  &  $4 \times 10^{7}$ & 400 & $-18.7$ & $-18.7$ & $1$ & 10 \\
        Fig. 4h & 3-dimensional design & $6.64 \times 10^8$ & 1728 & -15 & -40.3169 & 0.3720 & 100 \\
        Fig. 4h & 3-dimensional design  & $6.64 \times 10^8$ & 1728 & -15 & -31.8095 & 0.4716 & 100 \\
        Fig. 4h & 3-dimensional design & $6.64 \times 10^8$ & 1728 & -15 & -28.7625 & 0.5215 & 100\\
        Fig. 4h &3-dimensional design & $6.64 \times 10^8$ & 1728 & -15 & -25.8621 & 0.58 & 100\\
        \hline
    \end{tabular}\\
    \caption{\textbf{Simulation Parameters.}\\ The interaction energies $\epsilon_c, \epsilon_d$ and $\epsilon_f$ are expressed in units of $k_BT$. In our simulations we associate a large positive ``forbidden interaction'' $\epsilon_f$ to all interfaces that are neither of the crystalline nor of the defect type to prevent them from occurring. $A$ is the number of annealing steps per temperature.
    }
    \label{tab:simu_parameters}
\end{table}

\end{document}